\documentclass{aa}  
\usepackage{graphicx} 
\usepackage{xcolor}
\usepackage{verbatim} 
\usepackage{multirow} 
\usepackage{multicol} 
\usepackage{manyfoot}
\usepackage{subcaption}
\usepackage{appendix}
\usepackage[colorlinks=true]{hyperref}
\hypersetup{linkcolor=blue,citecolor=blue,filecolor=blue,urlcolor=blue}
\usepackage{txfonts}

\begin{document}

   \title{Spectral Energy Distribution Modeling of Broad Emission Line Quasars: From X-ray to Radio Wavelengths}

   \author{Avinanda Chakraborty\inst{1,2}, Maitreya Kundu\inst{2}, Suchetana Chatterjee\inst{2}, Swayamtrupta Panda\inst{3,4,\thanks{Gemini Science Fellow},\thanks{CNPq Fellow}}, Arijit Sar\inst{2},\ Sandra Jaison\inst{5},\  Ritaban Chatterjee\inst{2}}
          
   \institute{$^{1}$INAF – Osservatorio Astrofisico di Arcetri, Largo E. Fermi 5., 50125, Firenze, Italy\\
              \email{avinanda.chakraborty@inaf.it}\\
             $^{2}$School of Astrophysics, Presidency University, Kolkata, 700073, India\\
             $^{3}$International Gemini Observatory/NSF NOIRLab, Casilla 603, La Serena, Chile \\
             $^{4}$Laborat\'orio Nacional de Astrof\'isica, MCTI, Rua dos Estados Unidos, 154, Bairro das Na\c c\~oes, Itajub\'a, MG 37501-591, Brazil\\
             $^{5}$Department of Physics, Presidency University, Kolkata, 700073, India\\
             }

    \authorrunning{Chakraborty et al.}

  \abstract
{}
   { We study the differences in physical properties of quasar-host galaxies using an optically selected sample of radio loud (RL) and radio quiet (RQ) quasars (in the redshift range $0.15 \le z \le 1.9$) which we have further cross-matched with the VLA-FIRST survey catalog. The sources in our sample have broad H$\beta$ and Mg\textsc{ii} emission lines (1000 km/s $<$ FWHM $<$ 15000 km/s) with a subsample of high broad line quasars (FWHM $>$ 15000 km/s). We construct the broadband spectral energy distribution (SED) of our broad line quasars using multi-wavelength archival data and targeted observations with the \textit{AstroSat} telescope. }
   {We use the state-of-the-art SED modeling code \texttt{CIGALE v2022.0}  to model the SEDs and determine the best-fit physical parameters of the quasar host galaxies namely their star-formation rate (SFR), main-sequence stellar mass, luminosity absorbed by dust, e-folding time and stellar population age. }
   {We find that the emission from the host galaxy of our sources is between 20\%-35\% of the total luminosity, as they are mostly dominated by the central quasars. Using the best-fit estimates, we reconstruct the optical spectra of our quasars which show remarkable agreement in reproducing the observed SDSS spectra of the same sources. We plot the main-sequence relation for our quasars and note that they are significantly away from the main sequence of star-forming galaxies. Further, the main sequence relation shows a bimodality for our RL quasars indicating populations segregated by Eddington ratios.}
   {We conclude that RL quasars in our sample with lower Eddington ratios tend to have substantially lower star-formation rates for similar stellar mass. Our analyses, thus, provide a completely independent route in studying the host galaxies of quasars and addressing the radio dichotomy problem from the host galaxy perspective. }


   \maketitle
%

\section{Introduction}

   Quasars are the most luminous and the most distant members of the active galactic nuclei (AGN) population. Although, quasars were first discovered through their strong radio jet emission \citep{schmidt}, roughly 10\% of the quasars are considered as powerful radio sources termed as radio-loud quasars (RLQ) \citep[][]{sandage, stritt, schgr, kell, mill93, ive, jia, rcw}. The physical reason behind the existence of this radio dichotomy remains an unsolved problem in quasar physics \citep[][]{stritt, kell, mill, ive, whit, zam}.

Various factors have been proposed to explain the radio-loud (RL)--radio-quiet (RQ) divide, such as the mass and spin of the central black hole \citep{Ross}, the different physical origin of radio emission \citep{Behar}, accretion rates \citep{sikora, ham, marzi21} and overall spectral energy distribution \citep{Laor_etal_1997, Marziani_etal_2023Galax}. However, none of the above studies have been conclusive. One approach to address this radio dichotomy involves studying their host galaxy properties \citep[][]{sikora, lag, kim}. The hosts of quasars are often star-forming galaxies \citep[see review by][]{heck}. Synchrotron radiation from electrons accelerated to relativistic speeds in supernova remnants as well as free-free emission from HII regions are produced in star-formation regions \citep{con}. It has been found that radio continuum emission at low frequencies in low-luminosity quasars is consistent with being dominated by star-formation \citep{gur}. Further, in contrast with previous studies at lower redshifts \citep[e.g.,][]{McLDu}, \cite{fal} and \cite{kauff} showed that RL AGN appears to be found in denser environments than their RQ counterparts. This is in qualitative agreement with other works on higher redshift quasars \citep{Kalfountzou12}.

It was realized from early studies that the cosmic star-formation history \citep[e.g.,][]{madau, hop} and the evolving luminosity density of quasars \citep[e.g.,][]{boy, rich, croom} follow a similar trend \citep[e.g.,][]{france}. Supporting evidence of the presence of cold gas \citep[e.g.,][]{evan, scov, walter, emon}, dust \citep{arch, page, reu, stev} and young stars \citep[e.g.,][]{tad, baldi, her} in powerful RL quasars, contrary to the composition of normal ellipticals that usually host RL quasars \citep[e.g.,][]{dunlop}.

Many studies have also attempted multi-band optical photometry \citep[e.g.,][]{san} or spectroscopy \citep[e.g.,][]{tri10, tri12, kal} to determine the star-formation activity in quasar host galaxies. At lower redshift, the spectrophotometric data from SDSS for well-defined samples of radio galaxies has also been used to investigate differences in star-formation activity and environments of quasar host galaxies \citep[e.g.,][]{kauf3, kauf4, best5, best12}. It was found that the environment plays a relatively minor role if the radio-loudness is due to the physics of the central engine and how it is fuelled. However, the quasar properties may be connected with the star-formation in their host galaxies \citep[e.g.,][]{herb, croft, silkJ}. 

\begin{figure}
\centering
\includegraphics[width=\columnwidth]{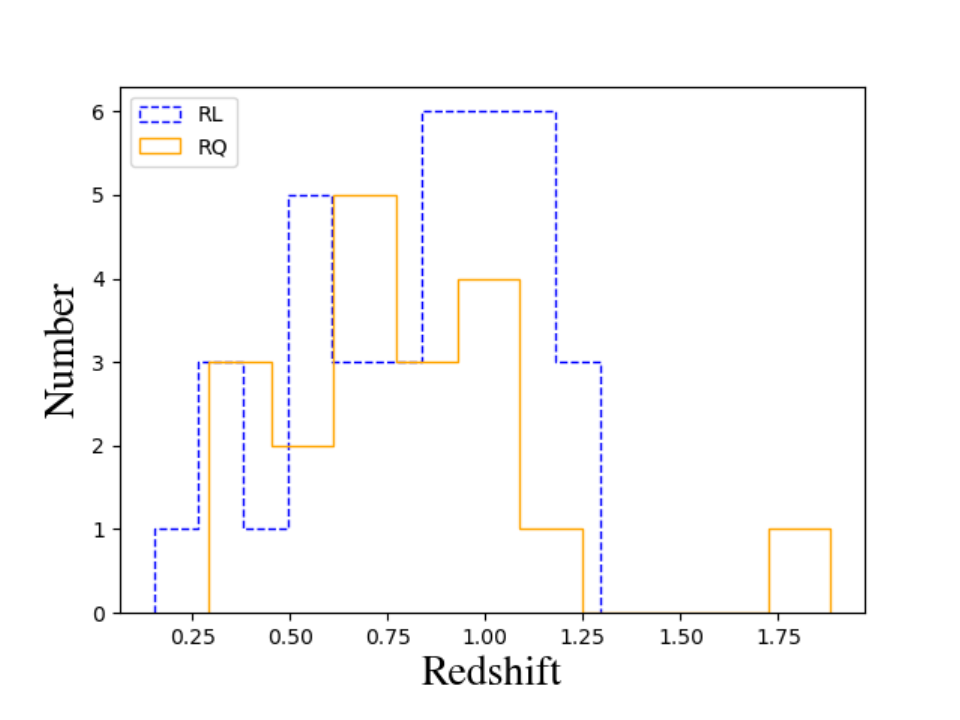}
\caption{Histograms of redshift for the non-HBL+HBL+\textit{AstroSat}-observed radio-loud (blue dashed) and radio-quiet (orange solid) sources in our sample. Our datasets are described in \S 2.1.} 
\label{fig:redshift}
\end{figure}

\begin{figure*}[h]
\begin{center}
\begin{tabular}{c}
\resizebox{6cm}{!}{\includegraphics{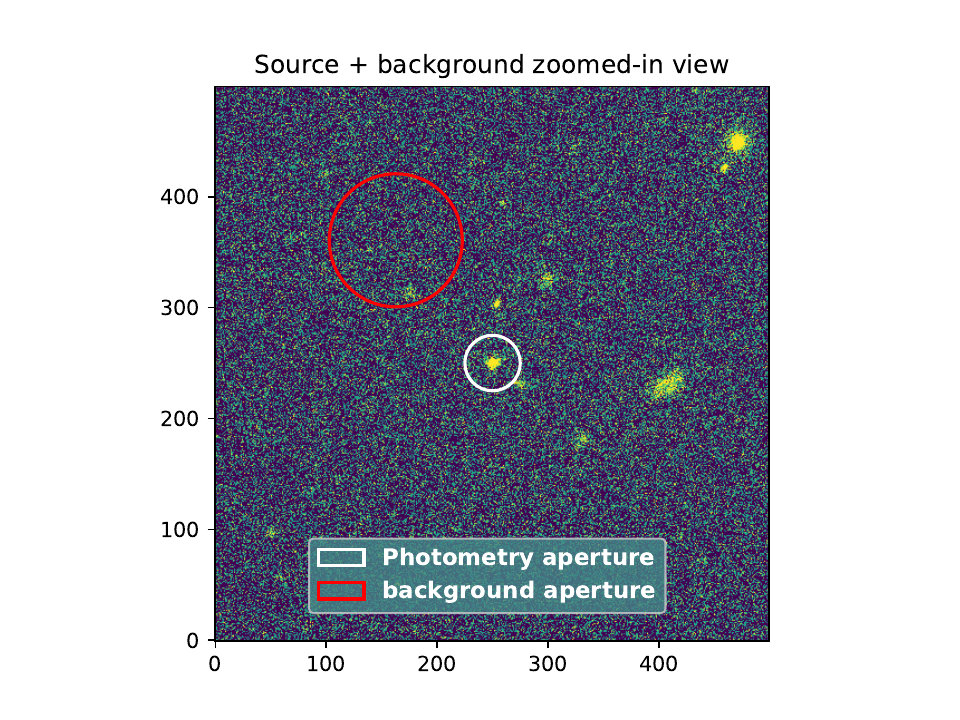}}
\resizebox{6cm}{!}{\includegraphics{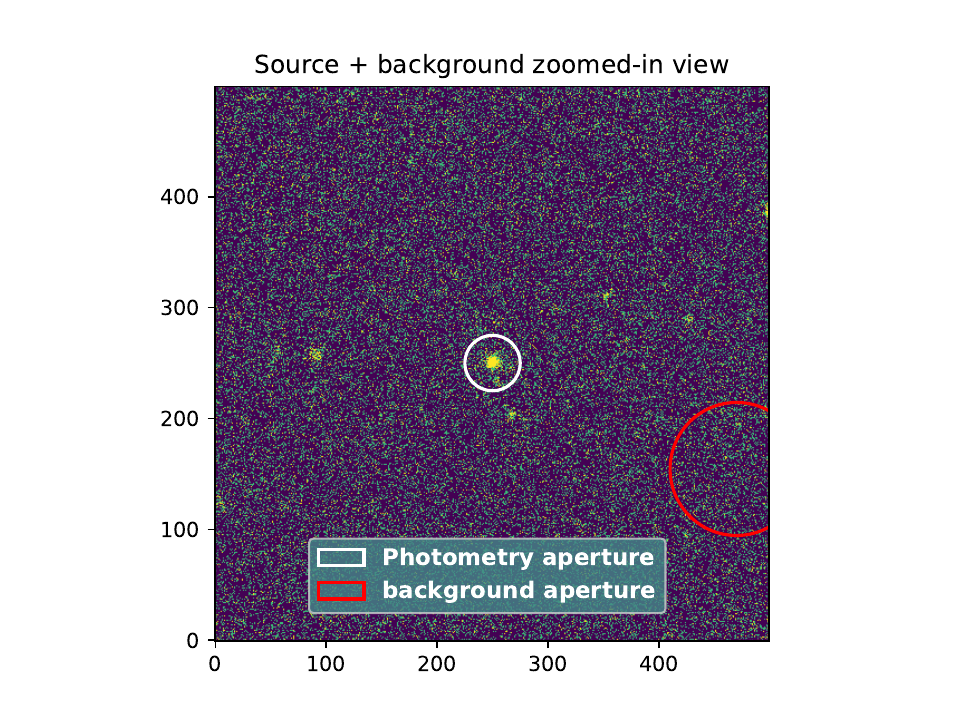}}
\resizebox{6cm}{!}{\includegraphics{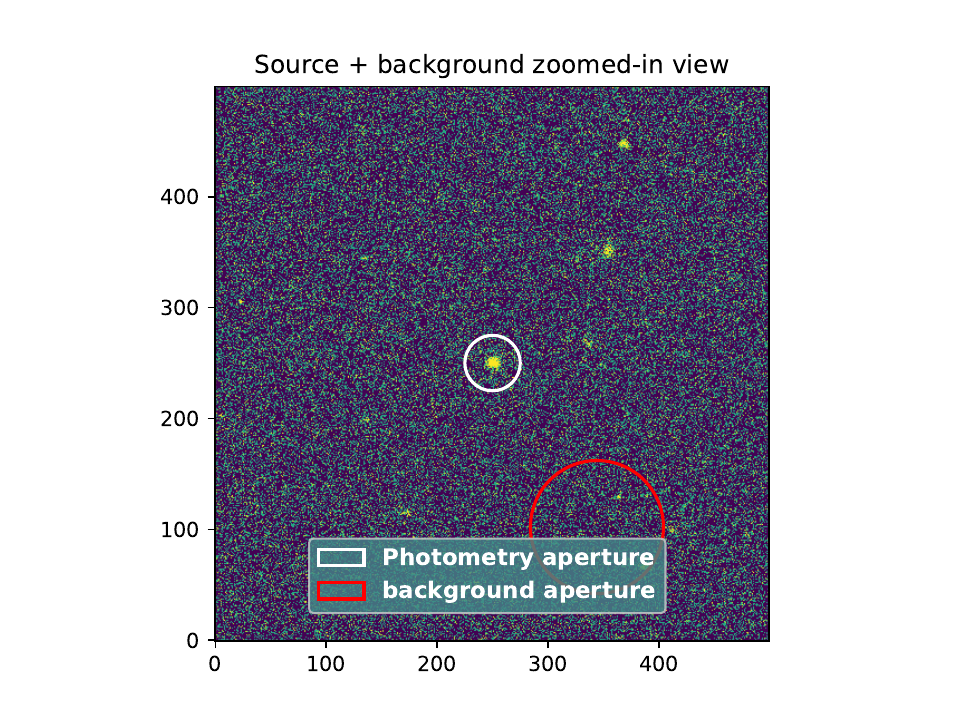}}\\
\resizebox{6cm}{!}{\includegraphics{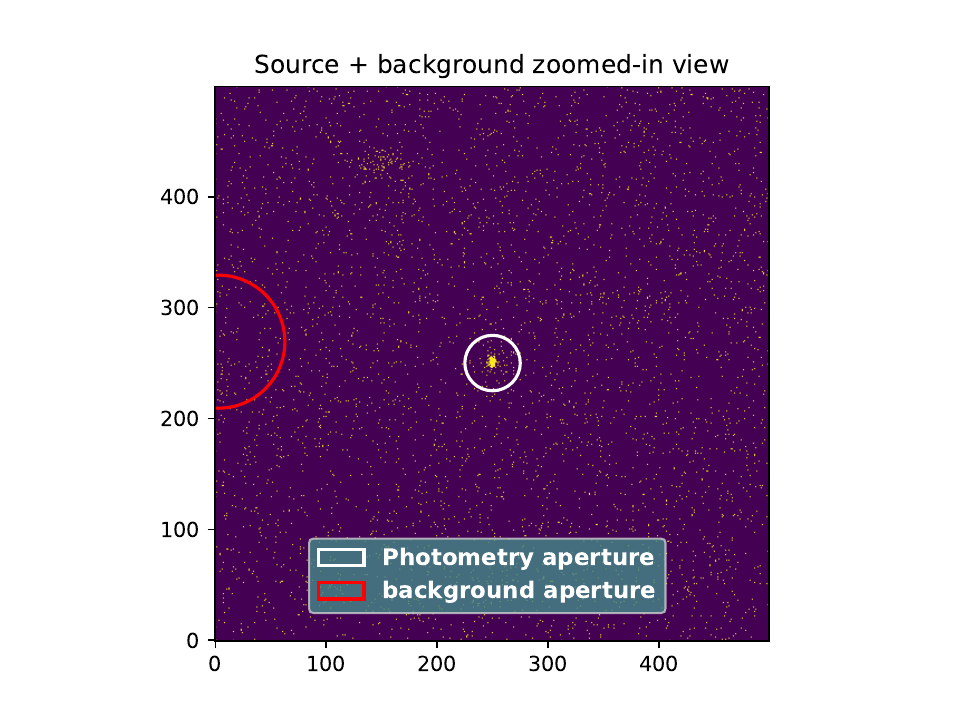}}
\resizebox{6cm}{!}{\includegraphics{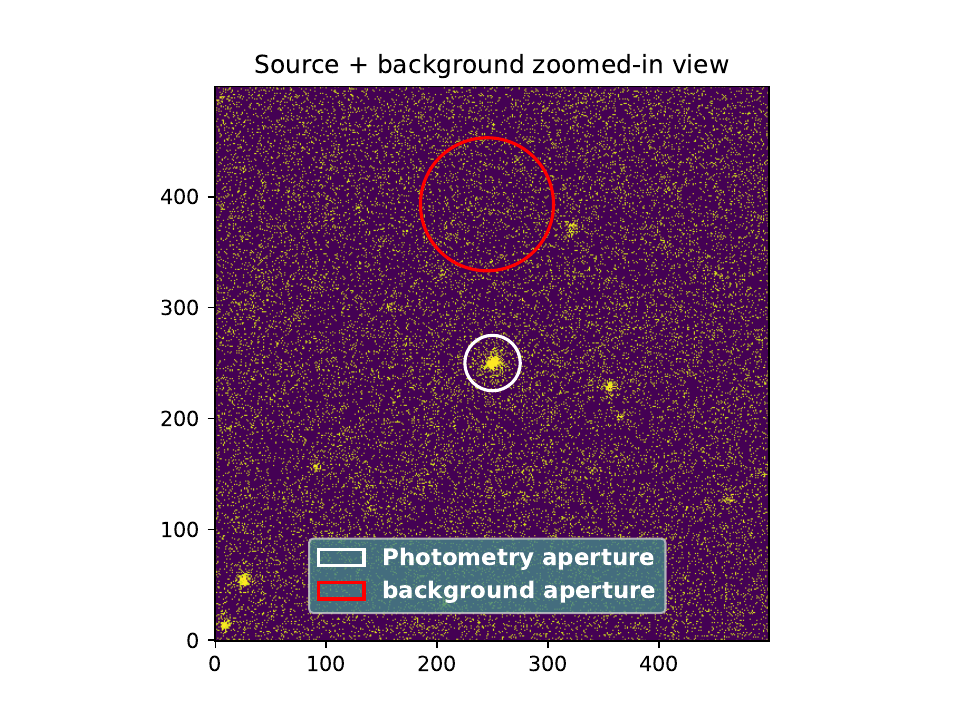}}
\resizebox{6cm}{!}{\includegraphics{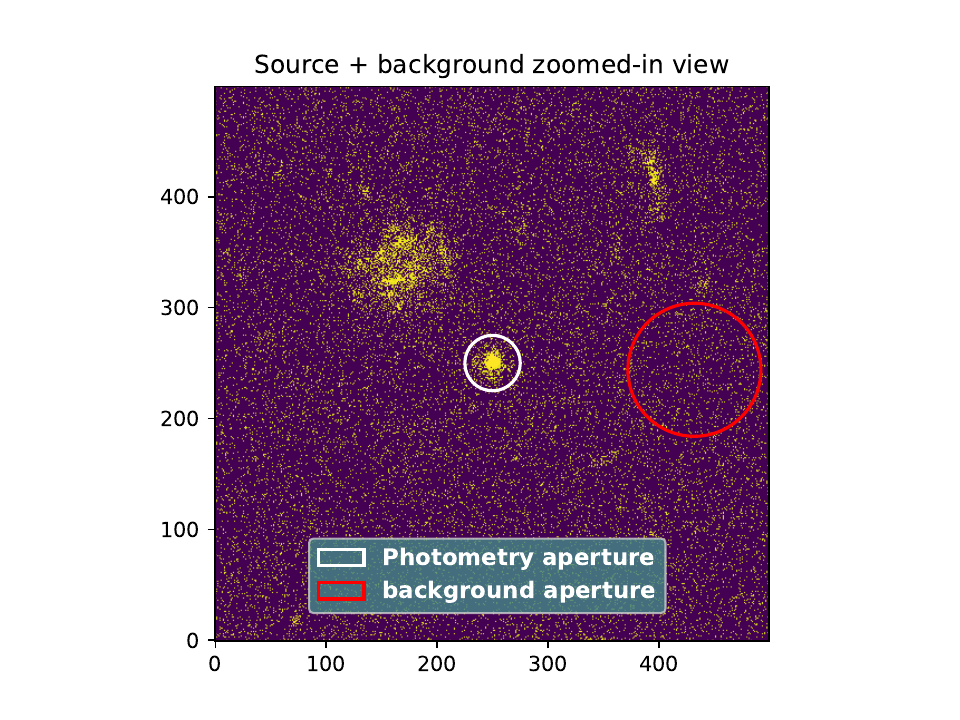}}
\end{tabular}
\caption{Photometry of the six sources observed with \textit{AstroSat}/UVIT. {\bf Top Panel: } Zoomed-in view of 3 RQ sources. {\bf Bottom Panel: } Zoomed-in view of 3 RL sources.} 
\label{fig:AstroSat photometry}
\end{center}
\end{figure*}
\begin{table*}[h]
    \caption{The six sources observed with \textit{AstroSat}/UVIT, using the ${\rm CaF}_2$ grating centered at $1481~\AA$ (the FUV Band). For more details about the analysis of these sources, see \S\ref{subsec:AstroSat}.} 
    \centering
    
    \begin{tabular}{c| c c c} 
    \hline \hline
        & Source Name & Magnitude (AB system) & $f_{\lambda}$ $[{\rm erg}\cdot{\rm cm}^{-2}\cdot{\rm s}^{-1}\cdot{\AA}^{-1}]$ \\
    \hline \hline
      \multirow{ 3}{*}{RQ}  & SDSS 075403.60+481428.0 & $18.983 \pm 0.191$ & $1.268 \times 10^{-15} \pm 1.168 \times 10^{-17}$ \\ 
       & SDSS 145331.47+264946.7 & $19.422 \pm 0.087$ & $8.459 \times 10^{-16} \pm 7.791 \times 10^{-18}$ \\ 
       & SDSS 161742.53+322234.3 & $18.555 \pm 0.059$ & $1.879 \times 10^{-15} \pm 1.731 \times 10^{-17}$\\ 
       \hline
       \multirow{ 3}{*}{RL} & SDSS 084957.04+234635.8 & $21.151 \pm 0.150$ & $1.721 \times 10^{-16} \pm 1.585 \times 10^{-18}$ \\ 
       & SDSS 085605.83+450520.0 & $20.558 \pm 0.095$ & $2.792 \times 10^{-16} \pm 2.737 \times 10^{-18}$\\ 
       &SDSS 163306.86+310641.0 & $20.796 \pm 0.115$ & $2.386 \times 10^{-16} \pm 2.198 \times 10^{-18}$\\ 
    \hline 
    \label{tab:AstroSat}
    \end{tabular}

\end{table*}

The complex physical interplay between the main baryonic components of galaxies like stars, their remnants, molecular, atomic, and ionized gas, dust, and supermassive black holes cause multi-wavelength emission ranging from $\gamma$–rays to the radio domain \citep[see e.g.][]{Harrison_2014PhDT, Netzer_2015_ARAA}. So the typical spectral energy distribution (SED) of a galaxy that covers a broad wavelength range, from X-ray to infrared (IR), contains the imprint of the baryonic processes that drove its formation and evolution across cosmic times. In other words, it is important to extract the information tightly woven into the SED of galaxies across a broad range of wavelengths to understand galaxy formation and evolution. Modeling the SED of galaxies is a heavily intricate problem as it involves AGN contribution as well. Galaxies with very different properties can have broadly similar SEDs and misinterpretation of the SED could lead to unrealistic physical conclusions. This is particularly the case when we do not consider the full SED. 

In our previous studies, using the Sloan Digital Sky Survey Data Release 7 quasar catalog we selected a sample of RL and RQ quasars through their broad H$\beta$ and Mg\textsc{ii}~ FWHM \citep{chakraborty21,chakraborty21b,chakraborty22}. Our results revealed that the radio loud fraction (RLF) increases with FWHM for both broad H$\beta$ and Mg\textsc{ii} samples. To understand this effect, we further selected a sub-sample with FWHM greater than $15,000$ km\,s$^{-1}$ (HBL hereafter) from both H$\beta$ and Mg\textsc{ii} samples and performed a systematic study of the different AGN properties, namely bolometric luminosity, black hole (BH) mass, optical continuum luminosity, and Eddington ratio. To search for possible reasons for higher RLF we further constructed and compared the composite SDSS spectra of the RL and RQ quasars for our H$\beta$ sample. We also compared the BH mass, Eddington ratio, bolometric luminosity, H$\beta$ FWHM, and $R_{Fe}$ (i.e., the flux ratio of the optical Fe {\sc ii} emission within 4434-4684\AA\ to the broad H$\beta$ emission) of our sample with the rest of the H$\beta$ broad emission line objects i.e., FWHM less than $15,000$ km\,s$^{-1}$ (non-HBL hereafter). We found that the luminosities are different but BH mass and Eddington ratios are similar for HBL RL and RQ. The accretion disc luminosity of the RL quasars in our HBL sample is higher, which indicates a connection between a brighter disc and a more prominent jet. By comparing them with the non-HBL broad emission line quasars, we found that the HBL sources have the lowest Eddington ratios in addition to having a very high RLF. We also found that the [O III] narrow emission line is stronger in the RL compared to the RQ quasars in our HBL sample.

    In addition to their intrinsic differences, we anticipate the presence of a radio jet to have a distinct effect on their host galaxies. In this work, we compare the physical properties of the RL and RQ quasar-host galaxies of broad H$\beta$ and Mg\textsc{ii} samples by modeling their spectral energy distributions (SED) with the \texttt{CIGALE v2022.0}\footnote{\bf \url{https://cigale.lam.fr/}} code \citep{yang22}. This analysis allows us to examine the radio dichotomy problem from the perspective of host galaxies of our RL and RQ quasars. The paper is organized as follows: In \S 2 we lay out the construction of our data set and discuss our methodology for modeling the broad-band SED of our sources. In \S 3 we report our results. Finally, we discuss our results and summarize our main conclusions in \S 4. 

\begin{sidewaystable*}
\caption{Data of RL quasars} 
\centering
\resizebox{25cm}{!}{%
\begin{tabular}{c c c c c c c c c c c c c c c c c c c c c c c c c c c c c} 
\hline 
\hline\\ 
SDSS object ID & Redshift & FWHM & VLA FIRST & VLA FIRST & GALEX & GALEX & GALEX & GALEX & SDSS & SDSS & SDSS & SDSS & SDSS & SDSS & SDSS & SDSS & SDSS & SDSS & WISE-1 & WISE-1 & WISE-2 & WISE-2 & WISE-3 & WISE-3 & WISE-4 & WISE-4 & X-ray & X-ray \\ [0.5ex]
& & [km/s] & flux & error & FUV flux & FUV error & NUV flux & NUV error & $u$-band & $u$-band & $g$-band & $g$-band & $r$-band & $r$-band & $i$-band & $i$-band & $z$-band & $z$-band & flux & error & flux & error & flux & error & flux & error & flux & error\\
& & & [mJy] & [mJy] & [mJy] & [mJy] & [mJy] & [mJy] & flux [mJy] & error [mJy] & flux [mJy] & error [mJy] & flux [mJy] & error [mJy] & flux [mJy] & error [mJy] & flux [mJy] & error [mJy] & [mJy] & [mJy] & [mJy] & [mJy] & [mJy] & [mJy] & [mJy] & [mJy] & [$10^{-5}$ mJy] & [$10^{-5}$ mJy]\\
\hline
\hline\\
024940.19-083426.4	& 1.1927 & 3739.1 & 49.67 & 7.0194 & 0.0114 & 0.0012 & 0.0982 & 0.0098 & 0.0809 & 0.0081 & 0.1084 & 0.0111 & 0.1684 & 0.0171 & 0.1695 & 0.0170 & 0.1765 & 0.0179 & 0.4785 & 0.0482 & 0.8602 & 0.0867 & 2.4317 & 0.2501 & 5.7645 & 0.6734 & 1.8421 & 0.2482 \\ 
\hline\\
030313.02-001457.4 & 0.6999 & 3091.9 & 761.31 & 107.6648 & 0.0581 & 0.0064 & 0.1247 & 0.0127 & 0.1136 & 0.0114 & 0.1581 & 0.0159 &  0.1909 & 0.0191 & 0.1823 & 0.0183 & 0.2204 & 0.0222 & 0.7598 & 0.0766 & 1.296 & 0.1307 & 4.3083 & 0.4374 & 11.6081 & 1.2435 & 1.8421 & 0.3091 \\
\hline\\
030458.96+000235.7 & 0.5637 & 6515.0 & 56.1 & 7.9337 & 0.047 & 0.0052 & 0.0832 & 0.0086 & 0.1253 & 0.0126 & 0.1828 & 0.0183 & 0.1808 & 0.0181 & 0.1999 & 0.0200 & 0.1855 & 0.0187 & 0.8911 & 0.0897 & 1.4041 & 0.1412 & 3.6636 & 0.3742 & 7.9134 & 0.9228 & 8.0309 & 0.8268\\ 
\hline\\
074417.47+375317.2 & 1.0669 & 7717.1 & 1244.87 & 176.05 & 0.0319 & 0.0032 & 0.1556 & 0.0156 & 0.1501 & 0.0151 & 0.19 & 0.0190 & 0.258 & 0.0258 & 0.2496 & 0.0249 & 0.234 & 0.0235 & 0.7216 & 0.0727 & 1.1657 & 0.1174 & 2.8517 & 0.2932 & 6.9946 & 0.8157 & 3.3193 & 0.3319\\ 
\hline\\ 
081002.69+502538.6 & 1.2057 & 7218.6 & 20.29 & 2.86 & 0.0425 & 0.0050 & 0.2466 & 0.0249 & 0.4907 & 0.0492 & 0.5186 & 0.0519 & 0.7475 & 0.0749 & 0.7863 & 0.0789 & 0.7178 & 0.0722 & 1.2703 & 0.1277 & 2.2813 & 0.2307 & 5.0618 & 0.5386 & 12.4383 & 1.3237 & 12.2002 & 2.198 \\
\hline\\
081108.77+453348.9 & 1.0165 & 2484.9 & 83.85 & 11.85 & 0.0166 & 0.0023 & 0.0421 & 0.0047 & 0.0606 & 0.0061 & 0.0806 & 0.0081 & 0.1006 & 0.0100 & 0.0953 & 0.0096 & 0.0987 & 0.0101 & 0.1449 & 0.0149 & 0.1547 & 0.0167 & 0.6086 & 0.0822 & 2.5349 & 0.4578 & 3.8944 & 0.6204 \\
\hline\\
081525.93+363515.0 & 1.0286 & 5609.7 & 783.52 & 110.80 & 0.0048 & 0.0008 & 0.053 & 0.0056 & 0.0762 & 0.0077 & 0.1015 & 0.0102 & 0.1489 & 0.0149 & 0.1343 & 0.0134 & 0.1401 & 0.0141 & 0.6125 & 0.0617 & 1.0323 & 0.1040 & 3.1268 & 0.3196 & 7.001 & 0.8238 & 3.0271 & 2.1532 \\ 
\hline\\
082231.53+231152.0 & 0.653 & 15398.4 & 29.45 & 2.95 & 0.0158 & 0.0022 & 0.1148 & 0.0119 & 0.1549 & 0.0155 & 0.2457 & 0.0246 & 0.2995 & 0.0299 & 0.3525 & 0.0353 & 0.3668 & 0.0369 & 1.7247 & 0.1733 & 2.6631 & 0.2674 & 4.7632 & 0.4818 & 10.7143 & 1.1921 & NA & NA \\ 
\hline\\
083740.24+245423.1 & 1.1254 & 2891.6 & 446.21 & 63.1029 & 0.0192 & 0.0028 & 0.1138 & 0.0118 & 0.1626 & 0.0163 & 0.1788 & 0.0179 & 0.2297 & 0.0230 & 0.1961 & 0.0196 & 0.2109 & 0.0212 & 0.2932 & 0.0297 & 0.4252 & 0.0436 & 1.3228 & 0.1493 & 4.066 & 0.6057 & 9.9188 & 1.2391 \\ 
\hline\\ 
084028.33+323229.2 & 1.1005 & 6464.3 & 3.11 & 0.4391 & 0.0113 & 0.0019 & 0.034 & 0.0038 & 0.1029 & 0.0103 & 0.1086 & 0.0109 & 0.126 & 0.0126 & 0.1167 & 0.0117 & 0.1075 & 0.0109 & 0.175 & 0.0179 &  0.2996 & 0.0308 & 0.5925 & 0.0831 & 2.7262 & 0.2936 & 0.1001 & 0.0927 \\
\hline\\
093113.87+600629.0 & 0.6271 & 17104.0 & 4.93 & 0.4955 & 0.0258 & 0.0026 & 0.053 & 0.0053 & 0.0417 & 0.0042 & 0.0694 & 0.0069 & 0.0736 & 0.0074 & 0.0857 & 0.0085 & 0.0876 & 0.0089 & 0.686 & 0.0689 & 0.9311 & 0.0937 & 1.3649 & 0.1429 & 2.1457 & 0.4202 & NA & NA \\
\hline\\
093241.14+530633.7 & 0.5969 & 2182.4 & 411.43 & 58.18 & 0.0581 & 0.0065 & 0.1247 & 0.0128 & 0.0672 & 0.0068 & 0.0959 & 0.0096 & 0.1001 & 0.0101 & 0.121 & 0.0121 & 0.1152 & 0.0117 & 0.3944 & 0.0397 & 0.5688 & 0.0573 & 1.6562 & 0.1709 & 5.3011 & 0.5899 & 3.7651 & 0.3765\\
\hline\\ 
094735.08+583046.4 & 0.9372 & 4270.3 & 171.93 & 24.31 & 0.0802 & 0.0093 & 0.2559 & 0.0259 & 0.248 & 0.0248 & 0.2714 & 0.0272 & 0.2992 & 0.0300 & 0.2724 & 0.0274 & 0.297 & 0.0298 & 0.769 & 0.0773 & 1.2653 & 0.1272 & 2.6056 & 0.2647 & 6.0752 & 0.6916 & 6.7311 & 0.7204 \\
\hline\\
095455.55+571952.7 & 0.9812 & 5596.0 & 77.14 & 10.91 & 0.0124 & 0.0021 & 0.0809 & 0.0085 & 0.0725 & 0.0073 & 0.0818 & 0.0082 & 0.1012 & 0.0102 & 0.0915 & 0.0092 & 0.0948 & 0.0096 & 0.2495 & 0.0252 & 0.4098 & 0.0415 & 1.1921 & 0.1262 & 3.2476 & 0.4439 & 2.8949 & 0.2895 \\
\hline\\
101703.49+592428.7 & 0.8503 & 3699.2 & 323.53 & 45.75 & 0.0982 & 0.0112 & 0.2377 & 0.0241 & 0.2475 & 0.0248 & 0.2932 & 0.0294 & 0.3003 & 0.0301 & 0.2821 & 0.0283 & 0.3003 & 0.0301 & 0.9816 & 0.0988 & 1.7163 & 0.1725 & 3.7558 & 0.3806 & 5.8125 & 0.6755 & 16.5367 & 1.9555 \\
\hline\\
102738.53+605016.4 & 0.332 & 38343.3 & 4.85 & 0.4875 & 0.1459 & 0.0155 & 0.2466 & 0.0249 & 0.2726 & 0.0274 & 0.3324 & 0.0333 & 0.3996 & 0.0401 & 0.4215 & 0.0423 & 0.5921 & 0.0593 & 1.4612 & 0.1469 & 1.9028 & 0.1913 & 3.2832 & 0.3351 & 5.7327 & 0.6894 & NA & NA \\
\hline\\
103250.73+631232.0 & 1.1753 & 6870.6 & 1.16 & 0.16 & 0.0062 & 0.0014 & 0.0302 & 0.0036 & 0.0259 & 0.0026 & 0.0349 & 0.0035 & 0.0521 & 0.0052 & 0.0553 & 0.0056 & 0.0617 & 0.0064 & 0.1494 & 0.0152 & 0.2161 & 0.0223 & 0.5581 & 0.0722 & 1.6397 & 0.3503 & 1.1727 & 0.6183 \\
\hline\\
143009.18-005319.2 & 0.8078 & 7169.2 & 3.45 & 0.4844 & 0.0938 & 0.0104 & 0.2051 & 0.0209 & 0.1067 & 0.0107 & 0.1647 & 0.0165 & 0.2222 & 0.0227 & 0.2452 & 0.0246 & 0.309 & 0.0310 & 0.7577 & 0.0763 & 0.9894 & 0.0997 & 2.1179 & 0.2236 & 4.617 & 0.6648 & 4.6849 & 0.6243 \\
\hline\\
143244.43-005915.1 & 1.0264 & 7067.8 & 16.53 & 2.34 & 0.0344 & 0.0044 & 0.1923 & 0.0196 & 0.2759 & 0.0277 & 0.3451 & 0.0469 & 0.4504 & 0.0485 & 0.455 & 0.0462 & 0.4293 & 0.0431 & 0.7371 & 0.0755 & 1.1277 & 0.1155 & 1.7262 & 0.1972 & 3.584 & 0.3860 & 5.9618 & 0.5962 \\
\hline\\
152544.71+354446.9 & 1.1007 & 5010.9 & 14.77 & 2.08 & 0.0124 & 0.0016 & 0.0421 & 0.0044 & 0.0666 & 0.0067 & 0.0759 & 0.0076 & 0.0957 & 0.0096 & 0.0826 & 0.0083 & 0.073 & 0.0075 & 0.1749 & 0.0178 & 0.3304 & 0.0339 & 0.9152 & 0.1246 & 3.0872 & 0.7237 & 3.5243 & 0.3524\\
\hline\\
152949.76+394509.6 & 1.0813 & 5382.9 & 114.87 & 16.24 & 0.0291 & 0.0035 & 0.0955 & 0.0099 & 0.055 & 0.0055 & 0.0602 & 0.0060 & 0.0825 & 0.0083 & 0.0838 & 0.0084 & 0.0669 & 0.0067 & 0.1888 & 0.0189 & 0.3075 & 0.0309 & 0.5354 & 0.0549 & 1.664 & 0.2221 & 4.9194 & 0.7071 \\
\hline\\
153404.87+482340.9 & 0.543 & 3193.8 & 308.72 & 43.66 & 0.0511 & 0.0083 & 0.1107 & 0.0121 & 0.0886 & 0.0089 & 0.1339 & 0.0134 & 0.1365 & 0.0137 & 0.1764 & 0.0177 & 0.1936 & 0.0196 & 1.0595 & 0.1076 & 1.4581 & 0.1479 & 2.4451 & 0.2675 & 3.681 & 0.7644 & 3.5063 & 0.3506 \\
\hline\\
153432.56+492049.1 & 1.2938 & 5846.5 & 62.75 & 8.8706 & 0.0127 & 0.0015 & 0.0555 & 0.0057 & 0.1262 & 0.0127 & 0.1014 & 0.0102 & 0.1447 & 0.0145 & 0.1478 & 0.0148 & 0.1286 & 0.0129 & 0.1995 & 0.0201 & 0.3925 & 0.0395 & 0.9879 & 0.1016 & 1.7962 & 0.2278 & 4.0278 & 0.4028 \\
\hline\\
153818.57+410548.3 & 0.4812 & 5049.0 & 50.37 & 7.12 & 0.0649 & 0.0078 & 0.1294 & 0.0129 & 0.1112 & 0.0112 & 0.1384 & 0.0139 & 0.1169 & 0.0118 & 0.1385 & 0.0139 & 0.1598 & 0.0161 & 0.6153 & 0.0619 & 0.966 & 0.0971 & 2.1953 & 0.2221 & 5.1899 & 0.5544 & 6.5631 & 0.6563 \\
\hline\\
160623.56+540555.7 & 0.8763 & 3752.1 & 124.7 & 12.53 & 0.0809 & 0.0088 & 0.2109 & 0.0230 & 0.2375 & 0.0239 & 0.307 & 0.0308 & 0.3248 & 0.0326 & 0.3048 & 0.0306 & 0.3154 & 0.0317 & 0.8324 & 0.0838 & 1.3734 & 0.1383 & 3.2591 & 0.3326 & 8.4326 & 0.9368 &  3.6681 & 0.3668 \\
\hline\\
160913.18+535429.6 & 0.9926 & 7413.3 & 43.82 & 6.19 & 0.0802 & 0.0081 & 0.1629 & 0.0163 & 0.1262 & 0.0127 & 0.1632 & 0.0164 & 0.2188 & 0.0219 & 0.2285 & 0.0229 & 0.227 & 0.0229 & 0.5384 & 0.0543 & 0.7575 & 0.0766 & 2.0545 & 0.2145 & 5.8286 & 0.9033 & 5.2152 & 0.5962 \\
\hline\\
161301.86+524749.5 & 0.5279 & 1650.4 & 1.35 & 0.19 & 0.057 & 0.0091 & 0.1067 & 0.0111 & 0.0985 & 0.0099 & 0.1108 & 0.0111 & 0.1208 & 0.0121 & 0.1338 & 0.0134 & 0.1318 & 0.0133 & 0.3561 & 0.0359 & 0.4303 & 0.0433 & 1.3941 & 0.1422 & 2.8998 & 0.3197 & 2.1089 & 0.4362 \\
\hline\\
162330.53+355933.1 & 0.8665 & 4960.7 & 266.83 & 37.73 & 0.0184 & 0.0019 & 0.092 & 0.0093 & 0.1247 & 0.0125 & 0.1644 & 0.0164 & 0.1683 & 0.0169 & 0.1474 & 0.0148 & 0.1552 & 0.0156 & 0.5469 & 0.0552 & 0.8982 & 0.0909 & 2.7359 & 0.2856 & 6.9496 & 0.9769 & 2.6791 & 0.2679 \\
\hline\\
162718.13+495511.8 & 0.9038 & 2956.6 & 211.21 & 29.86 & 0.035 & 0.0052 & 0.0839 & 0.0095 & 0.1223 & 0.0124 & 0.1265 & 0.0127 & 0.1406 & 0.0141 & 0.1262 & 0.0127 & 0.1369 & 0.0141 & 0.2805 & 0.0283 & 0.4248 & 0.0430 & 1.3326 & 0.1512 & 2.8468 & 0.5142 & 7.5200 & 0.8338 \\
\hline\\
163058.01+370733.4 & 0.8022 & 4980.0 & 3.59 & 0.50 & 0.0175 & 0.0018 & 0.0433 & 0.0043 & 0.063 & 0.0063 & 0.0754 & 0.0075 & 0.0752 & 0.0075 & 0.0667 & 0.0068 & 0.0773 & 0.0078 & 0.3311 & 0.0333 & 0.4287 & 0.0430 & 0.7542 & 0.0760 & 1.9354 & 0.2034 & 3.6876 & 0.6077 \\
\hline\\
163709.31+414030.8 & 0.7602 & 8788.7 & 7.14 & 1.0069 & 0.1706 & 0.0206 & 0.4325 & 0.0433 & 0.2968 & 0.0298 &  0.5021 & 0.0503 & 0.4769 & 0.0478 & 0.4337 & 0.0434 & 0.4747 & 0.0477 & 1.86 & 0.1872 & 2.9881 & 0.3008 & 6.0187 & 0.6224 & 10.3553 & 1.2206 & 21.7811 & 2.6339 \\
\hline\\
163955.97+470523.5 & 0.8596 & 4850.3 & 148.23 & 20.96 & 0.0039 & 0.0004 & 0.1067 & 0.0107 & 0.1288 & 0.0129 & 0.1768 & 0.0177 & 0.1883 & 0.0189 & 0.1667 & 0.0167 & 0.1767 & 0.0177 & 0.5043 & 0.0508 & 0.7568 & 0.0762 & 1.5542 & 0.1579 & 4.4214 & 0.4775 & 6.8999 & 0.7382 \\
\hline\\
164154.23+400033.0 & 1.0032 & 5686.0 & 5.06 & 0.7114 & 0.0425 & 0.0053 & 0.1406 & 0.0149 & 0.1932 & 0.0194 & 0.2401 & 0.0241 & 0.2721 & 0.0272 & 0.2682 & 0.0269 & 0.2871 & 0.0289 & 0.5062 & 0.0514 & 0.8884 & 0.0904 & 2.8074 & 0.3028 & 8.5185 & 1.4407 & 0.5167 & 0.0516 \\
\hline\\
170112.37+353353.3 & 0.5013 & 2386.8 & 54.03 & 7.64 & 0.031 & 0.0038 & 0.0608 & 0.0065 & 0.1264 & 0.0127 & 0.1582 & 0.0159 & 0.1389 & 0.0139 & 0.1545 & 0.0155 & 0.173 & 0.0174 & 0.6256 & 0.0636 & 1.038 & 0.1056 & 1.2494 & 0.1348 & 2.8104 & 0.4753 & 8.7748 & 0.8775 \\ 
\hline 
\hline
\label{tab:RL quasars}
\end{tabular}
}
\end{sidewaystable*}

\section{Methodology}
We now describe our methodology for the construction and modeling of the broad-band SEDs of our quasar sample. 

\begin{sidewaystable*}
\caption{Data of RQ quasars} 
\centering
\resizebox{25cm}{!}{%
\begin{tabular}{c c c c c c c c c c c c c c c c c c c c c c c c c c c} 
\hline 
\hline\\ 
SDSS object ID & Redshift & FWHM & GALEX & GALEX & GALEX & GALEX & SDSS & SDSS & SDSS & SDSS & SDSS & SDSS & SDSS & SDSS & SDSS & SDSS & WISE-1 & WISE-1 & WISE-2 & WISE-2 & WISE-3 & WISE-3 & WISE-4 & WISE-4 & X-ray & X-ray \\ [0.5ex]
& & [km/s] & FUV flux & FUV error & NUV flux & NUV error & $u$-band & $u$-band & $g$-band & $g$-band & $r$-band & $r$-band & $i$-band & $i$-band & $z$-band & $z$-band & flux & error & flux & error & flux & error & flux & error & flux & error\\
& & & [mJy] & [mJy] & [mJy] & [mJy] & flux [mJy] & error [mJy] & flux [mJy] & error [mJy] & flux [mJy] & error [mJy] & flux [mJy] & error [mJy] & flux [mJy] & error [mJy] & [mJy] & [mJy] & [mJy] & [mJy] & [mJy] & [mJy] & [mJy] & [mJy] & [$10^{-5}$ mJy] & [$10^{-5}$ mJy] \\
\hline
\hline\\
022804.36+010658.9 & 1.0244 & 5842.5 & 0.0445 & 0.0054 & 0.0895 & 0.0094 & 0.0908 & 0.0091 & 0.1097 & 0.0111 & 0.1405 & 0.0141 & 0.1289 & 0.0129 & 0.1319 & 0.0133 & 0.3777 & 0.0382 & 0.6371 & 0.0645 & 1.2425 & 0.1424 & 2.9673 & 0.5498 & 0.784165563 & 0.0784 \\ 
\hline\\
023025.03-004944.1 & 0.6535 & 10036.5 & 0.0111 & 0.0012 & 0.0488 & 0.0049 & 0.0383 & 0.0039 & 0.0563 & 0.0056 & 0.0575 & 0.0058 & 0.0605 & 0.0061 & 0.0734 & 0.0075 & 0.3078 & 0.0313 & 0.4327 & 0.0443 & 1.045 & 0.1245 & 2.8078 & 0.5317 & 4.36980163 & 0.5782 \\
\hline\\
023057.39-010033.5 & 0.65 & 3490.9 & 0.0356 & 0.0036 & 0.1076 & 0.0108 & 0.1113 & 0.0112 & 0.1494 & 0.0151 & 0.1364 & 0.0137 & 0.1432 & 0.0144 & 0.1415 & 0.0143 & 0.6467 & 0.0652 & 1.0219 & 0.1029 & 1.8754 & 0.1983 & 3.0114 & 0.5319 & 3.67034994 & 0.4224 \\
\hline\\
024105.83-081153.1 & 0.9785 & 7205.6 & 0.0059 & 0.0006 & 0.024 & 0.0024 & 0.0252 & 0.0026 & 0.0283 & 0.0028 & 0.0406 & 0.0041 & 0.0381 & 0.0039 & 0.0458 & 0.0052 & 0.1232 & 0.0128 & 0.2488 & 0.0259 & 0.385 & 0.0688 & 3.4831 & 0.3751 & 0.251272141 & 0.0306 \\
\hline\\
024135.69-000858.6 & 0.8224 & 2709.6 & 0.0185 & 0.0019 & 0.0445 & 0.0045 & 0.0555 & 0.0056 & 0.0682 & 0.0068 & 0.0694 & 0.0069 & 0.0661 & 0.0066 & 0.077 & 0.0079 & 0.3382 & 0.0341 & 0.5309 & 0.0539 & 0.9126 & 0.1051 & 7.3175 & 1.1385 & 2.19419228 & 0.3170 \\
\hline\\
024200.91+000021.0 & 1.1039 & 10864.4 & 0.0047 &0.0006 & 0.0603 & 0.0061 & 0.1218 & 0.0122 & 0.1289 & 0.0129 & 0.1711 & 0.0171 & 0.1486 & 0.0149 & 0.1275 & 0.0129 & 0.518 & 0.0518 & 0.951 & 0.0959 & 2.6418 & 0.2703 & 7.3175 & 0.8142 & 4.92309883 & 0.5055 \\
\hline\\
024214.99-000209.5 & 1.0103 & 9447.2 & 0.0076 & 0.0009 & 0.0213 & 0.0022 & 0.0268 & 0.0028 & 0.0316 & 0.0032 & 0.0401 & 0.0040 & 0.0404 & 0.0041 & 0.0443 & 0.0047 & 0.1623 & 0.0166 & 0.2374 & 0.0247 & 0.5137 & 0.0708 & 2.7974 & 0.5043 & 0.5723 & 0.0716 \\
\hline\\
024220.73-002059.8 & 0.6841 & 9210.7 & 0.0029 & 0.0005 & 0.0294 & 0.0029 & 0.0226 & 0.0023 & 0.0418 & 0.0042 & 0.0484 & 0.0048 & 0.0585 & 0.0058 & 0.0741 & 0.0076 & 0.3674 & 0.0371 & 0.4692 & 0.0477 & 1.0733 & 0.1178 & 3.206 & 0.3453 & 1.67494542 & 0.4775 \\
\hline\\
025959.68+004813.6 & 0.8926 & 7712.5 & 0.0062 & 0.0010 & 0.0265 & 0.0029 & 0.0357 & 0.0036 & 0.051 & 0.0051 & 0.0614 & 0.0061 & 0.0593 & 0.0059 & 0.0739 & 0.0075 & 0.2654 & 0.0269 & 0.3807 & 0.0390 & 1.3204 & 0.1459 & 2.1536 & 0.4901 & 0.562450807 & 0.0562 \\
\hline\\
030315.67-000702.1 & 0.7025 & 5466.2 & 0.0319 & 0.0039 & 0.1028 & 0.0105 & 0.0837 & 0.0084 & 0.1343 & 0.0135 & 0.1444 & 0.0145 & 0.155 & 0.0156 & 0.177 & 0.0178 & 0.8716 & 0.0877 & 1.1497 & 0.1159 & 2.0697 & 0.2176 & 28.7847 & 3.1002 & 4.75601785 & 0.9258\\ 
\hline\\
031640.43-000955.3 & 0.6371 & 4343.4 & 0.0126 & 0.0023 & 0.0302 & 0.0036 & 0.0218 & 0.0023 & 0.0357 & 0.0036 & 0.0372 & 0.0037 & 0.0458 & 0.0046 & 0.0498 & 0.0052 & 0.212 & 0.0217 & 0.2766 & 0.0287 & 0.4676 & 0.0762 & 2.7464 & 0.2958 & 0.4850 & 0.0485 \\
\hline\\
031828.90-001523.1 & 1.9845 & 3258.2 & 0.0068 & 0.0016 & 0.0344 & 0.0040 & 0.1436 & 0.0144 & 0.1703 & 0.0171 & 0.2023 & 0.0203 & 0.222 & 0.0223 & 0.2575 & 0.0259 & 0.3394 & 0.0343 & 0.6896 & 0.0697 & 2.8386 & 0.2917 & 6.0808 & 0.7593 & 25.584 & 2.5585 \\
\hline\\
074408.40+375841.1 & 0.8803 & 3575.4 & 0.0258 & 0.0026 & 0.0879 & 0.0088 & 0.1358 & 0.0136 & 0.172 & 0.0172 & 0.2023 & 0.0202 & 0.2095 & 0.0209 & 0.243 & 0.0244 & 0.8463 & 0.0853 & 1.7563 & 0.1766 & 4.3642 & 0.4451 & 10.3744 & 1.1351 & 0.102586043 & 0.0103 \\
\hline\\
075058.20+421616.9 & 0.9381 & 1923.7 & 0.0105 & 0.0019 & 0.0347 & 0.0040 & 0.0271 & 0.0027 & 0.0337 & 0.0034 & 0.0429 & 0.0043 & 0.0409 & 0.0041 & 0.0446 & 0.0046 & 0.1729 & 0.0177 & 0.2848 & 0.0292 & 0.789 & 0.0950 & 3.3324 & 0.3589 & 0.0974 & 0.0744 \\
\hline\\
083137.40+414853.2 & 0.5529 & 18755.5 & 0.0126 & 0.0023 & 0.0363 & 0.0043 & 0.0347 & 0.0035 & 0.0587 & 0.0058 & 0.061 & 0.0061 & 0.0929 & 0.0093 & 0.114 & 0.0115 & 1.5875 & 0.1595 & 2.1968 & 0.2208 & 4.4906 & 0.4559 & 9.2889 & 1.0369 & NA & NA \\
\hline\\
093556.85+621249.7 & 0.4079 & 15634.0 & 0.0242 & 0.0025 & 0.0581 & 0.0058 & 0.0525 & 0.0052 & 0.0771 & 0.0077 & 0.1093 & 0.0109 & 0.1393 & 0.0139 & 0.2047 & 0.0206 & 1.4412 & 0.1449 & 2.4968 & 0.2208 & 6.6604 & 0.6705 & 13.2911 & 1.3802 & NA & NA \\
\hline
\hline
\label{tab:RQ quasars}
\end{tabular}
}
\end{sidewaystable*}

\subsection{Data Sets}
\label{subsec:Data Sets}
Our main quasar sample is drawn from the SDSS \citep{york} Data Release 7 \citep{abaz} quasar catalog \citep{shen}. It consists of 105,783 quasars brighter than $M_{i}$ = -22.0 which are spectroscopically confirmed and taken from $\sim$ 9,380 sq. deg. of the sky. The catalog consists of quasars that have reliable redshifts and have at least one emission line (H$\beta$ and Mg\textsc{ii}) with FWHM greater than 1000 km\,s$^{-1}$. The flux limit for the main spectroscopic sample is $i < 19.1$, hence the majority of quasars are brighter than $i \approx 19$ within the redshift limit $z < 1.9$, which include broad H$\beta$ and Mg\textsc{ii} line samples. We have chosen to use DR7 because it contains all the BH masses and includes line fits with multiple Gaussian functions for the broad H$\beta$ and Mg\textsc{ii} lines. Except for C\textsc{iv} emission line, different procedures to measure the broad H$\beta$ and Mg\textsc{ii} have been adopted \citep{Shen08, M&D04}, and different fiducial scaling relations to compute H$\beta$- and Mg\textsc{ii}-based virial BH masses are used. 

\begin{table*} 
\centering 
\caption{Model parameters for the RL sources (non-HBL+HBL+\textit{AstroSat}-detected sources) in our sample. For parameters not listed here, we have used the default \texttt{CIGALE v2022.0} values. The model parameters for RQ sources are identical, except the \texttt{radio} module has been omitted in those cases.} 

\resizebox{\textwidth}{!}{
\begin{tabular}{|l|c|c|} 
\hline 
    \textbf{Parameter} & \textbf{Symbol} & \textbf{Values} \\ 
\hline \hline 
    \multicolumn{3}{|c|}{\textsc{star-formation History}} \\ 
    \multicolumn{3}{|c|}{Delayed star-formation model: ${\rm SFR} \propto t \exp(-t / \tau)$ [\texttt{sfhdelayed}]} \\
\hline 
    Stellar e-folding time [Gyr] & $\tau$ & 0.1, 0.5, 1, 5 \\ 
    Stellar age [Gyr] & $t_{\rm age}$ & 0.5, 1, 3, 5, 7 \\ 
    
\hline 
    \multicolumn{3}{|c|}{\textsc{Simple Stellar Population (SSP)}} \\ 
    \multicolumn{3}{|c|}{\citet{bru} Model [\texttt{bc03}]} \\ 
\hline 
    Initial Mass Function (IMF) & --- & \citet{cha} \\ 
    Metallicity & $Z$ & 0.02 \\ 

\hline
    \multicolumn{3}{|c|}{\textsc{Dust Attenuation}} \\ 
    \multicolumn{3}{|c|}{\citet{calz} Model [\texttt{dustatt\_modified\_starburst}]} \\ 
\hline 
    Color excess of nebular lines [mag] & $E(B - V)$ & 0.05, 0.1, 0.2, 0.3, 0.4, 0.5, 0.7, 0.9 \\ 

\hline
    \multicolumn{3}{|c|}{\textsc{Dust Emission}} \\ 
    \multicolumn{3}{|c|}{\citet{dale} Model [\texttt{dale2014}]} \\ 
\hline 
    Slope in $dM_{\rm dust} \propto U^{-\alpha} dU$ & $\alpha$ & 2 \\ 

\hline
    \multicolumn{3}{|c|}{\textsc{AGN Emission (UV to IR)}} \\ 
    \multicolumn{3}{|c|}{Clumpy Torus Model: SKIRTOR \citep{sta16} [\texttt{skirtor2016}]} \\
\hline
    Average edge-on optical depth at 9.7 $\mu$m & $\tau_{9.7}$ & 7 \\ 
    Power-law exponent for the radial gradient of dust density & $p$ & 1.0 \\ 
    Index setting dust density gradient with polar angle & $q$ & 1.0 \\ 
    Angle between the equatorial plane and edge of the torus & $\theta$ & $40^{\circ}$ \\ 
    Ratio of outer to inner radius & $R$ & 20 \\ 
    AGN Inclination & $i$ & $30^{\circ}$, $60^{\circ}$, $70^{\circ}$, $80^{\circ}$, $90^{\circ}$ \\ 
    Disk Spectrum & --- & Broken power-law \citep{Schartmann2005} \\ 
    Power-law index modifying the optical slope of the disk & $\delta$ & $-1, -0.8, -0.6, -0.4, -0.2, 0, 0.2, 0.4, 0.6, 0.8, 1$ \\ 
    AGN's contribution to IR Luminosity & ${\rm frac}_{\rm AGN}$ & 0--0.9 (step 0.1), 0.99 \\ 
    Polar dust color excess & $E(B - V)_{\rm PD}$ & 0, 0.01, 0.02, 0.05, 0.1, 0.15, 0.2, 0.3, 0.4, 0.5, 0.6 \\ 

\hline 
    \multicolumn{3}{|c|}{\textsc{Radio Emission} [\texttt{radio}]} \\ 
\hline
    SF radio-IR correlation parameter & $q_{\rm IR}$ & 2.4, 2.5, 2.6, 2.7 \\ 
    SF power-law slope & $\alpha_{\rm SF}$ & 0.8 \\ 
    Radio-loudness parameter & $R_{\rm AGN}$ & 0.01, 0.02, 0.05, 0.1, 0.2, 0.5, ..., 1000, 2000, 5000, 10000 \\ 
    AGN power-law slope & $\alpha_{\rm AGN}$ & 0.7 \\ 

\hline 
    \multicolumn{3}{|c|}{\textsc{X-ray Emission} [\texttt{xray}]} \\ 
\hline 
    Photon-index of AGN intrinsic X-ray spectrum & $\Gamma$ & 1.8 \\ 
Power-law slope connecting rest-frame $L_{\nu}$ at 2500 \AA~and 2 keV & $\alpha_{\rm ox}$ & $-1.9, -1.8, -1.7, -1.6, -1.5, -1.4, -1.3, -1.2, -1.1$ \\ 
    Maximum deviation from the $\alpha_{\rm ox}$-$L_{\nu, 2500 \AA}$ relation & $|\Delta \alpha_{\rm ox}|_{\rm max}$ & 0.2 \\ 
    AGN X-ray angle coefficients & $(a_1, a_2)$ & (0, 0), (0.5, 0), (1, 0), (0.33, 0.67) \\ 
\hline 
\label{table:CIGALE parameters}
\end{tabular}}
\end{table*}

\cite{shen} cross-matched the quasar catalogue of SDSS DR7 with the FIRST survey \citep{whit}. The quasars having only one FIRST source within 5" are classified as core-dominated radio sources and those having multiple FIRST sources within 30" are classified as lobe-dominated. These two categories are together termed as RL quasars by \cite{shen}. Quasars with only one FIRST match between 5" and 30" are classified as FIRST non-detections and are classified as RQ quasars. The SDSS sample contains 105,783 quasars of which 99,182 have FIRST counterparts. Among these, 9,399 are RL and 88,979 are RQ quasars. 

Further, these RL and RQ samples are cross-matched with WISE \citep{wri} for infrared counterparts. The WISE cross-matching leaves the sample size for RL and RQ quasars to be 959 and 7252 respectively. Next, the sample is cross-matched with GALEX \citep{mart} for UV counterparts where 186 and 74 counterparts were found for RL and RQ quasars respectively. Among these sources 3 RL and 2 RQ sources are HBL. Lastly, for X-ray counterparts, our sample was cross-matched with Chandra \citep{evans10}, ROSAT \citep{Truemper82}, Swift \citep{tueller08}, and XMM-Newton \citep{bianchi09}. The cross-matching with the X-ray catalogs was conducted within a radius of 3 arcseconds of our sources \citep{svobodaetal17}. This sample (non-HBL hereafter) consists of 31 RL and 14 RQ quasars in the redshift range $0.33<z<1.88$, with at least one broad emission line (H$\beta$ or Mg\textsc{ii}). After cross-matching with X-rays our sample do not contain any HBL source. However, we added 3 RL and 2 RQ HBL sources to our sample (HBL hereafter) without the X-ray data.  Including the HBL sources, our sample consists of 34 RL (Table \ref{tab:RL quasars}) and 16 RQ (Table \ref{tab:RQ quasars}) quasars. We note that we also cross-matched our sources with the 2MASS catalog \citep{skrutskieetal06} for near-infrared coverage. We found 17 matches consisting of both the RL and RQ populations. For the 17 2MASS matches we performed the SED analysis with and without the 2MASS data point. Our results indicated that the derived host galaxy properties with and without the 2MASS data are statistically similar. Adding the 2MASS sources would have significantly reduced our sample size and hence we did not include the 2MASS analysis in this work. 

\begin{figure*}
\begin{center}
\begin{tabular}{c} 
\resizebox{6 cm}{!}{\includegraphics{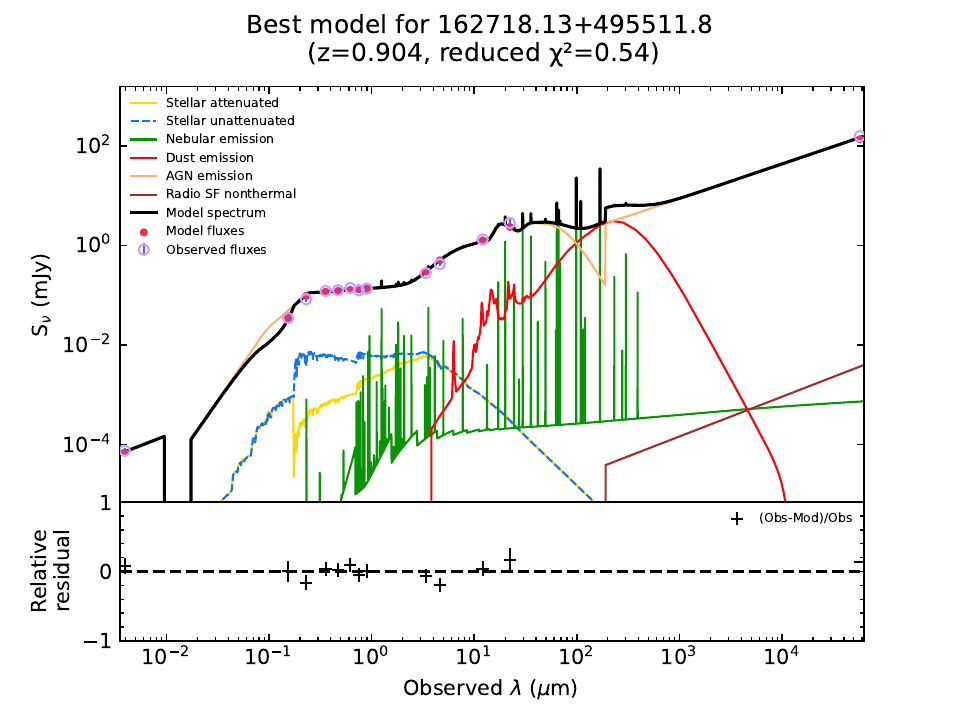}} 
\resizebox{6cm}{!}{\includegraphics{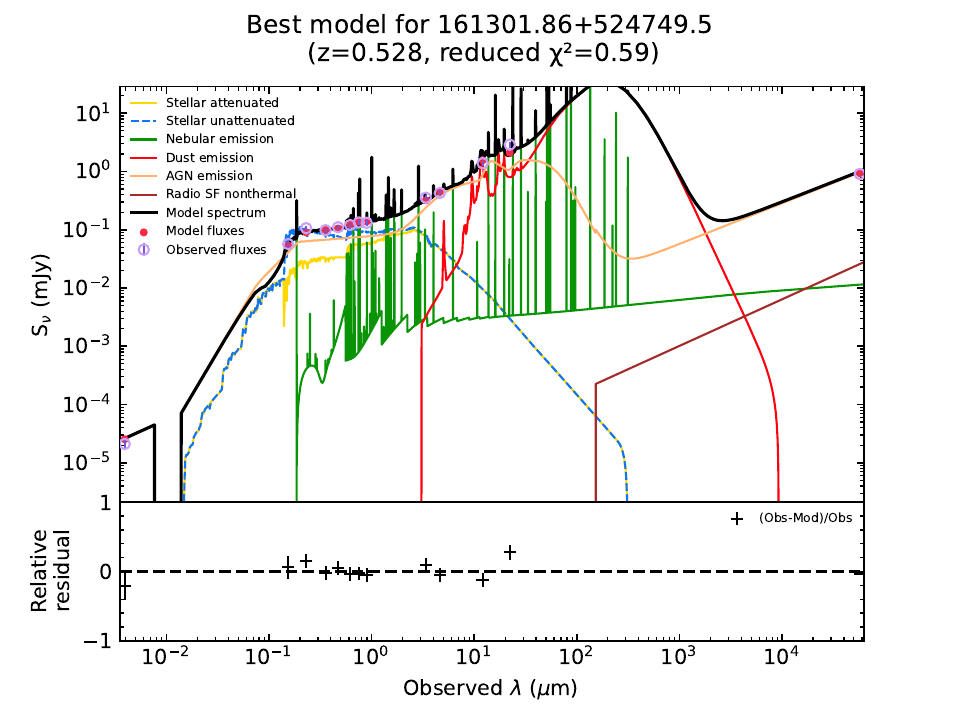}} 
\resizebox{6cm}{!}{\includegraphics{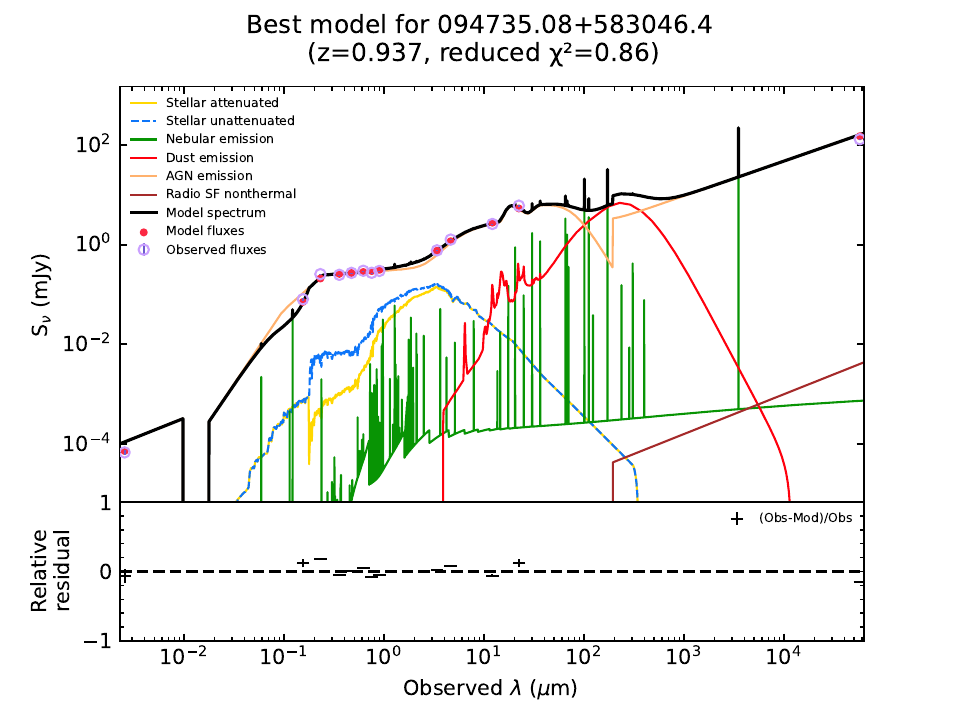}} \\ 
\resizebox{6cm}{!}{\includegraphics{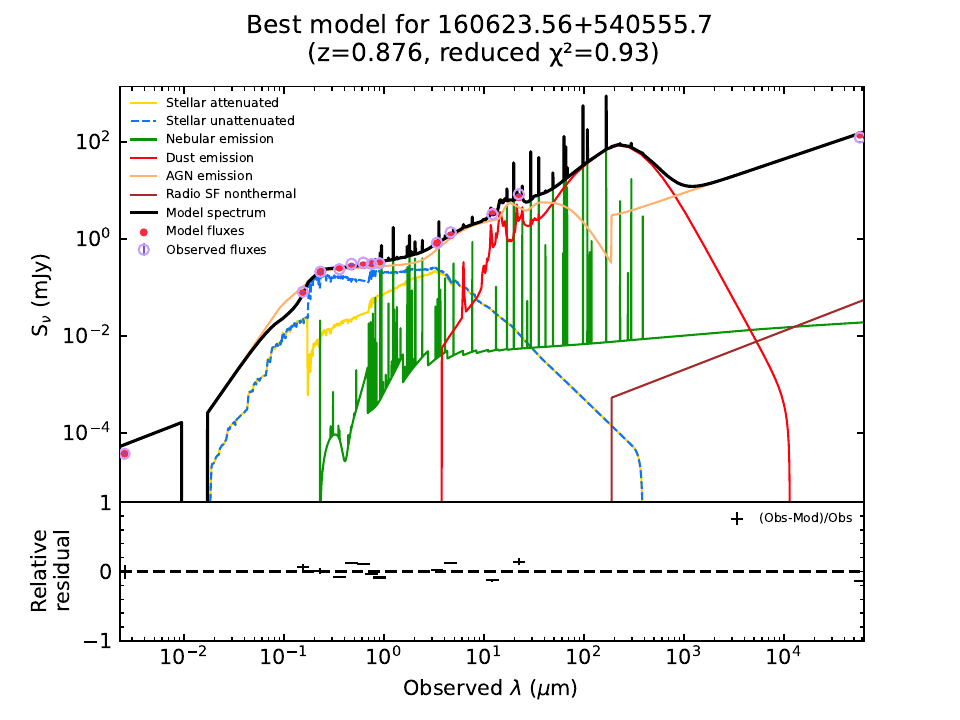}} 
\resizebox{6cm}{!}{\includegraphics{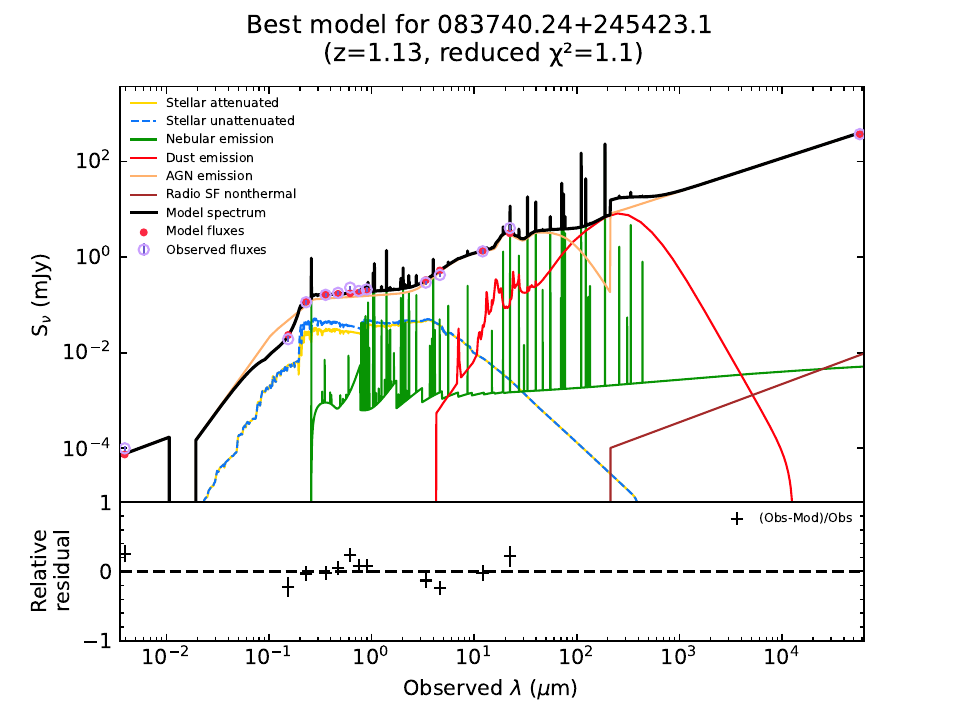}} 
\resizebox{6cm}{!}{\includegraphics{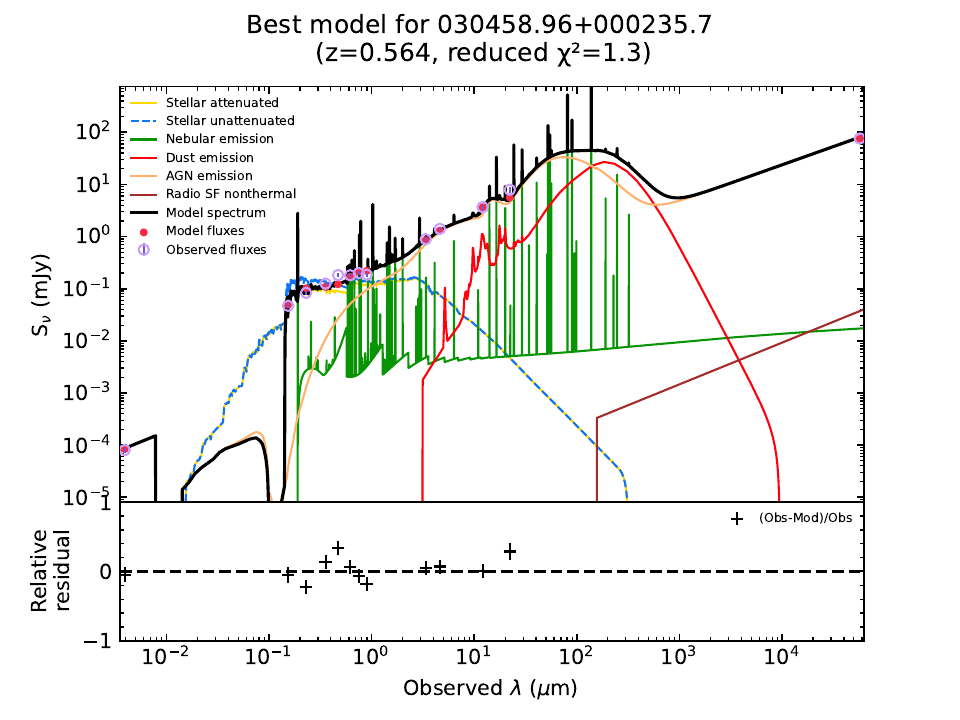}} 
\end{tabular}
\caption{Examples of best-fit SEDs of non-HBL radio-loud quasar sources, constructed using the \texttt{CIGALE v2022.0} code. Reduced $\chi^2$ values calculated by \texttt{CIGALE v2022.0} code for individual SEDs are provided in the figures.}
\label{fig:RL_SEDs}
\end{center}
\end{figure*}

\subsubsection{\textit{AstroSat} Data} 
\label{subsec:AstroSat}

While constructing the broad-band SED, we found that some of our sources have near-UV (NUV) but no far-UV (FUV) observations in GALEX. For those sources, we performed observations of a sample of 3 RL and 3 RQ quasars in the FUV band using the Ultraviolet Imaging Telescope (UVIT) onboard \textit{AstroSat} (Proposal ID: A11\textunderscore047). These 6 sources were particularly selected based on their low exposure time requirement for observation. After including the 6 \textit{AstroSat} sources, our final sample consists of 37 RL and 19 RQ sources ($0.15<z<1.88$, Figure \ref{fig:redshift}). We briefly discuss the UVIT data reduction here. 

For the analysis of our {\it AstroSat} sources, we obtained the Level 1 (L1) UVIT data from the \textit{AstroSat} data archive\footnote{\url{https://astrobrowse.issdc.gov.in/astro_archive/archive/Home.jsp}}, which we processed using the \textsc{CCDLAB} pipeline \citep{postma17}, to generate orbitwise cleaned images. \textsc{CCDLAB} applies various detector and satellite corrections to the L1 data such as corrections for the flat field, centroiding bias, pointing drift, etc. Using \textsc{CCDLAB} we aligned and merged the orbitwise images into a final image for each observation. Astrometry of the final images using \textsc{CCDLAB} was performed for two sources (SDSS 085605.83+450520.0 and SDSS 085605.83+450520.0). 

For the other four observations, identification of the source of interest was done by comparing the UVIT field of view with the same field in NUV from the GALEX All-sky Imaging Survey using SIMBAD. We then performed aperture photometry on the final images using standard Python tools, \texttt{astropy} and \texttt{photutils}. For all the sources we extracted the source counts from circular regions with radii of 25 pixels centred on the source. Background counts were extracted from nearby circular regions with 60-pixel radii. Finally, we converted the background corrected source count rates to respective flux ($f_{\lambda}$) and magnitude values using the UVIT zero point flux and magnitude information provided by \cite{tandon20}. The observed flux values are corrected for galactic extinction with the Milky Way extinction curve with $R_V = 3.1$ following \cite{Cardelli1988}, using the \texttt{extinction.ccm89} package. The images and other details of our sources are provided in Figure \ref{fig:AstroSat photometry} and Table \ref{tab:AstroSat}.

\subsection{The Code: \texttt{CIGALE v2022.0}}
The \texttt{CIGALE v2022.0} code \citep{yang20} is a modular code written in the Python language, the purpose of which is to model the spectra of galaxies from FUV to radio wavelengths. Based on the earlier \texttt{CIGALE} code \citep{boq} it can also estimate the physical properties of a galaxy, for example, star-formation rate (SFR), total stellar mass in the galaxy, contribution of dust luminosity, etc. The code performs the modeling of galaxy spectra through several steps, \textit{viz.}, input processing, input parameter specification, model computation, and analysis.

\begin{figure*}
    \centering
        \includegraphics[scale = 0.37]{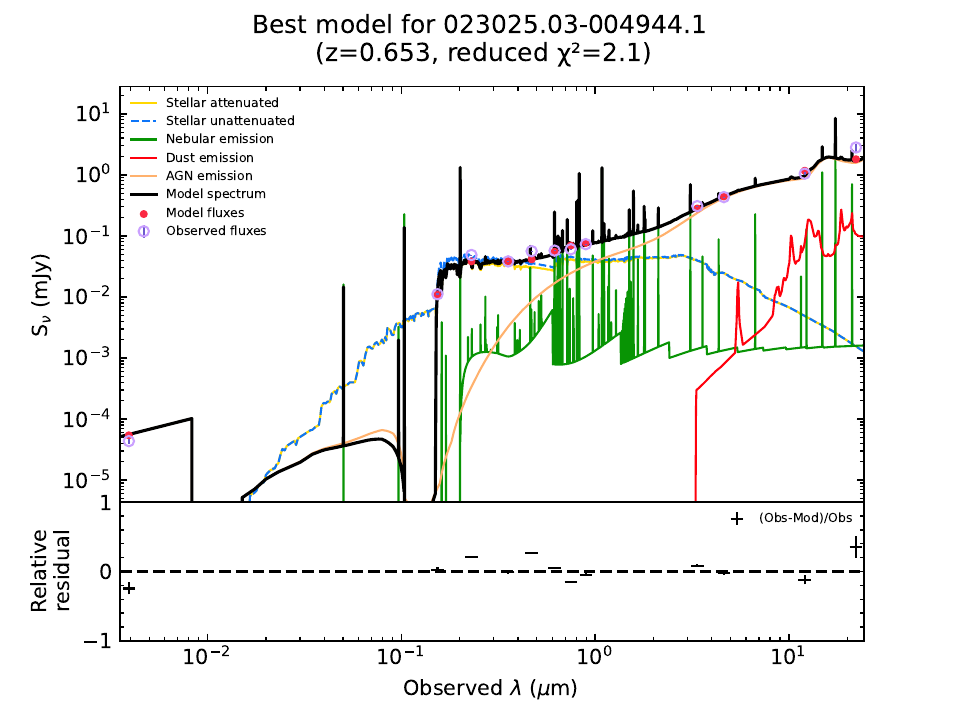} 
        \includegraphics[scale = 0.37]{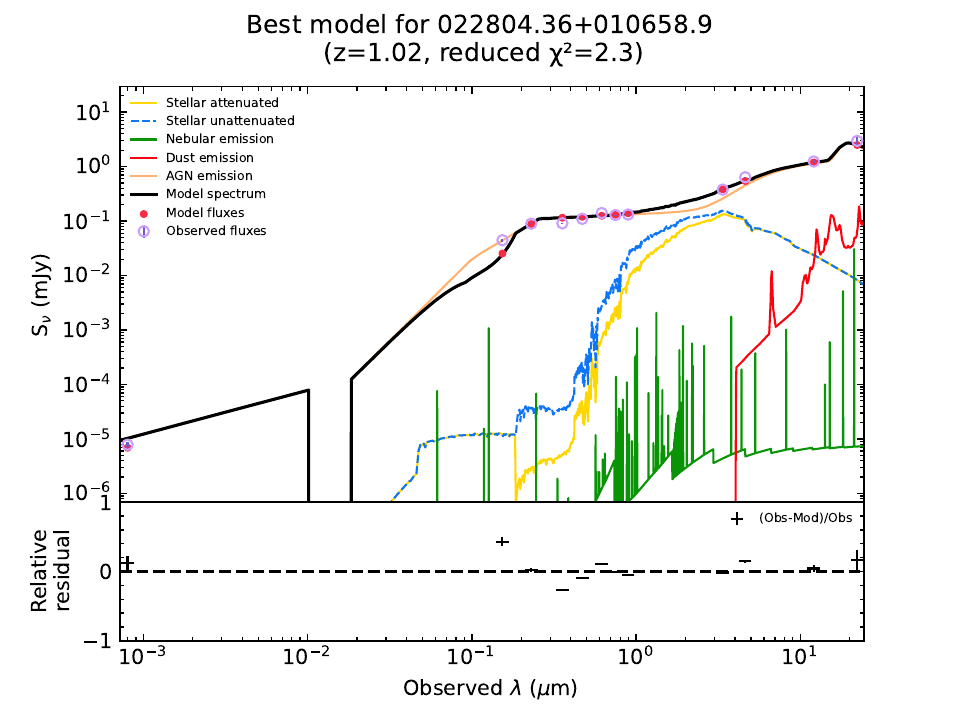} 
        \includegraphics[scale = 0.37]{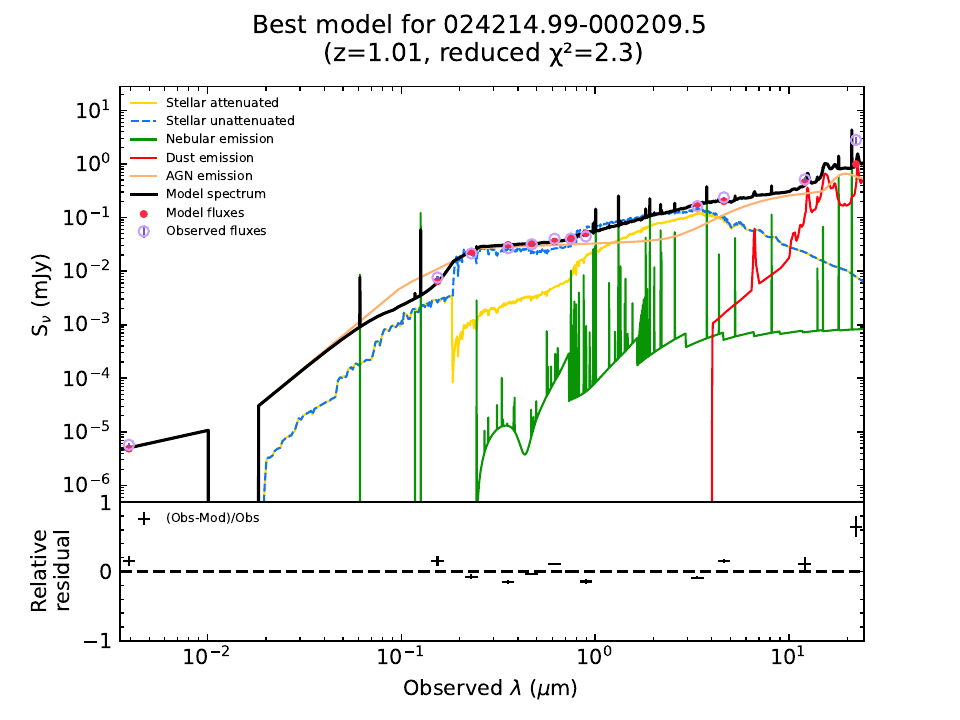} \\
        \includegraphics[scale = 0.37]{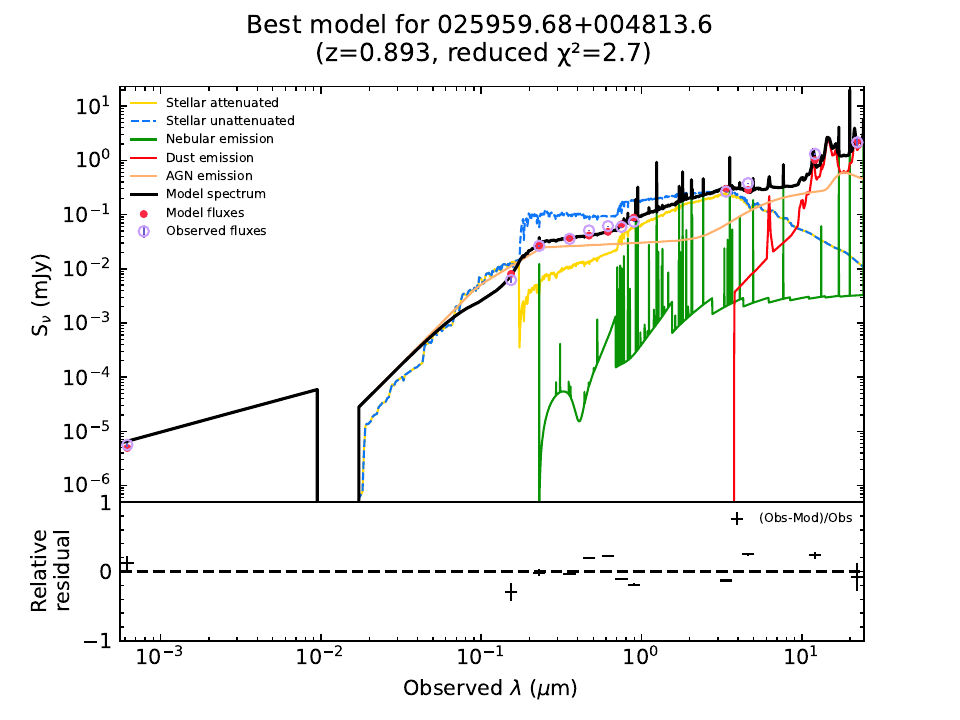} 
        \includegraphics[scale = 0.37]{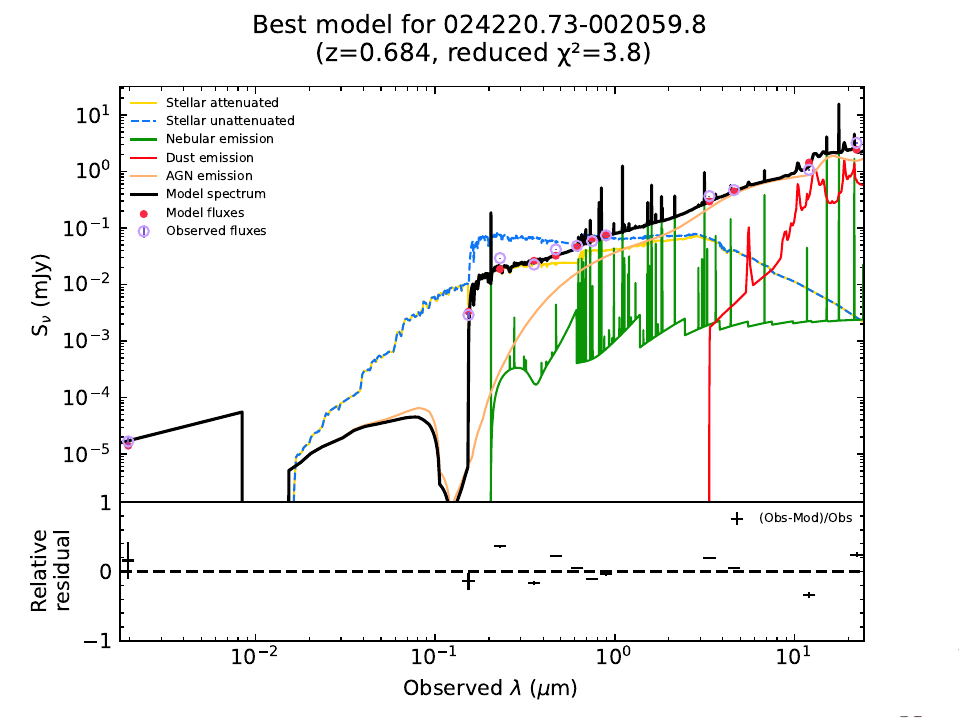} 
        \includegraphics[scale = 0.37]{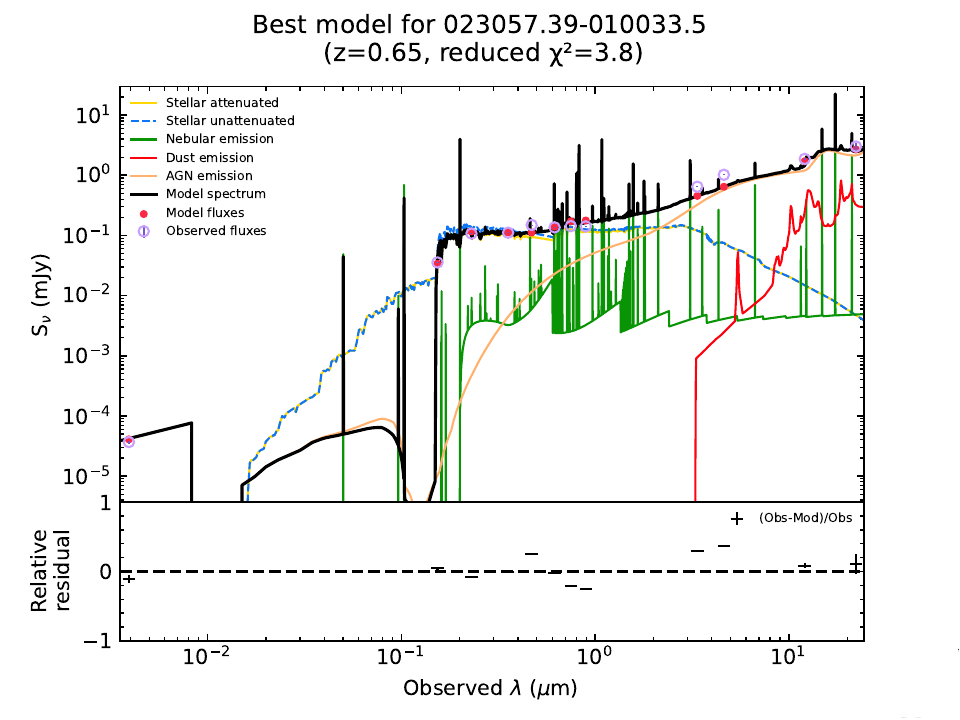}
    \caption{Examples of best-fit SEDs of non-HBL radio-quiet quasars sources, constructed using the \texttt{CIGALE v2022.0} code. Reduced $\chi^2$ values calculated by \texttt{CIGALE v2022.0} code for individual SEDs are provided in the figures.}
    \label{fig:RQ_SEDs}
\end{figure*}

Firstly, the code checks the input file for input data (e.g., fluxes, errors, redshifts, physical properties), which modules to consider, and their parameter values. After processing and normalizing the input data, the code uses this data file to determine the photometric bands/filters to use, among other operations (which are skipped if only model generation is required).

The code first computes the star-formation history (SFH) of the galaxy, from which it computes the stellar spectrum and single stellar population (SSP) models. It then models the nebular emission (both line and continuum) caused by Lyman continuum photons and applies dust attenuation and emission models to compute the luminosity contributed/diminished by the dust. Furthermore, it can also model the emission of an active galactic nucleus, and incorporate the effects of redshift and absorption by the intergalactic medium (IGM). 

The SFR can be modeled using several available models, notable among them are the double-exponential SFR, the delayed SFR, the periodic SFR, etc. It can also read and process SFH from files, using the \texttt{sfhfromfile} module, for modeling arbitrarily complex SFHs. To determine the stellar spectrum, a SSP library is adopted, in addition to the SFH computed in the previous step. The code offers two popular SSP libraries: that of \citealt[(module \texttt{bc03})]{bru} and \citealt[(module \texttt{m2005})]{mar}. Each of them is available for several metallicities, and two initial mass functions: \cite{cha} and \cite{salp}. 

The nebular emission is important since it allows one to probe into recent star-formation through hydrogen lines and radio continuum. In addition, we can estimate gas metallicity from the metal lines. \texttt{CIGALE v2022.0} uses nebular templates based on \cite{inoue}, which have been generated with \texttt{CLOUDY 13.01} \citep{fer98, fer13}. These templates depend on ionization fraction ($U$), and metallicity ($Z$, assumed to be the same as stellar metallicity). The electron density is set at 100 cm$^{-3}$ and is taken to be a constant which is a fiducial value assumed for the density. 
To take into account the absorption of Lyman continuum photons by dust and their escape from the galaxy without ionization, the nebular emission is downscaled following \citet{inoue}.

The effects of dust attenuation and emission are further applied to these emissions, based on the principle that the energy absorbed by the dust in the UV-NIR range is re-emitted in the mid to far-IR range. The emission of the dust can be modeled, among other ways, by assuming a power-law variation of the radiation field intensity ($U$) with the dust mass ($M_d$): 
\begin{equation}
\frac{dM_d(U)}{dU} \propto U^{-\alpha}
\end{equation} 
where $\alpha$ is a parameter of the model. 

Attenuation due to dust is calculated following \cite{calz} starburst attenuation curve (module \texttt{dustatt\_modified\_staburst}), extended with the \cite{lei} curve between the Lyman break and 150 nm. \cite{dale} model (module \texttt{dale2014}) has been implemented for the IR SED of the dust heated by stars, without the AGN component. Continuum and line nebular emission produced in HII regions (module \texttt{nebular}) have been considered. Redshifts and the absorption of short wavelength radiation by the IGM (module \texttt{redshifting}) have also been taken care of. 

Furthermore, the radio synchrotron emission and AGN emission can also be modeled with several available modules in the code. The radio synchrotron emission is modeled by two parameters: a power-law slope $\alpha$ and a radio-IR correlation $q_{\rm IR}$ \citep{hel}. 

\begin{figure*}
\begin{center}
\begin{tabular}{c}
\resizebox{6cm}{!}{\includegraphics{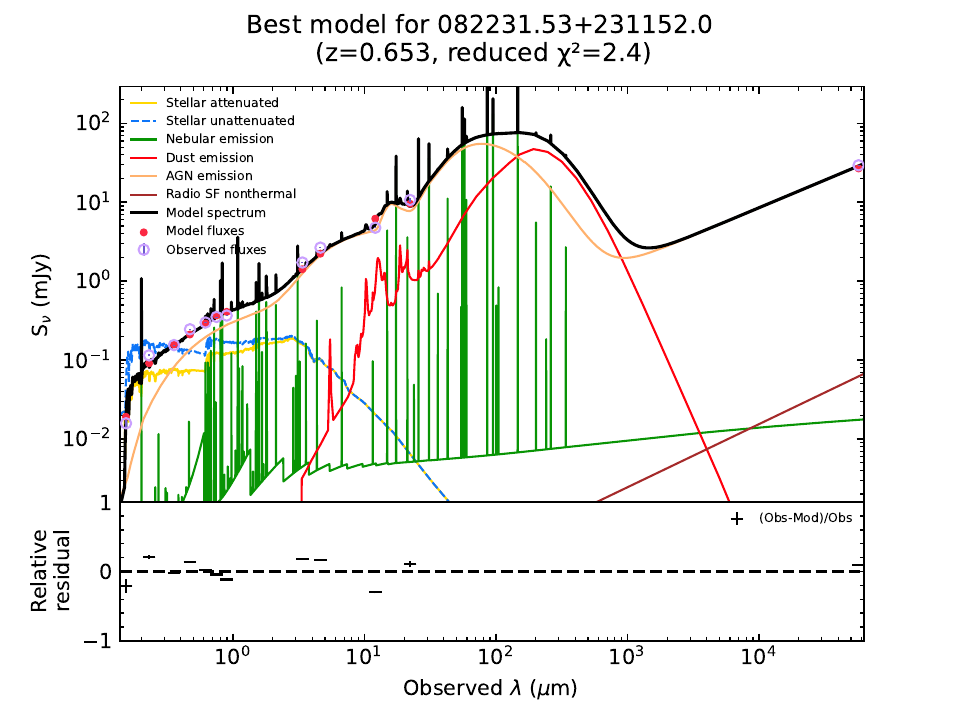}} 
\resizebox{6cm}{!}{\includegraphics{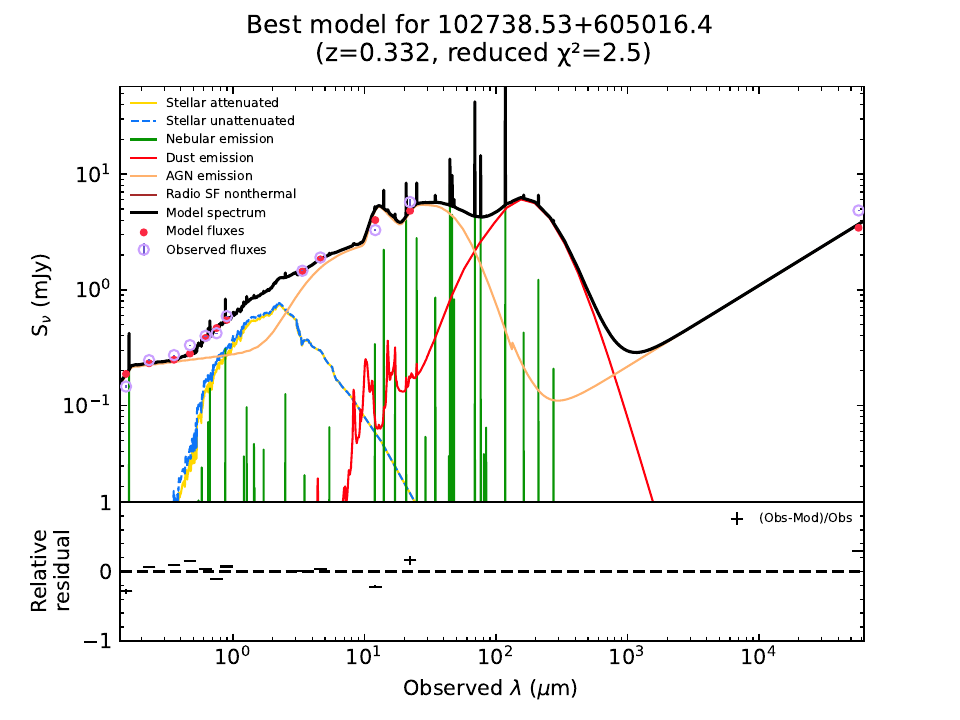}}
\resizebox{6cm}{!}{\includegraphics{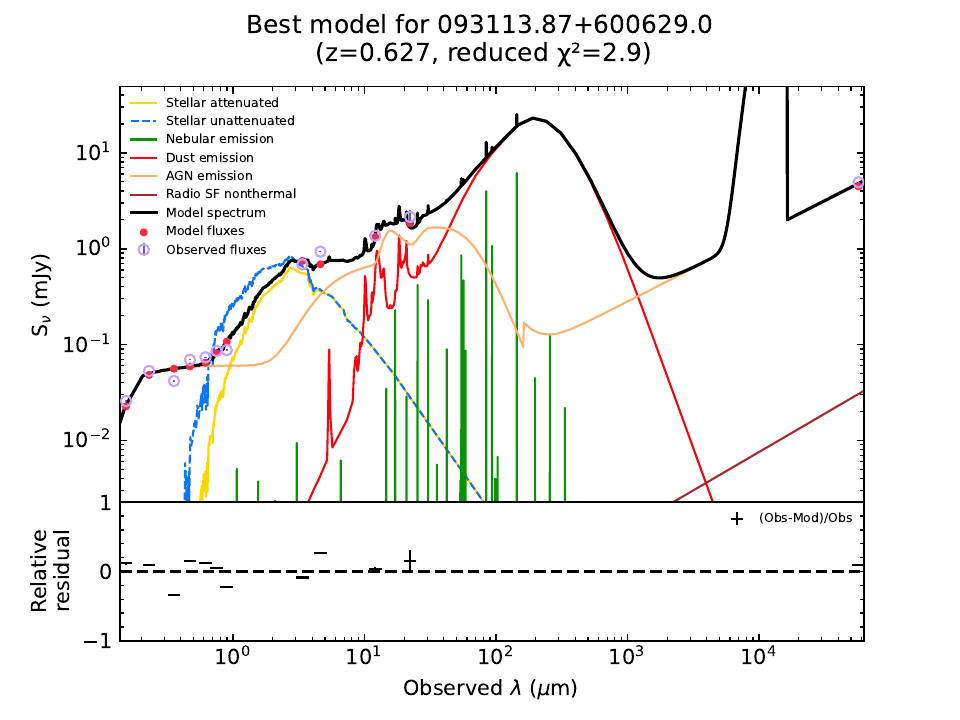}} \\
\resizebox{6cm}{!}{\includegraphics{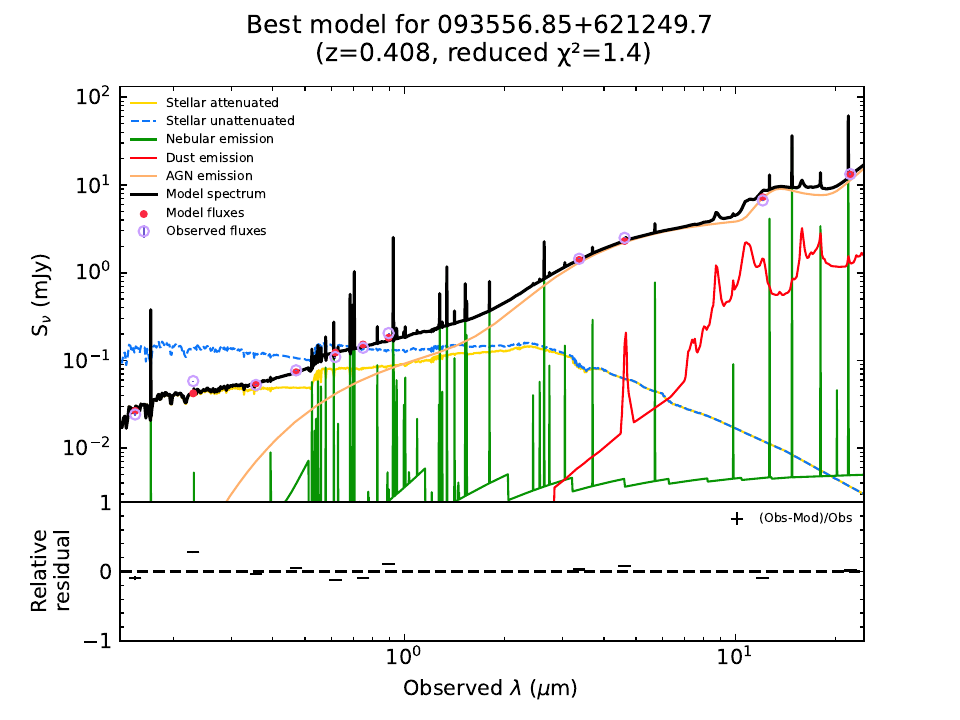}}
\resizebox{6cm}{!}{\includegraphics{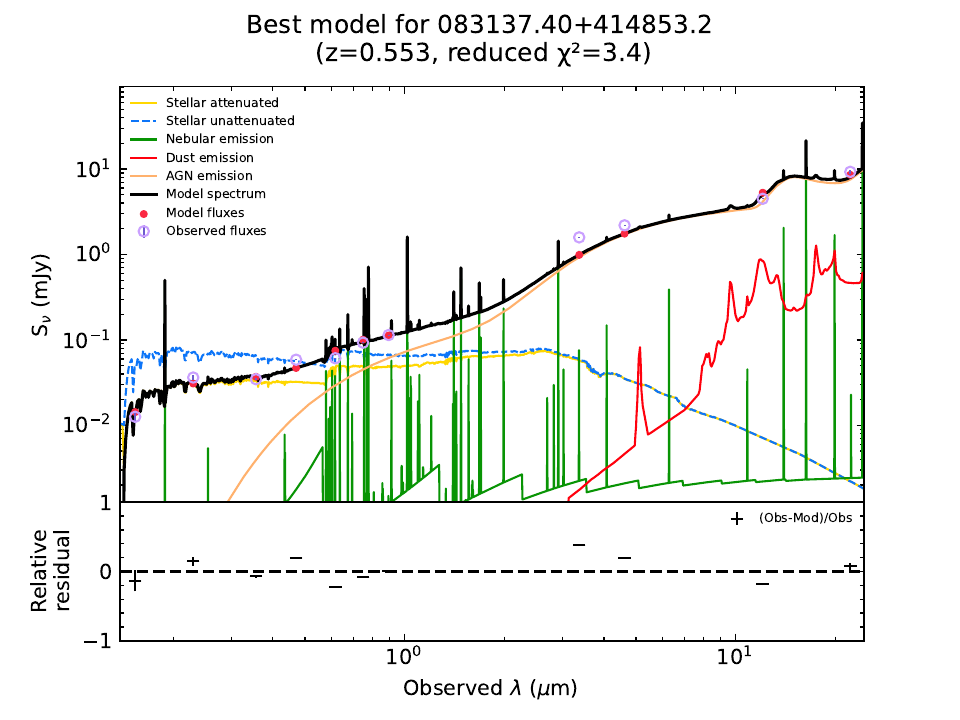}}
\end{tabular}
\caption{Best-fit SEDs of HBL sources. Reduced $\chi^2$ values calculated by \texttt{CIGALE v2022.0} code for individual SEDS are provided in the figures. {\bf Top Panel:} RL sources. {\bf Bottom Panel:} RQ sources.}
\label{fig:HBL_SEDs}
\end{center}
\end{figure*}

\subsubsection{AGN Model}
\texttt{CIGALE v2022.0} is potentially very accurate for the characterization of galaxies that host AGN \citep{Best23, Pacifici23} as it incorporates AGN models that can account for direct AGN light contributions and infrared emission arising from AGN heating of the dust. It also predicts the AGN X-ray emission \citep{yang20}. The inclusion of AGN models gives \texttt{CIGALE v2022.0} a significant advantage over
other SED fitting codes when fitting the SEDs of galaxies that have a significant AGN contribution. It allows more robust estimation of host galaxy parameters, and also a mechanism to identify and classify AGN within the sample. 

The SKIRTOR (module \texttt{skirtor2016}) templates \citep{sta12, sta16} are used for AGN emission modeling. SKIRTOR, a clumpy two-phase torus model considers an anisotropic and constant disk emission. A detailed description of how the SKIRTOR model is being implemented in \texttt{CIGALE v2022.0} is given in \cite{yang20}. ${\rm frac}_{\rm AGN}$ which is the AGN fraction, is defined as the ratio of the AGN IR emission to the total galaxy IR emission. A dust screen absorption and a grey-body emission are modeled by polar dust component ($E_{B-V}$). For our work, we adopt the Small Magellanic Cloud extinction curve (SMC) \cite{pre}. Re-emitted grey-body dust is parameterized with an emissivity index of 1.6 and a temperature of 100 K. For the radio emission modelling, the slope of the power-law synchrotron emission related to star-formation and that of the power-law AGN isotropic radio emission is considered to be 0.8 and 0.7 respectively \citep{randall12, tiwari19}. The radio-loudness parameter of AGN is defined as $R_{AGN}= L_{\nu,5GHz}/L_{\nu,2500A}$, where $L_{\nu,2500A}$ is the 2500\AA\ intrinsic disk luminosity measured at a viewing angle of $30^{\circ}$. Moreover, the ratio of the total star-forming IR luminosity (mostly in FIR) and the corresponding radio synchrotron luminosity at 21 cm (1.4 GHz) is defined as $q_{IR}$.

The code fits the models to the observational data, determines the likelihoods of the models, and uses those likelihoods to estimate the physical properties. Here, the `analysis modules' of the code are used to estimate the physical parameters of interest. The code can utilize two analysis modules to do this, \textit{viz}, \texttt{savefluxes}, which generates a grid of models and saves the outputs, and \texttt{pdf\_analysis}, which in addition to generating a grid of models, fits them to observational data to estimate physical parameters. The fluxes are observer-dependent and are computed after calculating a given model; The fluxes are obtained in bands by integrating the model spectrum through a corresponding filter. 

\begin{figure*}[hbt!]
    \centering
    {\resizebox{6 cm}{!}{\includegraphics{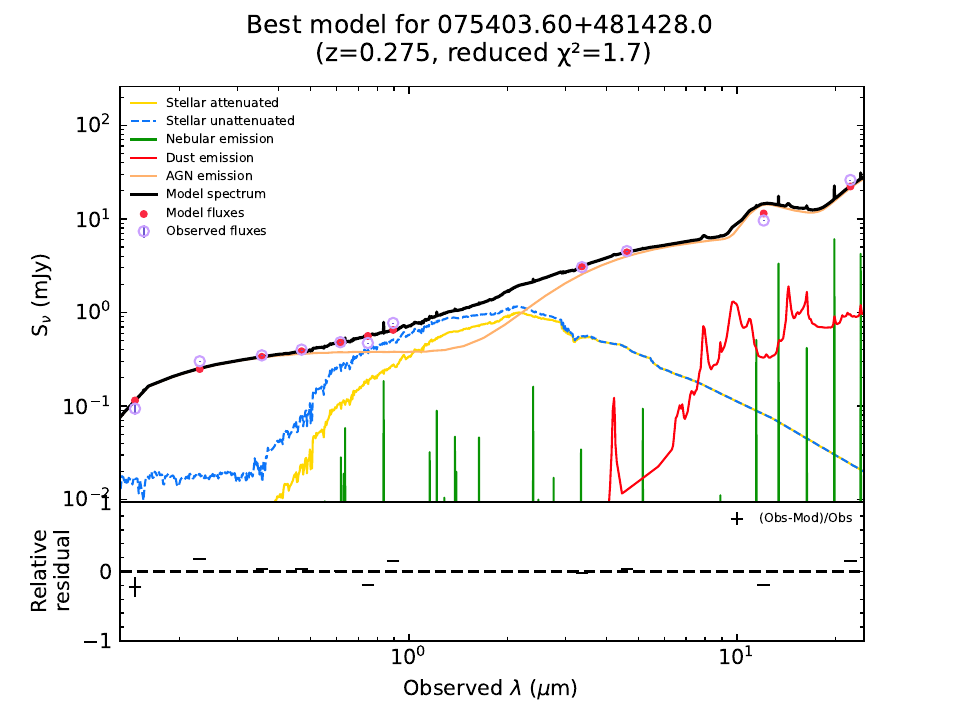}}} 
    {\resizebox{6 cm}{!}{\includegraphics{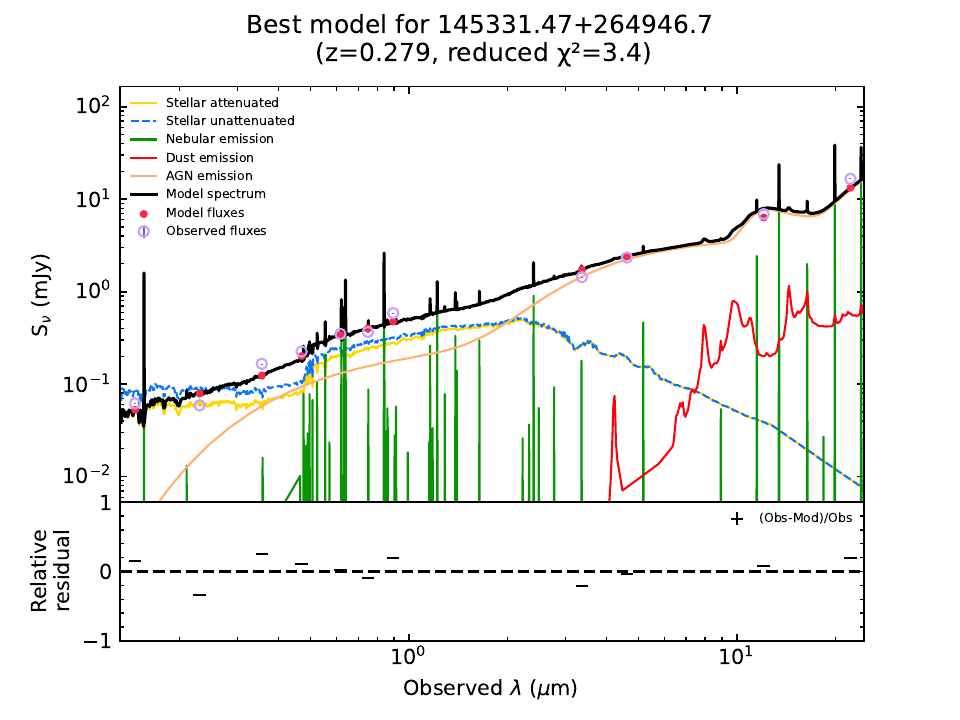}}} 
    {\resizebox{6 cm}{!}{\includegraphics{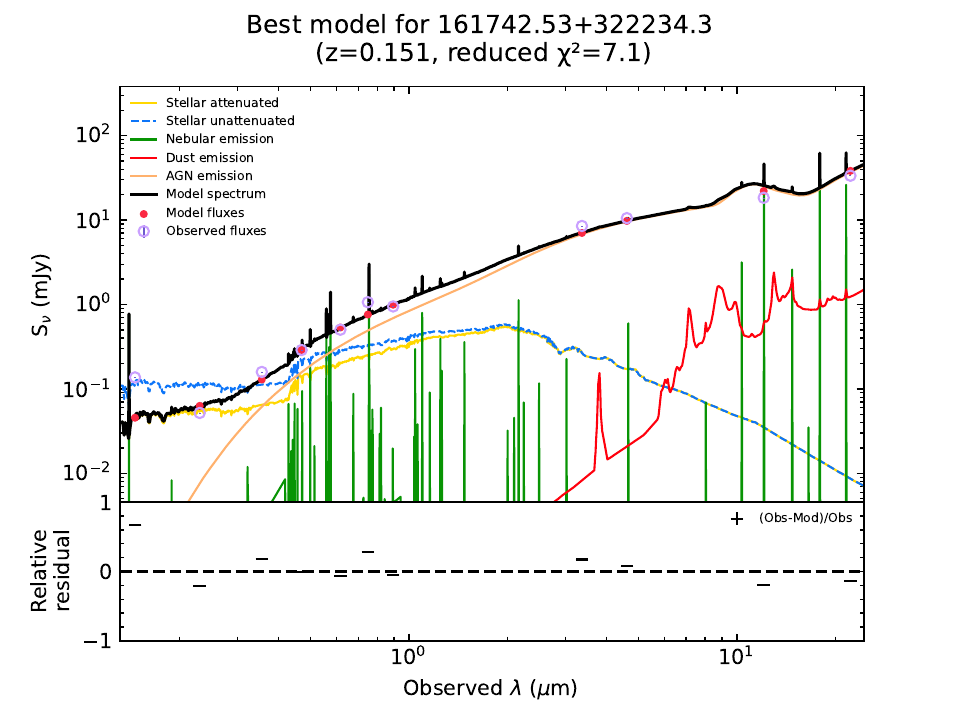}}} \\ 
    {\resizebox{6 cm}{!}{\includegraphics{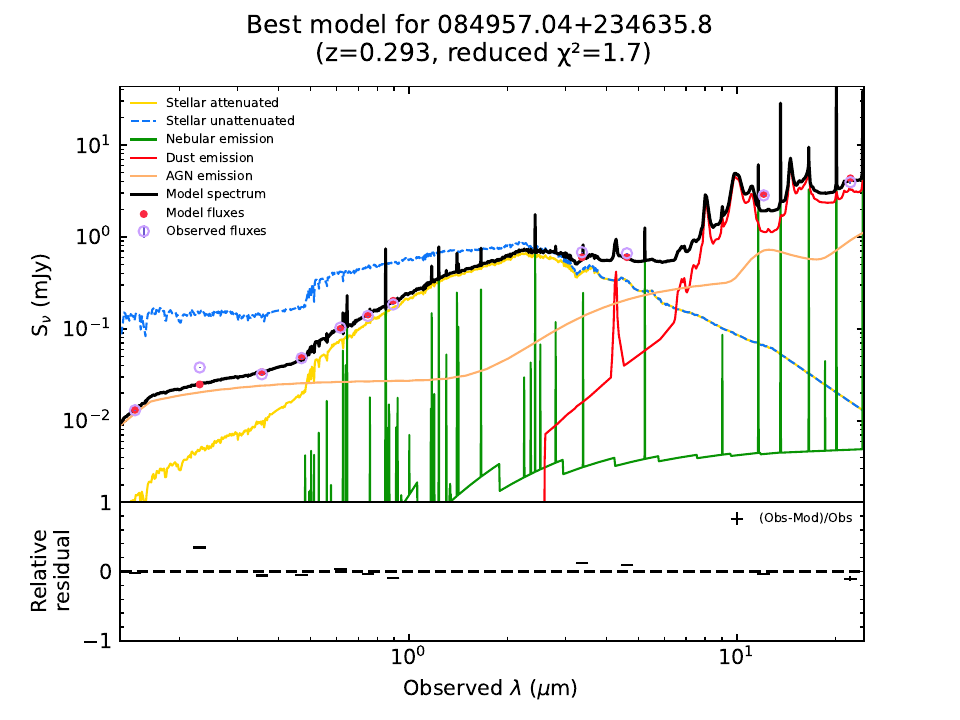}}} 
    {\resizebox{6 cm}{!}{\includegraphics{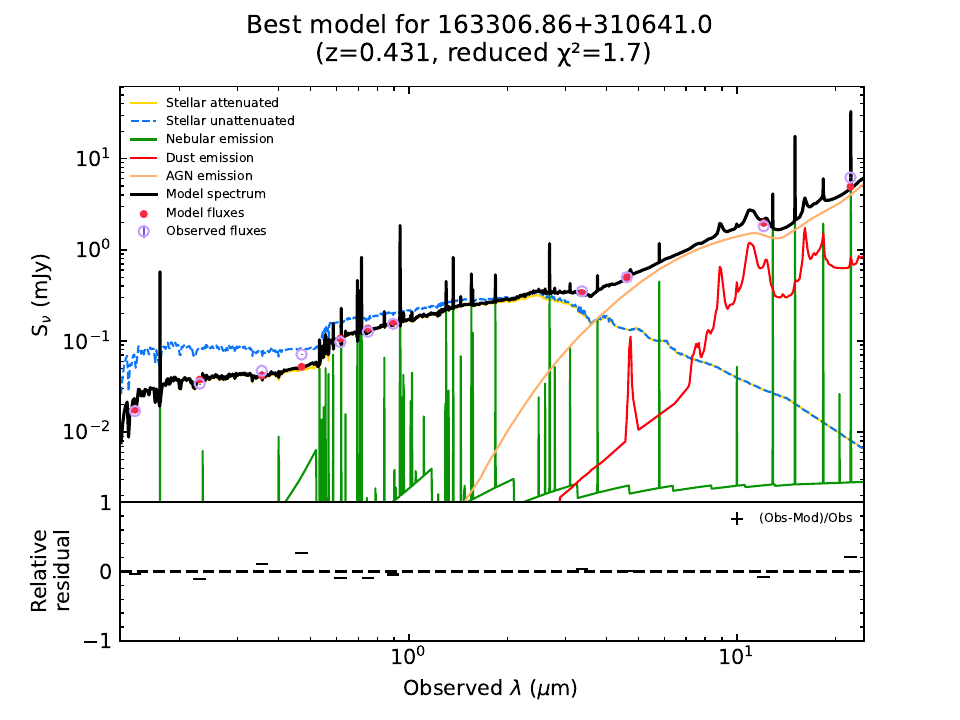}}} 
    {\resizebox{6 cm}{!}{\includegraphics{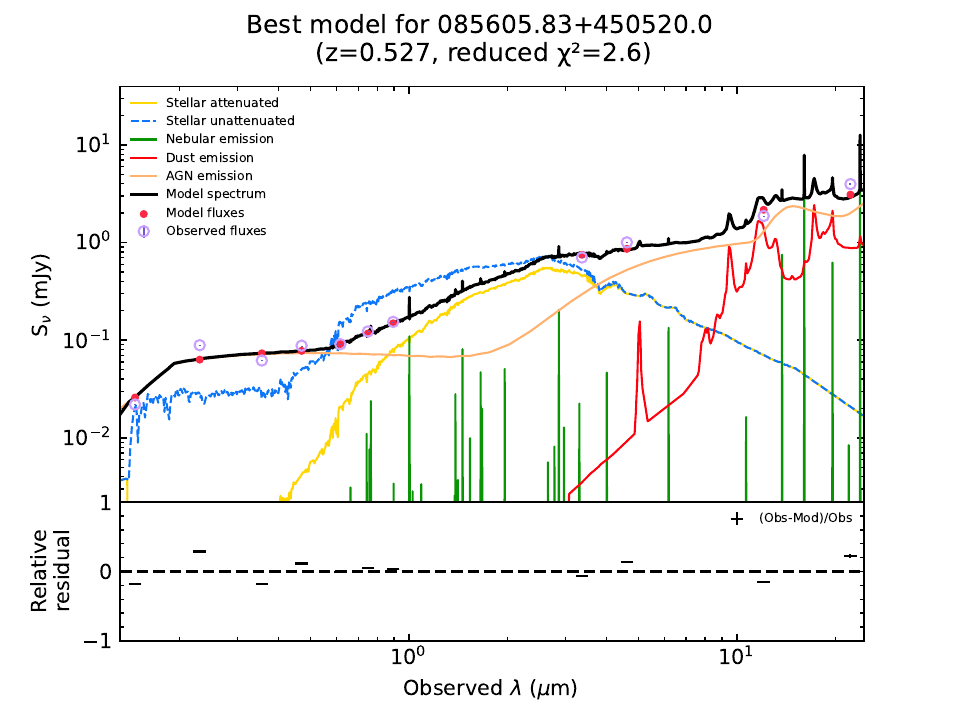}}} \\
    
    \caption{Best-fit SEDs of the sources observed with \textit{AstroSat}/UVIT without the X-ray data constructed using the \texttt{CIGALE v2022.0} code. Reduced $\chi^2$ values provided by \texttt{CIGALE v2022.0} for individual SEDs are shown in the figures. {\bf Top Panel: } RL sources. {\bf Bottom Panel: } RQ sources.}
    \label{fig:AstroSat_w_FUV}
\end{figure*} 

As discussed before, SED fitting using \texttt{CIGALE v2022.0} depends on the input parameter values and choice of models. For our analysis, a delayed star-formation history (SFH) model (module \texttt{sfhdelayed}) is applied to build the galaxy component. The model also includes a constant ongoing star-formation not allowed to be longer than 50Myr. Using the \cite{bru} single stellar population templates (module \texttt{bc03}) with the initial mass function (IMF) of \cite{salp} stellar emission is modeled and metallicity is taken to be equal to the solar value (0.02). The modules and input parameters we use in our analysis are presented in Table \ref{table:CIGALE parameters}.

\begin{figure*}
\begin{center}
\begin{tabular}{c}
\resizebox{6.5cm}{!}{\includegraphics{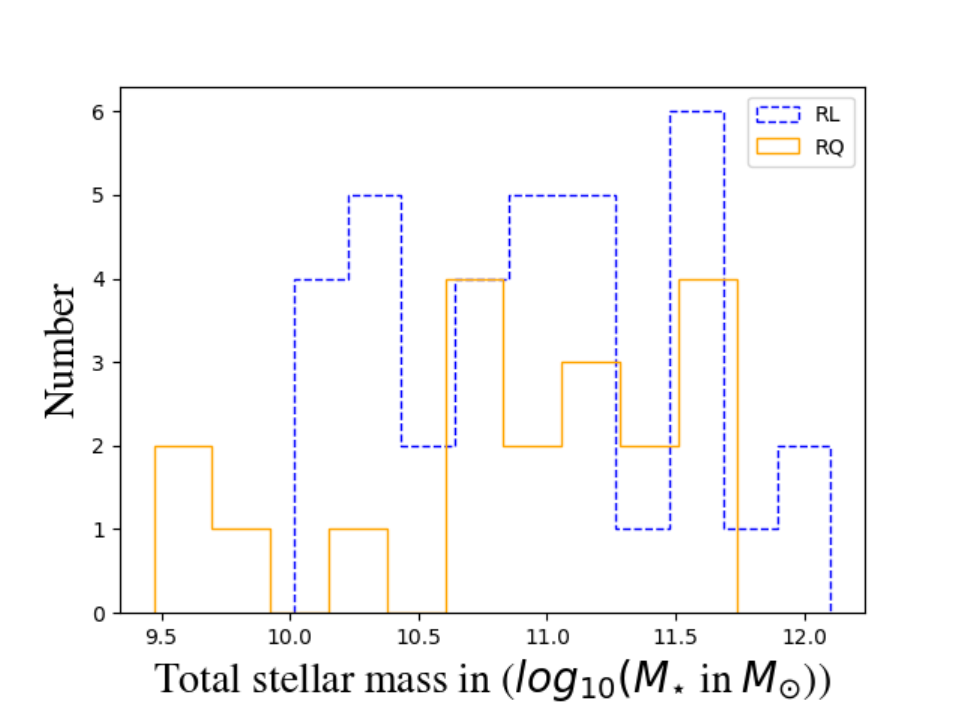}}
\resizebox{6.5cm}{!}{\includegraphics{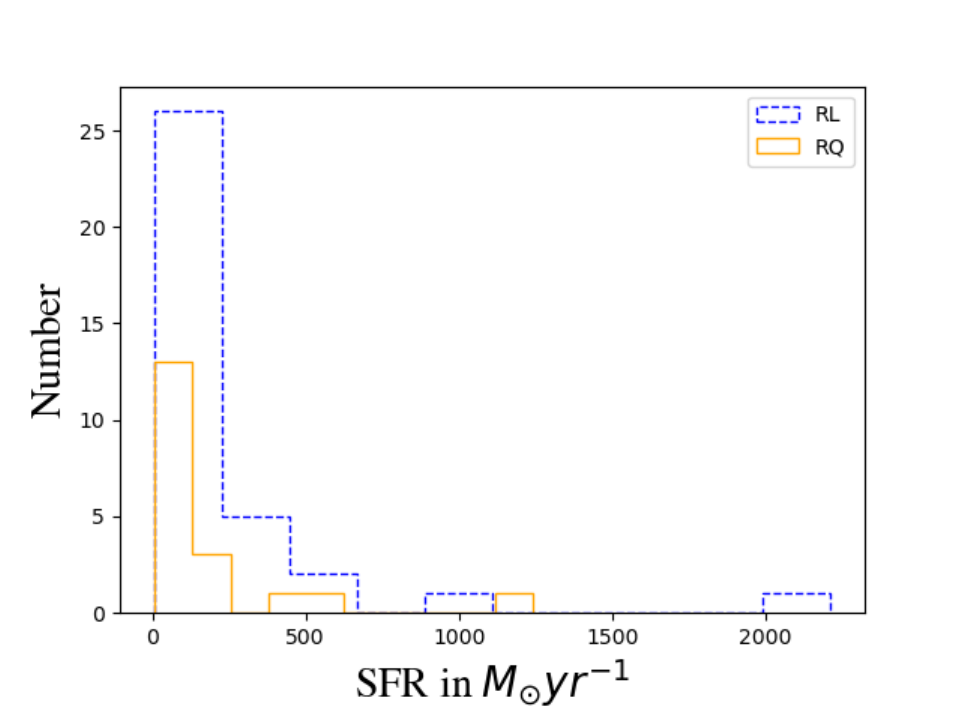}}
\resizebox{6.5cm}{!}{\includegraphics{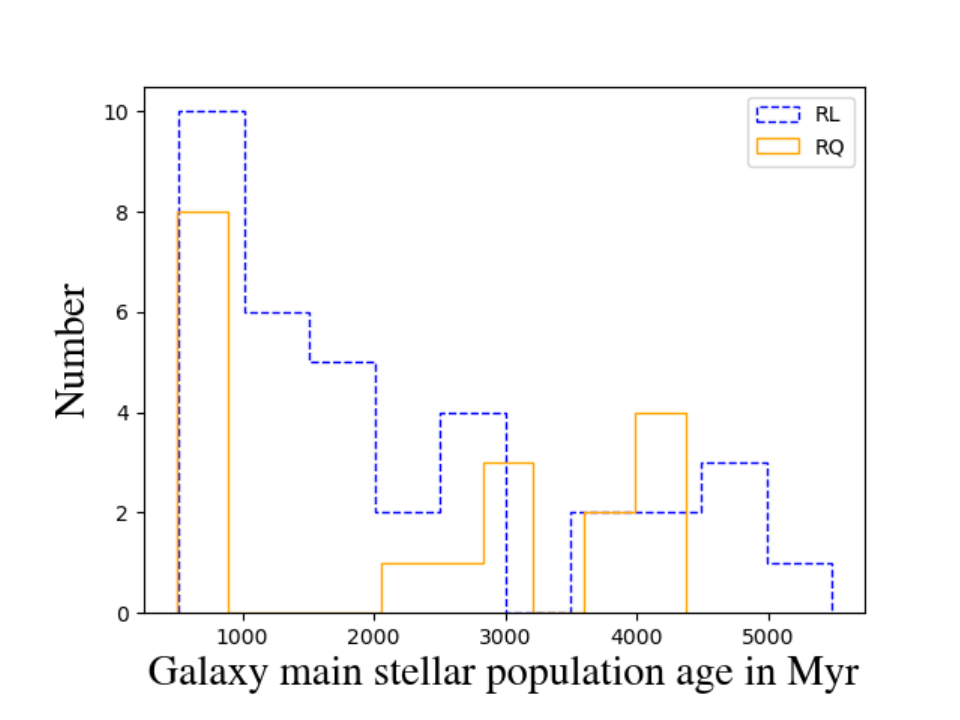}}\\
\resizebox{6cm}{!}{\includegraphics{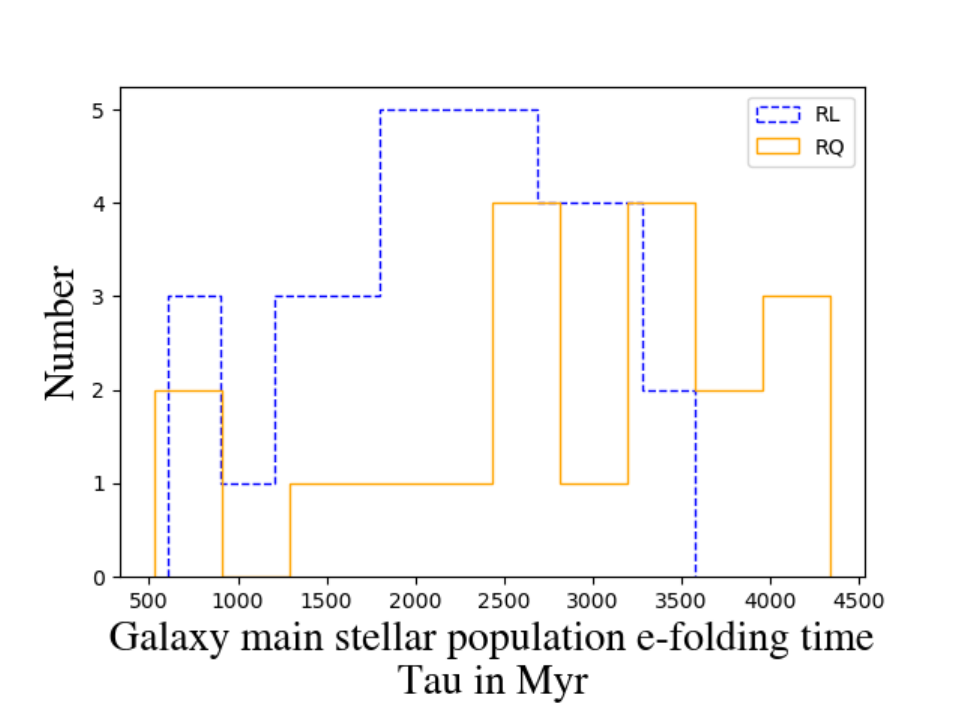}}
\resizebox{6cm}{!}{\includegraphics{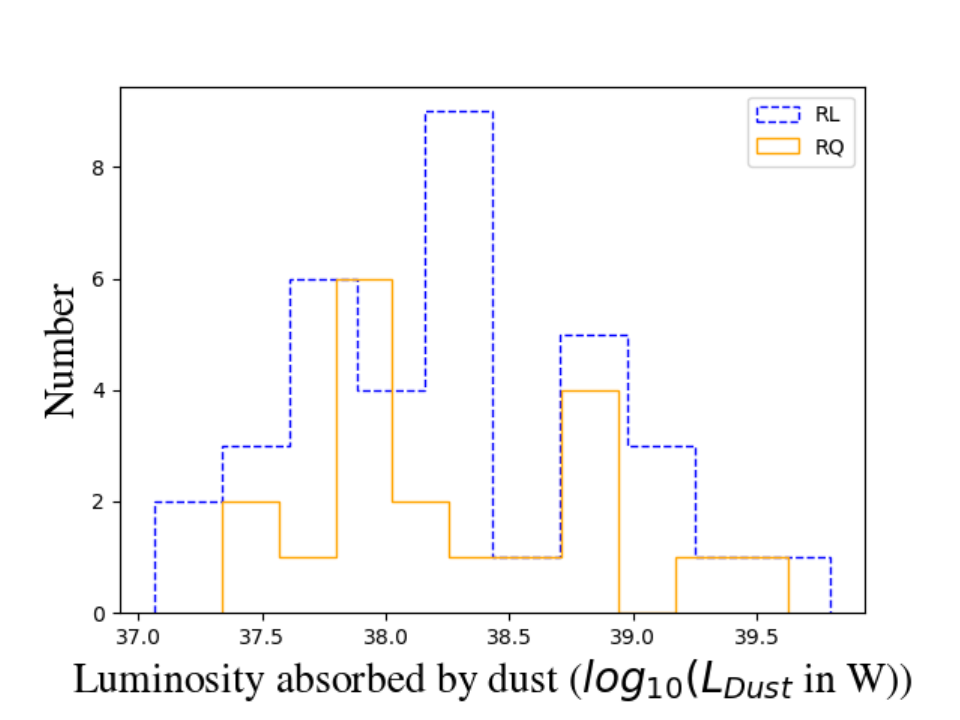}}
\end{tabular}
\caption{{\bf Top Left Panel:} Orange solid and blue dashed lines denote the histograms of stellar mass ($M_{\star}$) for the host galaxies of radio-quiet and radio-loud quasars of the non-HBL+HBL+\textit{AstroSat}-observed samples. {\bf Top Middle Panel: }Same as the top left panel but for the SFR distributions. {\bf Top Left Panel: }Same as the top left panel but for the stellar population age distributions. {\bf Bottom Left Panel: }Same as the top left panel but for the e-folding time distributions. {\bf Bottom Right Panel: }Same as the top left panel but for the dust luminosity distributions. The mean values of the parameters for our RL and RQ sources are provided in Table \ref{tab:galaxy_properties}.} 
\label{fig:Galaxy properties}
\end{center}
\end{figure*}

\begin{figure*}
\begin{center}
\begin{tabular}{c}
\resizebox{9cm}{!}{\includegraphics{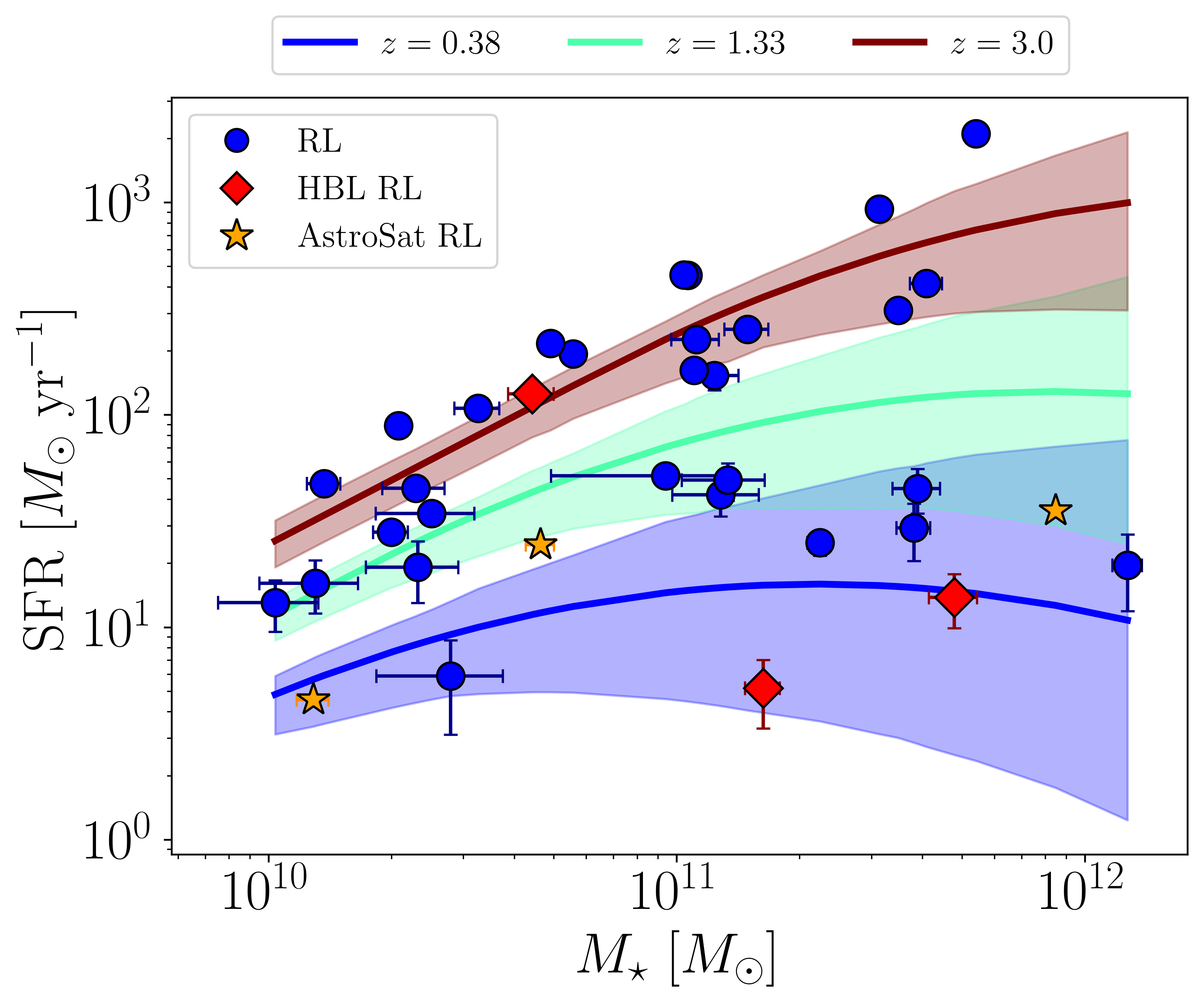}}
\resizebox{9cm}{!}{\includegraphics{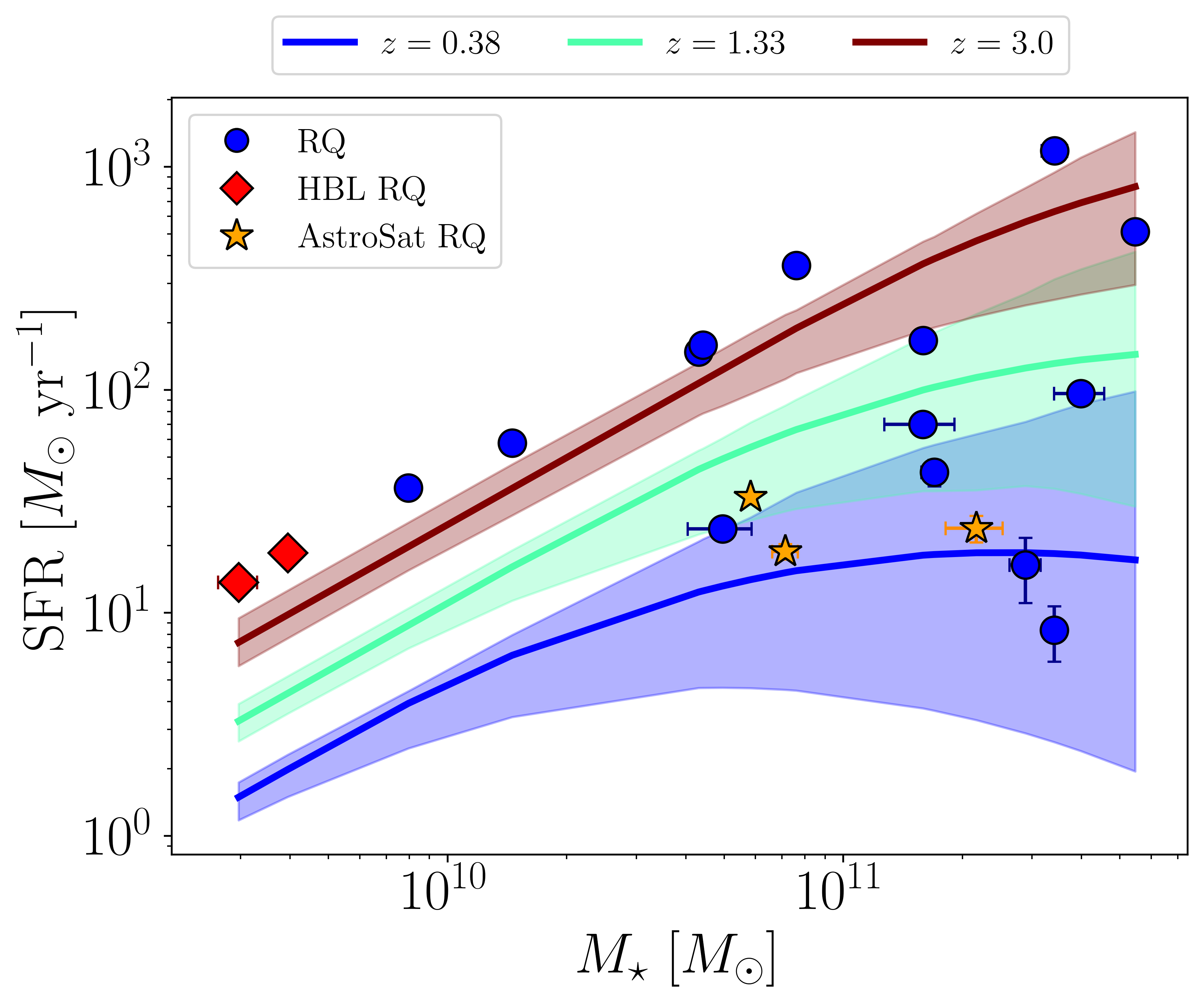}}\\
\resizebox{9cm}{!}{\includegraphics{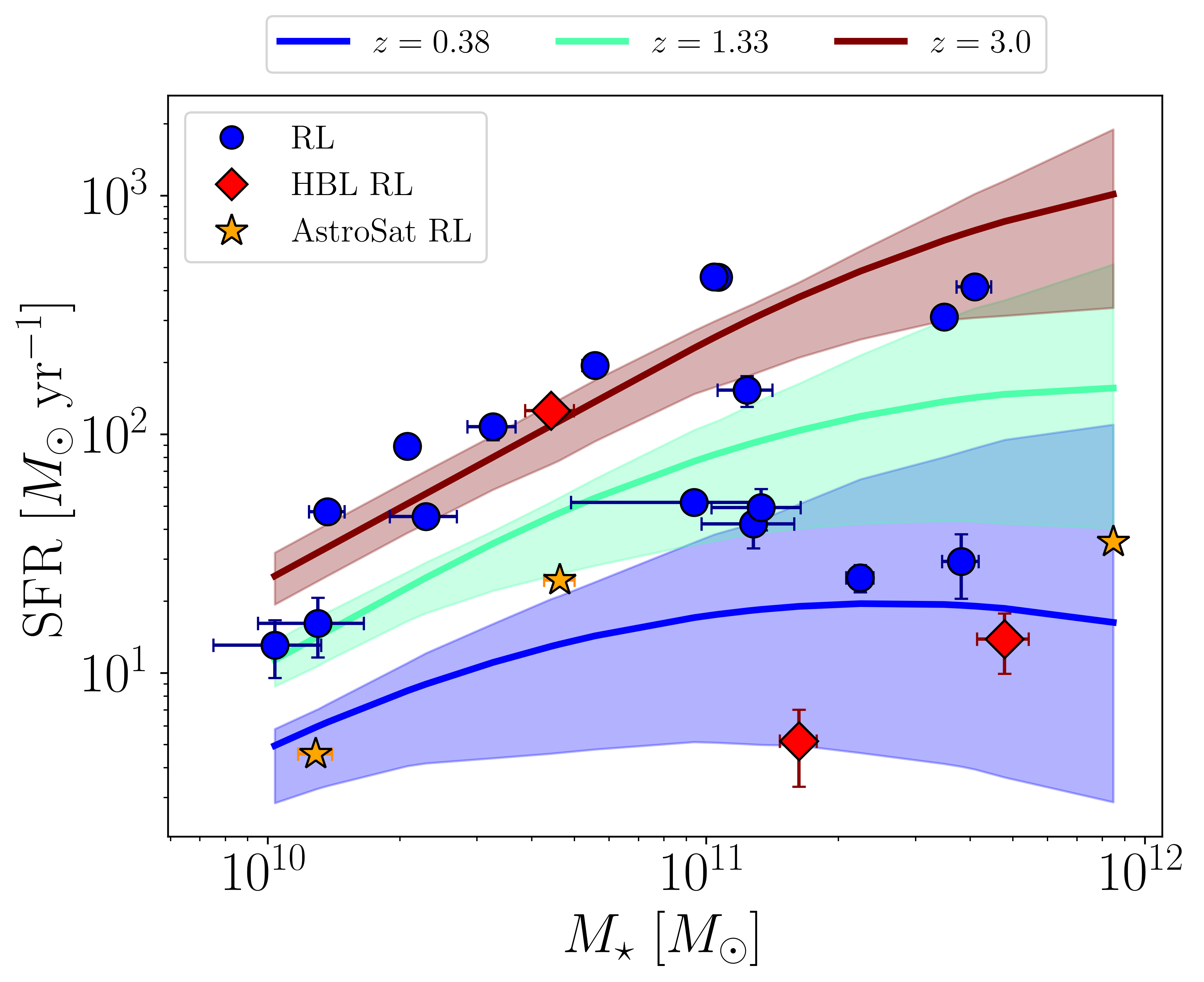}}
\resizebox{9cm}{!}{\includegraphics{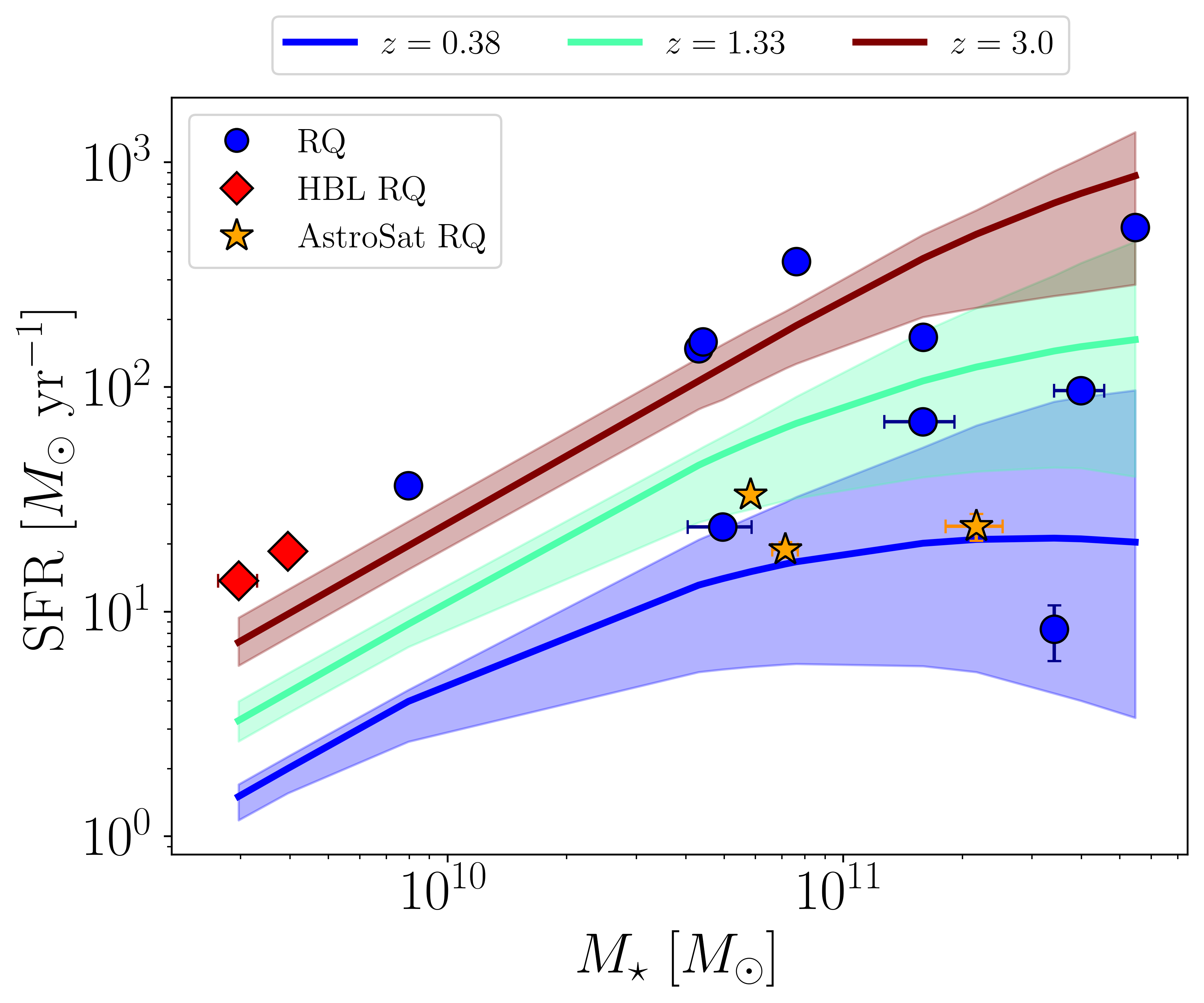}}\\
\end{tabular}
      \caption{The variation of stellar mass ($M_{\star}$) of the host galaxies with their corresponding SFR. Overplotted solid lines are analytical galaxy main-sequence (MS) relations from \citet{Schreiber2015}, at different redshifts with corresponding 1-$\sigma$ uncertainty contours. We note that the highest redshift of our sample is $z \sim 1.8$ (see Fig.\ \ref{fig:redshift}), and thus, the $z = 3$ trend lies outside our sample's redshift range. The majority of the sources in our sample lie significantly away from the galaxy MS and tend to follow the high redshift ($z = 3$) trend. In Fig.\ \ref{fig:MS_z} we color-code our sources with redshifts.
      {\bf Top Left Panel: } For entire RL non-HBL, HBL and \textit{AstroSat}-observed sources. {\bf Top Right Panel: } For entire RQ non-HBL, HBL and \textit{AstroSat}-observed sources. {\bf Bottom Left Panel: } For RL non-HBL, HBL and \textit{AstroSat}-observed sources that follow the criteria of \citet{Mountrichas21}. {\bf Bottom Left Panel: } For RQ non-HBL, HBL and \textit{AstroSat}-observed sources that follow the criteria of \citet{Mountrichas21}. The observed bimodality in the MS for RL quasars is evident. See Fig.\ \ref{fig:Eddington_ratio} and discussions in \S 4. }
    \label{fig:S15_noZ}
    \end{center}
\end{figure*} 

\begin{figure*}
\begin{center}
\begin{tabular}{c}
\resizebox{9cm}{!}{\includegraphics{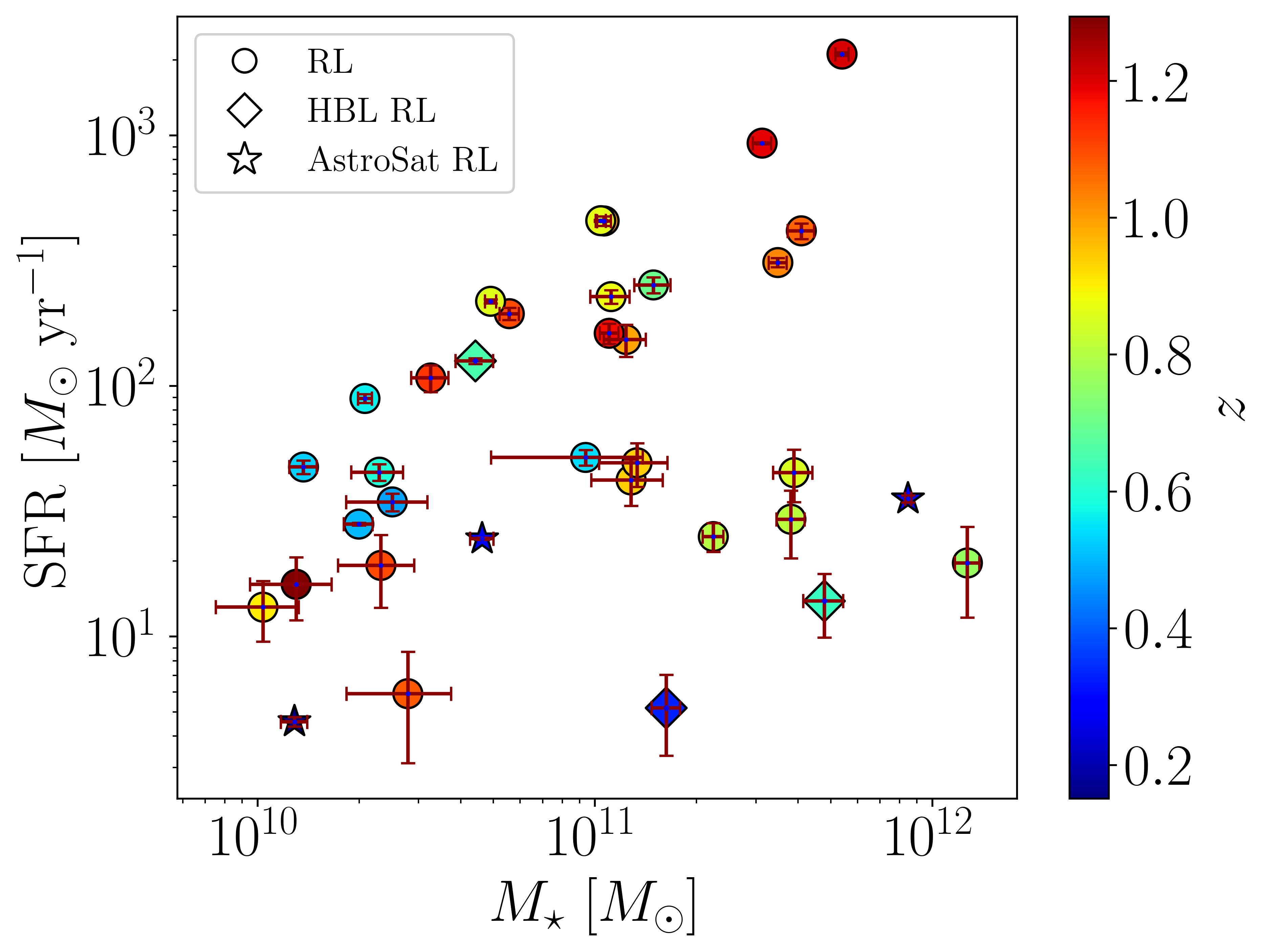}}
\resizebox{9cm}{!}{\includegraphics{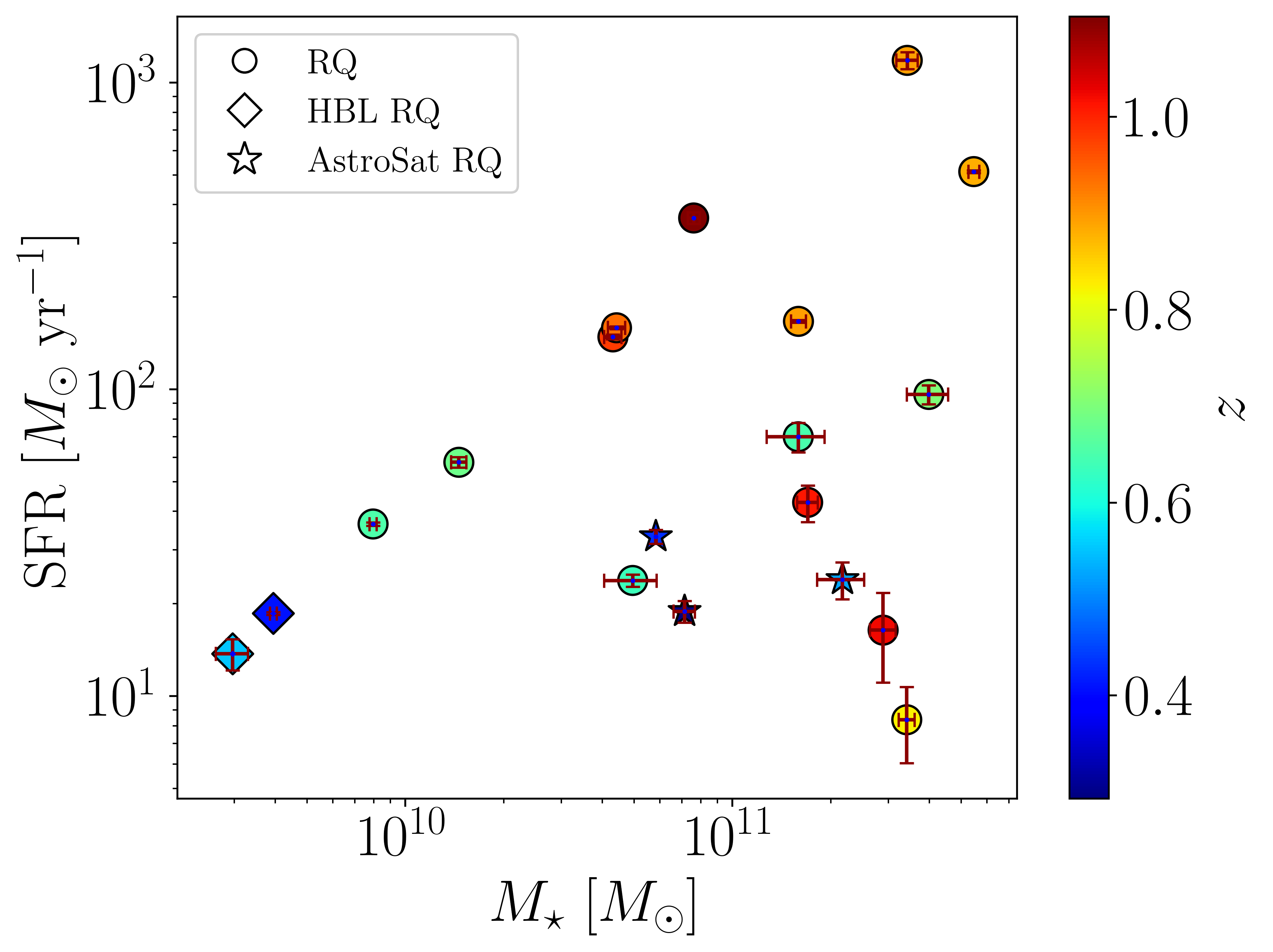}}\\
\resizebox{9cm}{!}{\includegraphics{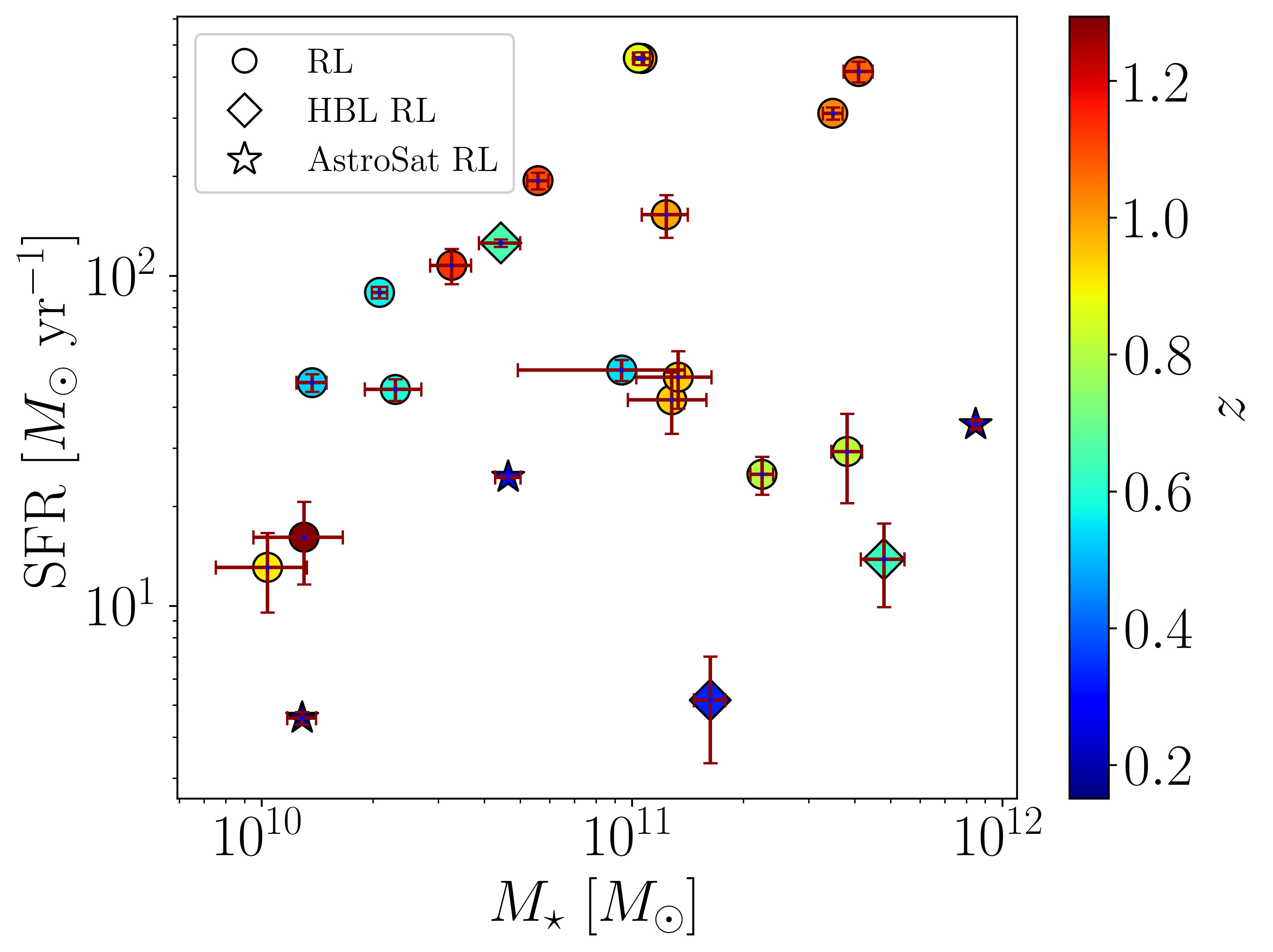}}
\resizebox{9cm}{!}{\includegraphics{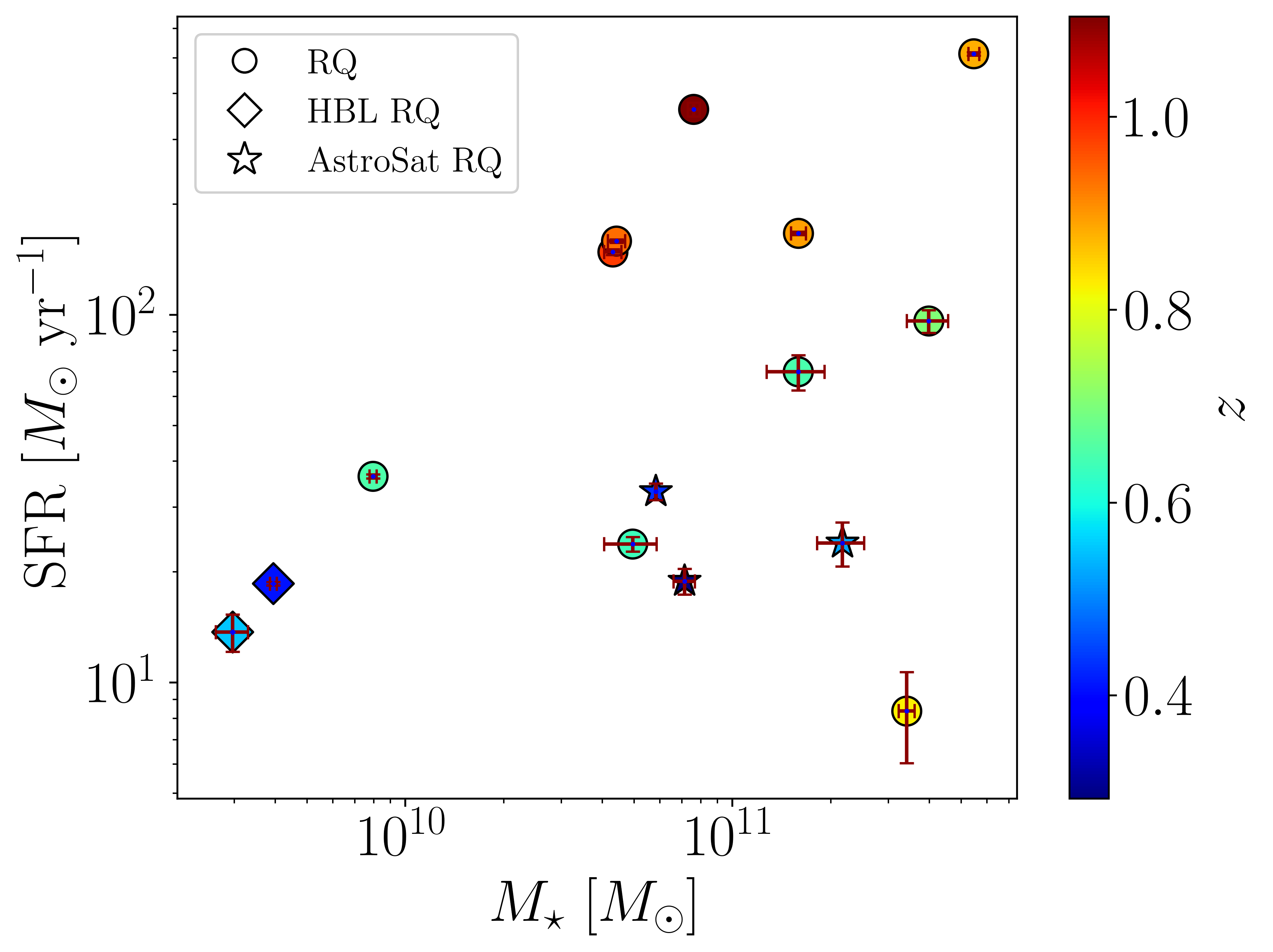}}\\
\end{tabular}
      \caption{The variation of stellar mass ($M_{\star}$) of the host galaxies with their corresponding SFR for our sources (Fig.\ \ref{fig:S15_noZ}). The color bars represent the redshift of each source. {\bf Top Left Panel: } For entire RL non-HBL, HBL and \textit{AstroSat}-observed sources. {\bf Top Right Panel: } For entire RQ non-HBL, HBL and \textit{AstroSat}-observed sources. {\bf Bottom Left Panel: } For RL non-HBL, HBL and \textit{AstroSat}-observed sources that follow the criteria of \citet{Mountrichas21}. {\bf Bottom Right Panel: } For RQ non-HBL, HBL and \textit{AstroSat}-observed sources that follow the criteria of \citet{Mountrichas21}. We do not observe any significant redshift effect in the main-sequence relation.}
    \label{fig:MS_z}
    \end{center}
\end{figure*} 

\begin{figure}
    \centering
    \includegraphics[width = \columnwidth]{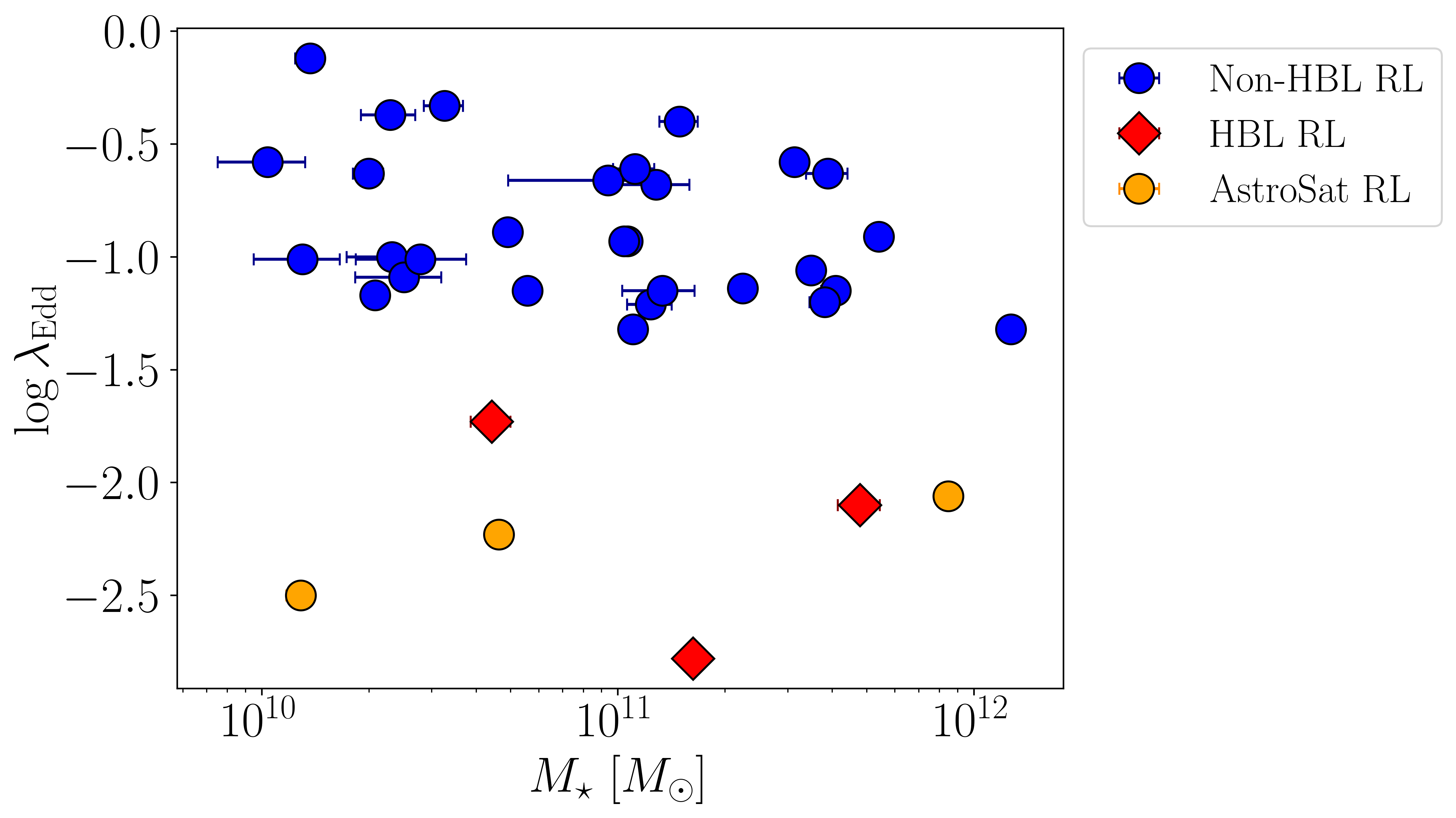} 
    \includegraphics[width = \columnwidth]{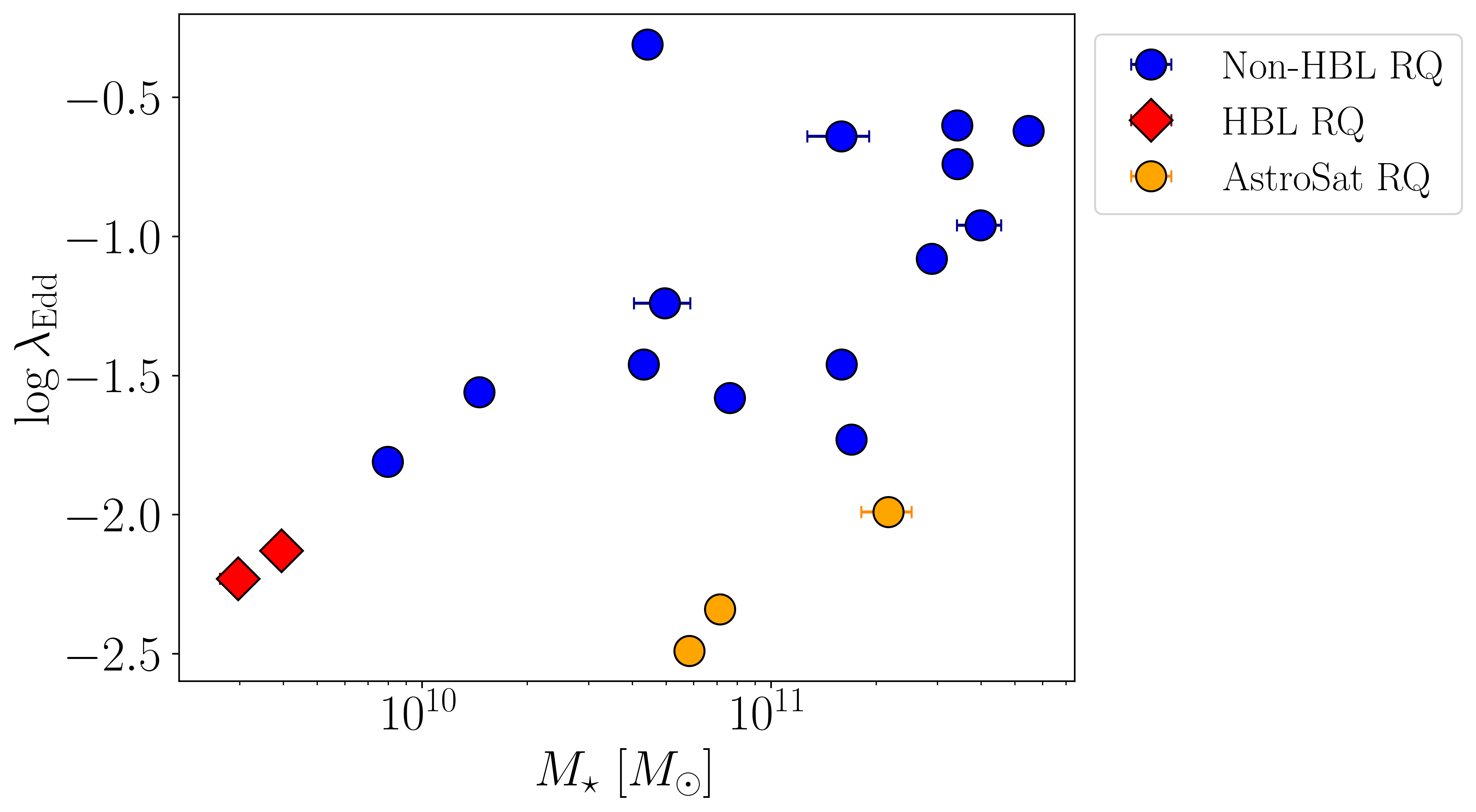}     
    \caption{The variation of stellar mass ($M_{\star}$) with the Eddington ratio ($\lambda_{\rm Edd}$) of the central quasar. {\bf Top Panel: } For RL non-HBL, HBL and \textit{AstroSat}-observed sources. {\bf Bottom Panel: } For RQ non-HBL, HBL and \textit{AstroSat}-observed sources. See, Fig.\ \ref{fig:S15_noZ} and \S\ref{subsec:discussion} for discussions.}
    \label{fig:Eddington_ratio}
\end{figure}

\section{Results}
As mentioned before, our RL and RQ samples have five detections in the optical band ($u$, $g$, $r$, $i$, and $z$ photometric bands from SDSS), four detections in the infrared band ($W1$, $W2$, $W3$, and $W4$ bands from WISE), two UV detections (FUV and NUV bands of GALEX), and one X-ray detection (Chandra/ROSAT/Swift/XMM) with their respective errors. In addition to that our RL quasars have one radio detection (FIRST 1.4 GHz band). \cite{shen} do not include errors for FIRST detections, so we have assumed an error of 10\% in the radio flux. Figure \ref{fig:redshift} represents the redshift distributions of our final RL and RQ samples.

\subsection{SED modeling using \texttt{\texttt{\texttt{CIGALE v2022.0}}}}
  Based on our input parameters described in Table \ref{table:CIGALE parameters} we model the SEDs of our sources described in Table \ref{tab:RL quasars} and Table \ref{tab:RQ quasars}. We note that the model parameters have been included after extensive parameter studies which followed analysis of previous studies \citep[e.g.,][]{Mountrichas21, mountrichas22a, mountrichas23a, Mountrichas24}.
  Figures \ref{fig:RL_SEDs} and \ref{fig:RQ_SEDs} represent some example SEDs of our non-HBL RL and RQ sources used in our analysis. Figure \ref{fig:HBL_SEDs} represents\ SEDs of HBL RL and RQ sources and Figure \ref{fig:AstroSat_w_FUV} represents the SEDs of RL and RQ sources observed with \textit{AstroSat}. For HBL sources and the sources observed with \textit{AstroSat}, we did not find any X-ray counterpart (see \S\ref{subsec:Data Sets}). For the SED fitting of our \textit{AstroSat} sources we have used a box-car filter centered around 1481 \AA~as the response function for the ${\rm CaF_2}$ grating of \textit{AstroSat}/UVIT FUV observations of these six sources.

\subsection{Extraction of Host Galaxy Properties}
\label{subsec:Host_Galaxy_Properties}
From the best-fit SED, we computed the stellar population age, stellar mass, star-formation rate (SFR), dust luminosity and e-folding time of our RL and RQ quasar samples. Figure \ref{fig:Galaxy properties} represents distributions of the host galaxy properties for our RL and RQ sources (non-HBL-SED+HBL+\textit{AstroSat}-observed). From the best-fit values, we note that the stellar masses are similar for both RL (mean $10.95 \pm 0.097$ $M_{\odot}$) and RQ (mean $10.86 \pm 0.15$ $M_{\odot}$) quasars although the stellar masses extend to lower values for RQ quasar host galaxies (see top-left panel of Fig.\ \ref{fig:Galaxy properties}). The luminosity absorbed by dust (for RL mean is $38.28 \pm 0.11$ W and for RQ mean is $38.29 \pm 0.14$ W), the e-folding time (for RL mean is $2190.7 \pm 129.96$ Myr and for RQ mean is $2266.6 \pm 249.33$ Myr), stellar population age (for RL mean is $2080.0 \pm 246.29$ Myr and for RQ mean is $2850.7 \pm 356.69$ Myr) and the SFR (for RL mean is $199.15 \pm 67.88$ $M_{\odot}$/yr and for RQ mean is $163.70 \pm 67.52$ $M_{\odot}$/yr) are similar for both the populations. The mean values are provided in Table \ref{tab:galaxy_properties}. We note that the values remain statistically similar for all the subclasses together and for each subclass separately however some striking features are evident in the stellar mass and star-formation rate properties for our HBL sources which we discuss later. 

To check for the consistency of our modeling we fitted the SED with the optical+IR+UV, optical+UV, and optical+IR data points. Our results show that excluding the UV leads to an underestimation of the stellar mass while excluding the IR data affects the best-fit values of the SFR. We note that the UV emission traces young stellar populations and for higher redshift sources SDSS u-band might allow observations of radiation emitted by young stars. Hence, it is likely that the absence of UV data might primarily affect SED fitting measurements at lower redshifts. To test for this, we randomly selected RL and RQ quasars at different redshifts and performed the SED analysis with and without the UV coverage. We did notice that the effect is stronger at lower redshifts, but the offset (which is always an underestimation of stellar mass) does not exhibit any systematic pattern. Previous studies using X-ray AGN have found that the absence of UV coverage has a negligible impact on SED analysis, even at low redshifts, in terms of SFR calculations \citep[e.g][]{Koutoulidis22, mountrichas22a}. Since our UV coverage substantially reduces the sample size, we wish to perform a more extensive analysis of our sources for studying host galaxy correlations in future work (Chatterjee et al. 2025, in preparation). As expected, the exclusion of the radio and the X-ray data has almost no effect on the inferred stellar mass or SFR. We are thus confident about the robustness of our fits. 

 We also note that \citet{mountrichas23a} found that the star-formation history (SFH) parameters are highly degenerate. For example, the degeneracy between stellar age and e-folding time has been discussed. To test for this we adopted the method proposed by \citet{mountrichas23a} where we include the Hdelta and Dn4000 spectral parameters along with our initial values of other parameters to generate mock fluxes for our sources. We then extract the best-fit host galaxy parameters from these mock catalogs. We find the best-fit values from the mock catalogs to be statistically identical with our derived best-fit values validating the robustness of our fits.  

 To check for the dependence of host galaxy properties with redshift and FWHM (of either the H$\beta$ or Mg\textsc{ii} lines), we divided the entire sample into two equally splitted redshift bins (the lower redshift bin for $z \le 0.9$ and the higher redshift bin for $z > 0.9$) and 3 FWHM bins ($1500 \le {\rm FWHM} < 5000~{\rm km}\cdot{\rm s}^{-1}$; $5000 \le {\rm FWHM} < 8000~{\rm km}\cdot{\rm s}^{-1}$; and ${\rm FWHM} > 8000~{\rm km}\cdot{\rm s}^{-1}$). We do not find either of the host galaxy properties to be dependent on redshift or FWHM, however, we note that the sample sizes are not significant to rule out any such dependence.

To further test for the quality of our fits, we follow the prescription of \citet{Mountrichas21} by checking the consistency between the Bayesian and best-fit values of host galaxy parameters obtained from \texttt{CIGALE v2022.0}. A large difference between the Bayesian values (which considers the entire parametric grid of all allowed models, with a weight of $\exp(-\chi^2 / 2)$ for each model) and the best-fit values, indicates that a Gaussian probability distribution function (PDF) cannot correctly reproduce the observed data. As per \citet{Mountrichas21}, we consider the following criteria: $\frac{1}{5} \le \frac{{\rm SFR}_{\rm best}}{{\rm SFR}_{\rm Bayes}} \le 5 $ and $\frac{1}{5} \le \frac{{M_{*,}}_{\rm best}}{{M_{*,}}_{\rm Bayes}} \le 5 $. We see that from our final sample, 54\% of the radio-loud and 75\% of the radio-quiet sources satisfy these criteria (M21 sub-sample hereafter). We further note that sources, for which the best-fit parameter values of the host galaxy contain large uncertainties are majorly the ones that do not satisfy the above criterion. 

In Fig.\ \ref{fig:S15_noZ} we plot the $M_{\star}-SFR$ relations for our full as well as the M21 sub-sample. As noted, the highest and the lowest redshifts of our sources are 1.8 and 0.15 respectively, while the median redshift is about 1.3. The solid lines in all panels represent the main-sequence relation for three redshifts (.38: blue, 1.33: cyan, and 3.0: brown) from \citet{Schreiber2015}. The choice of $z= 0.38$ comes from the lower limit in redshifts of our sources barring a few of the {\it Astrosat} sources. The top and bottom rows are for the full and the M21 sub-samples respectively. We clearly see that our RL and RQ quasars are off from the \citet{Schreiber2015} relation. In fact, a large number of our quasar sources follow the $z = 3.0$ main-sequence relation for star-forming galaxies. When we consider our HBL sources (red diamonds) we note a distinct feature. Our HBL RL sources tend to have higher stellar mass with lower SFRs while for the RQ ones we obtain lower stellar mass with higher SFRs. In addition, we observe a clear hint of dichotomy in the $M_{\star}-{\rm SFR}$ relation for our RL sources at the higher stellar mass range. The dichotomous trend is more evident for the M21 sub-sample (bottom left panel of Fig.\ \ref{fig:S15_noZ}). 

Simple power-law fits to the data show that for our RL sources ${\rm SFR} \propto M_{\star}^{-0.4 \pm 0.02}$ and ${\rm SFR} \propto M_{\star}^{0.5 \pm 0.01}$ when we consider the full sample and the non-HBL (blue dots in Fig.\ \ref{fig:S15_noZ}) sample respectively. For the RQ sources ${\rm SFR} \propto M_{\star}^{-0.2 \pm 0.01}$ and ${\rm SFR} \propto M_{\star}^{0.2 \pm 0.01}$ respectively for the full and the non-HBL sample. The trends do not change for the M21 sub-sample. We thus observe that while the mean values of stellar mass and SFR are similar, the average SFR for RL scales higher with stellar mass compared to the RQ population. In a companion paper (Chatterjee et al. 2024, in prep) we perform a detailed study of the host galaxy correlations of our sample using the best-fit values derived from the SED modeling. 

To check for any trend in the observed main-sequence relation with redshifts we plot the main-sequence relation in Fig.\ \ref{fig:MS_z} but are now color-coded by the redshift of the sample. We note that our HBL quasars have a distribution of redshifts, yet the SFR tends to be low for the HBL-RL sources. A somewhat similar trend is also observed for our \textit{AstroSat} sources. We note that our HBL and \textit{AstroSat} sources did not have any X-ray counterparts. It was also noted in \citet{chakraborty22} that the HBL sample had the lowest Eddington ratio. To test for that in Fig.\ \ref{fig:Eddington_ratio} we plot the Eddington ratios of our sources as a function of their derived stellar mass. A clear hint of dichotomy in the Eddington ratio distribution is observed in our RL quasars. We further note that our HBL and \textit{AstroSat} sources have lower Eddington ratios than our non-HBL sources. Thus we clearly observe a connection between SFR and Eddington ratio. Quasars with lower accretion rates tend to have lower SFR in their host galaxies, which is a physically plausible scenario, hinting toward the unavailability of cold gas in these galaxies.  

\begin{figure*}
\begin{center}
\begin{tabular}{c}
\resizebox{7cm}{!}{\includegraphics{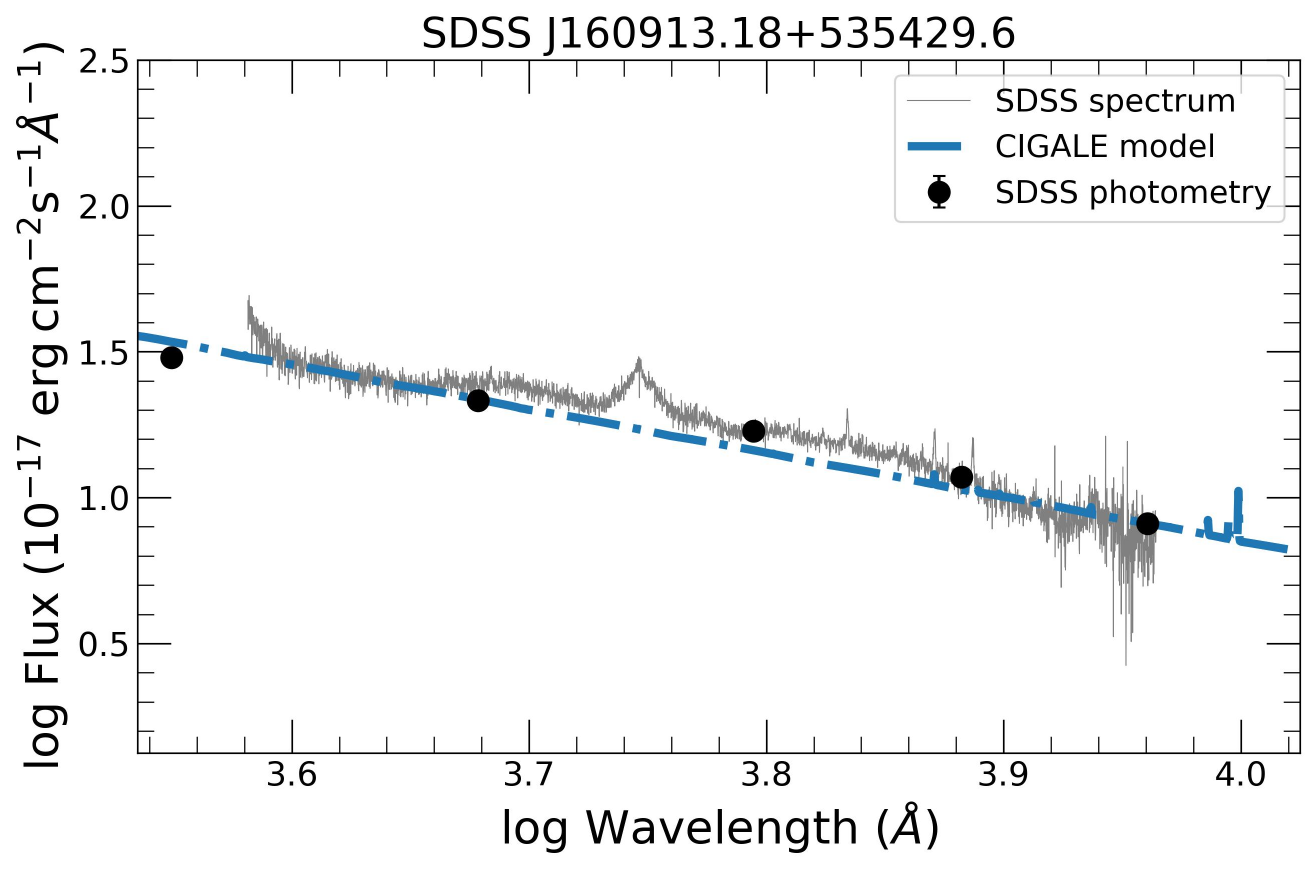}}
\hspace{0.2cm}
\resizebox{7cm}{!}{\includegraphics{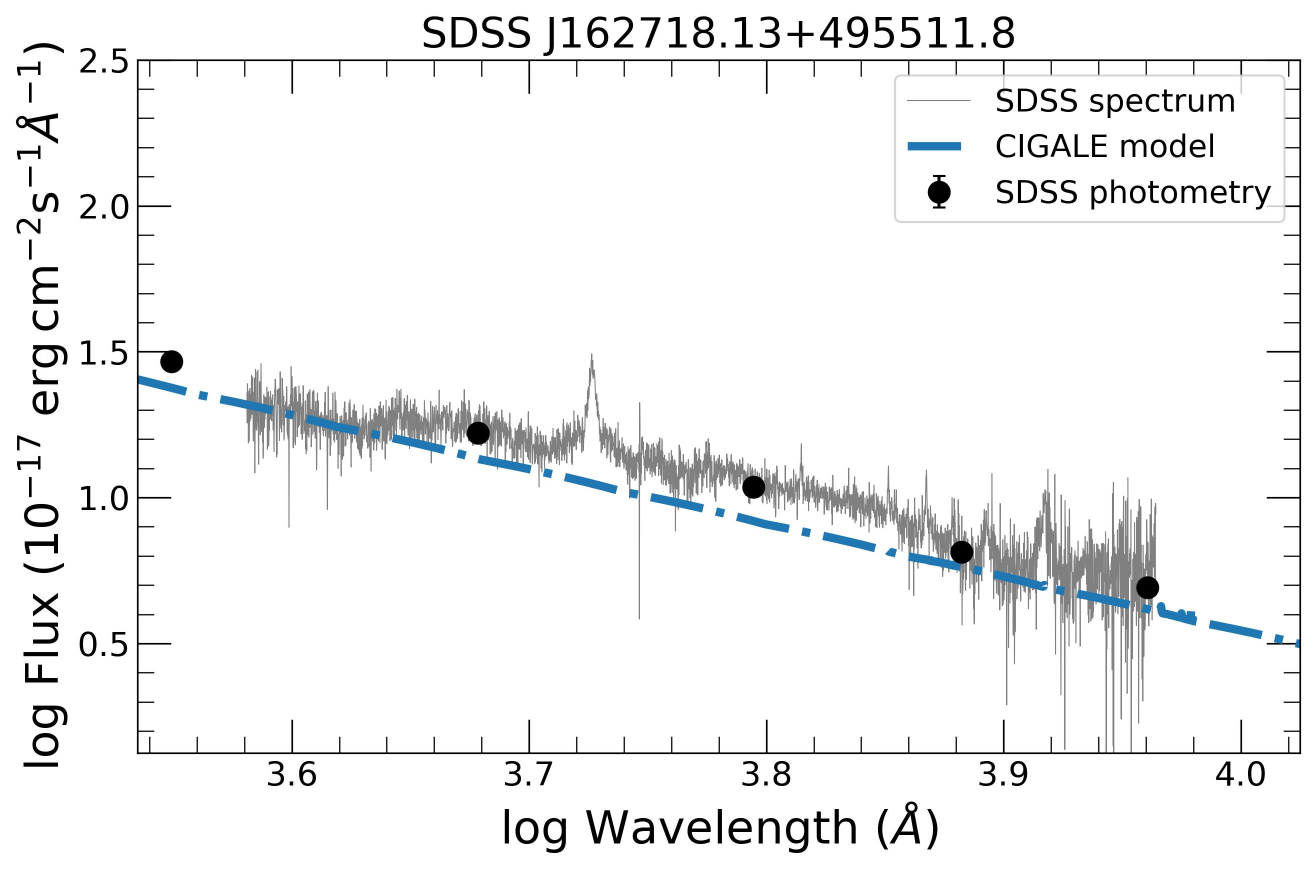}}\\
\resizebox{7cm}{!}{\includegraphics{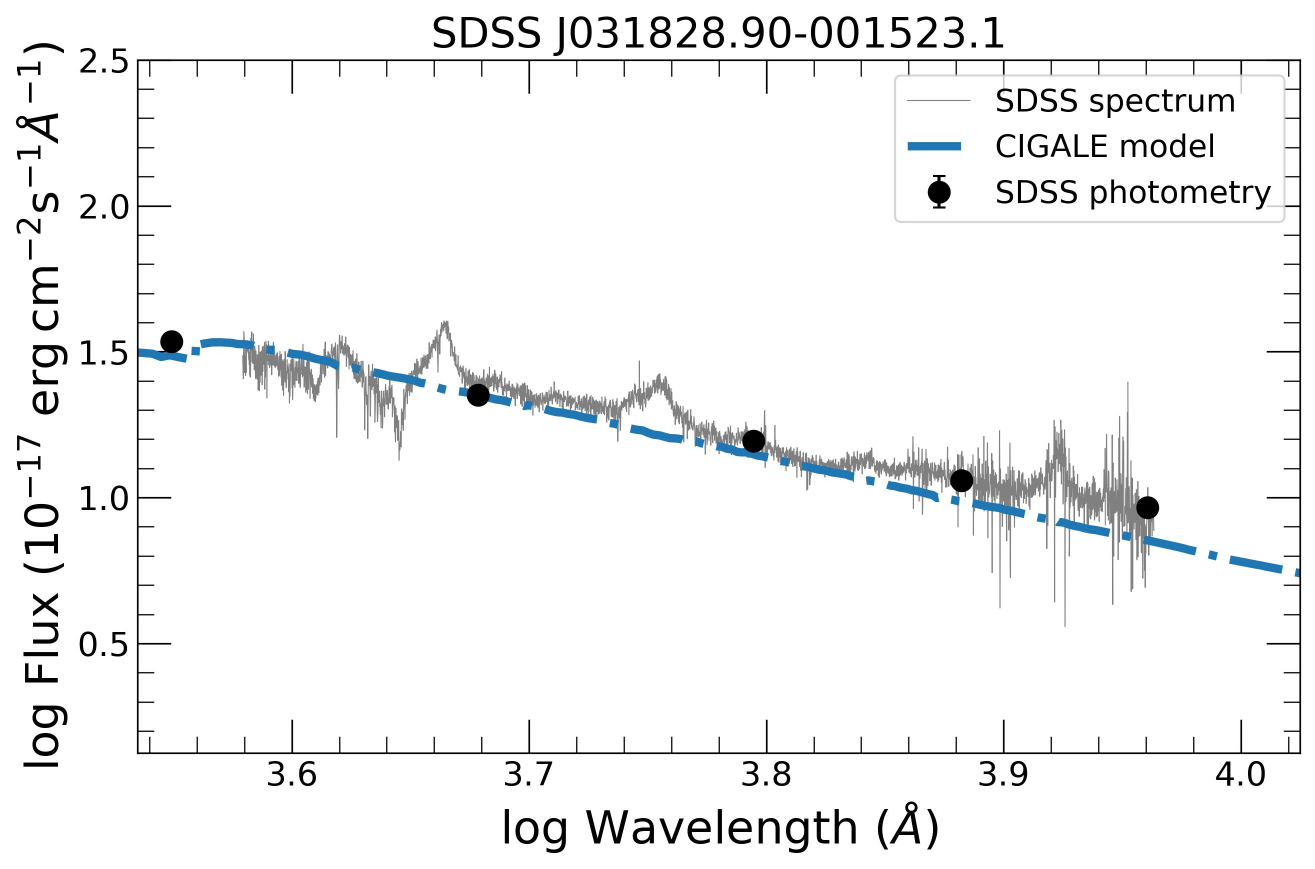}}
\hspace{0.2cm}
\resizebox{7cm}{!}{\includegraphics{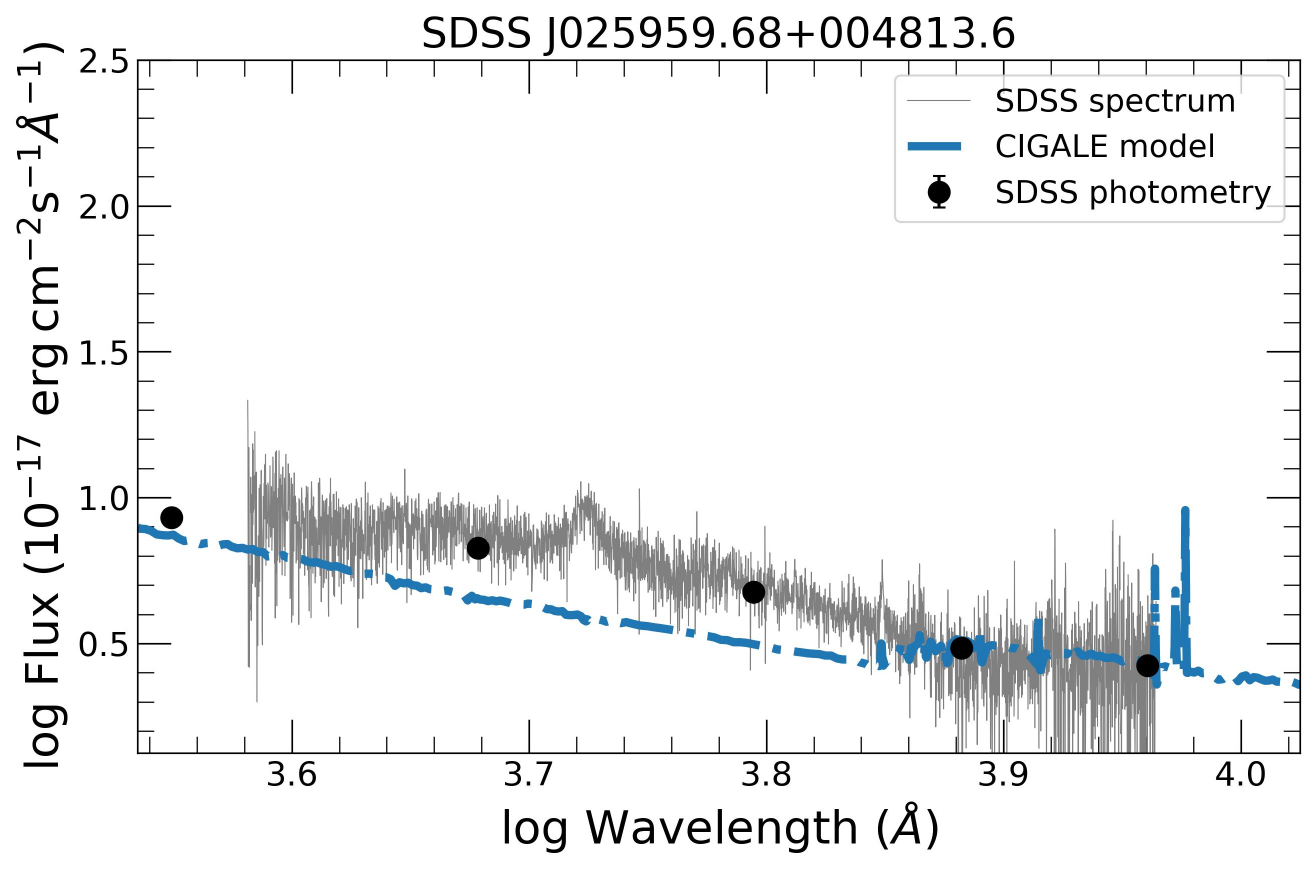}}\\
\resizebox{7cm}{!}{\includegraphics{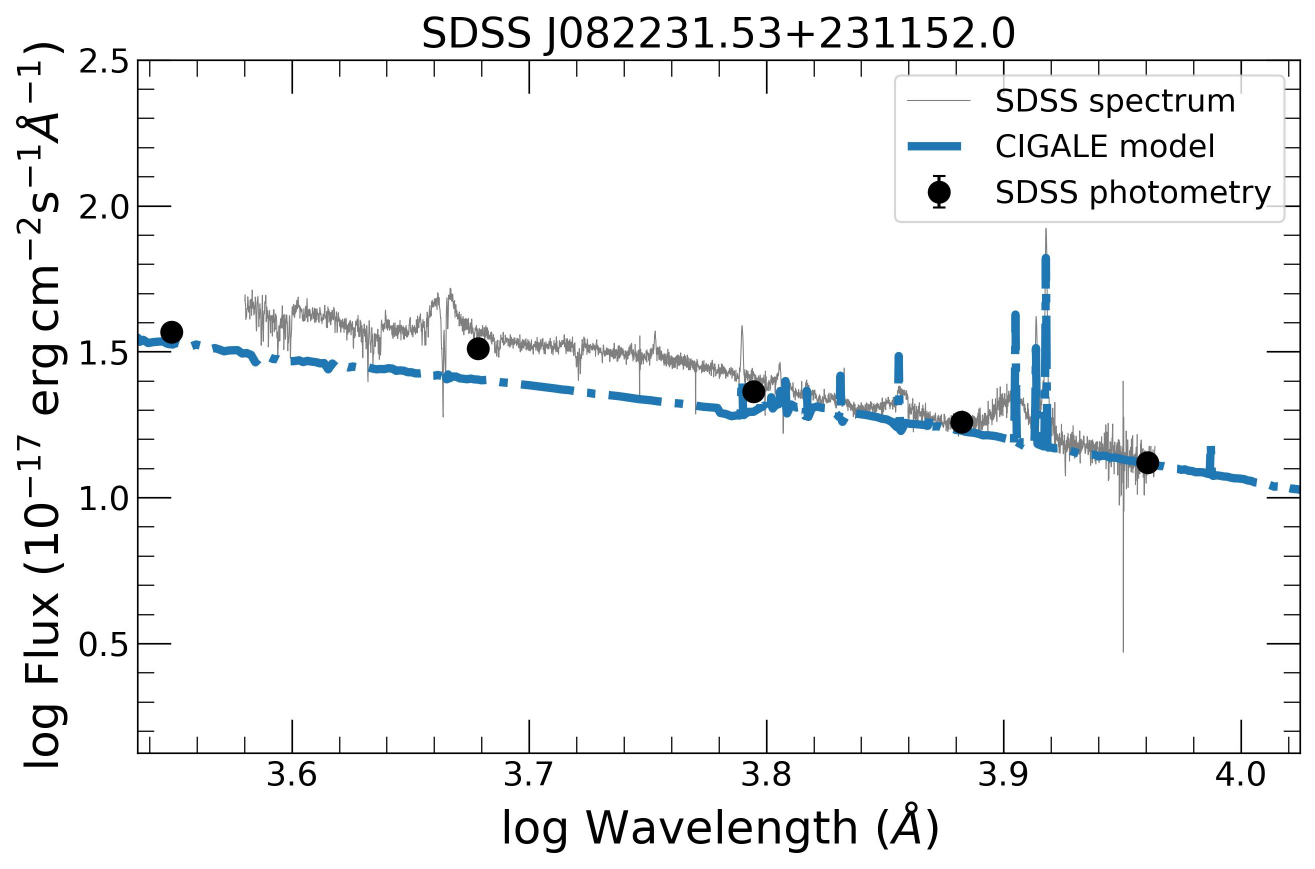}}
\hspace{0.2cm}
\resizebox{7cm}{!}{\includegraphics{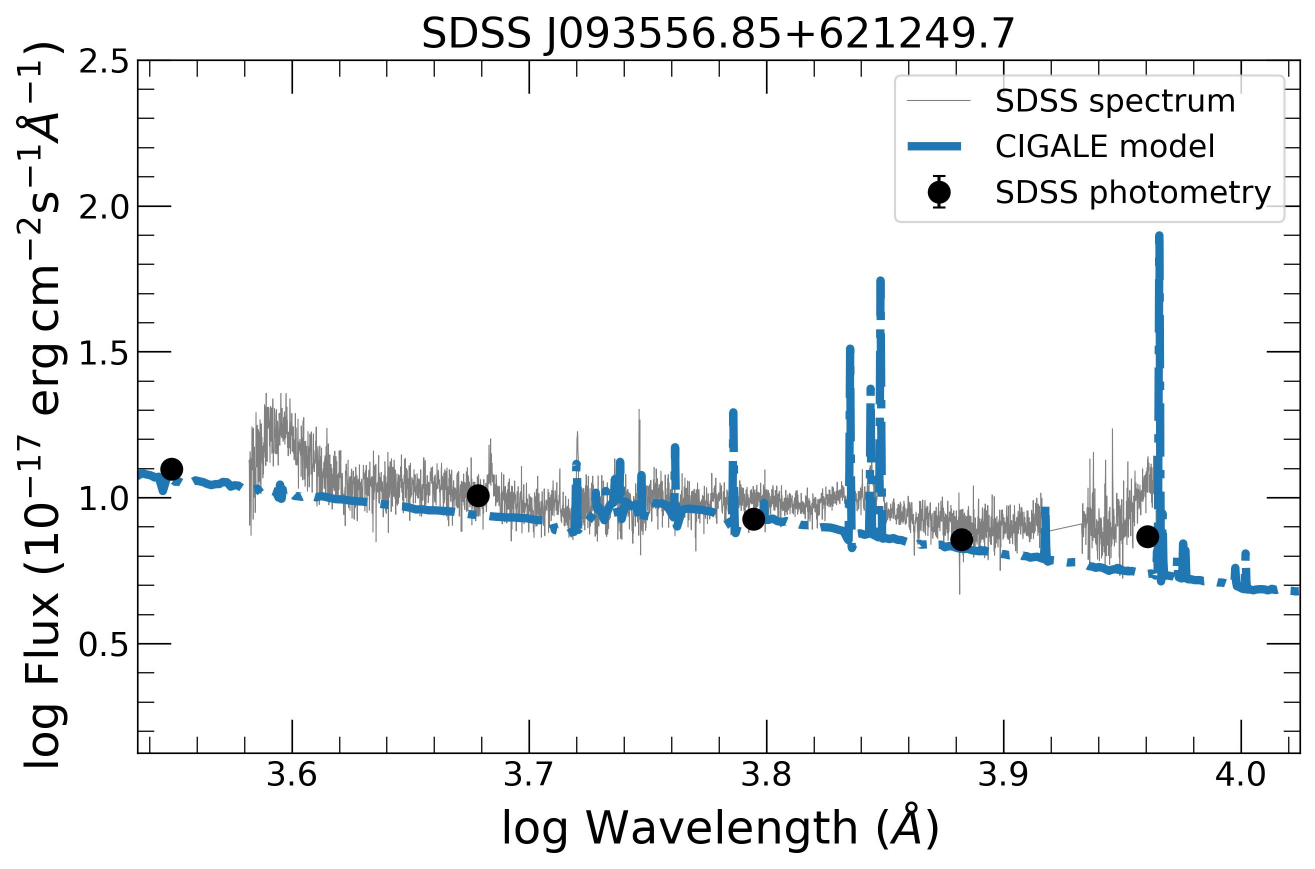}}\\
\end{tabular}
\caption{Examples of source spectra from SDSS overplotted with the best-fit SEDs from \texttt{CIGALE v2022.0} in the optical range. {\bf Top Panel: }For two non-HBL radio-loud quasars from our sample. {\bf Middle Panel: }For two non-HBL radio-quiet quasars from our sample. {\bf Bottom Left Panel: } For one HBL radio-loud quasar from our sample. {\bf Bottom Right Panel: }For one HBL radio-quiet quasar from our sample. Our SED modeling based on the photometric data is successful in reproducing the quasar spectra.}
\label{fig:spectra}
\end{center}
\end{figure*} 


\begin{figure}
    \centering
    \includegraphics[width = \columnwidth]{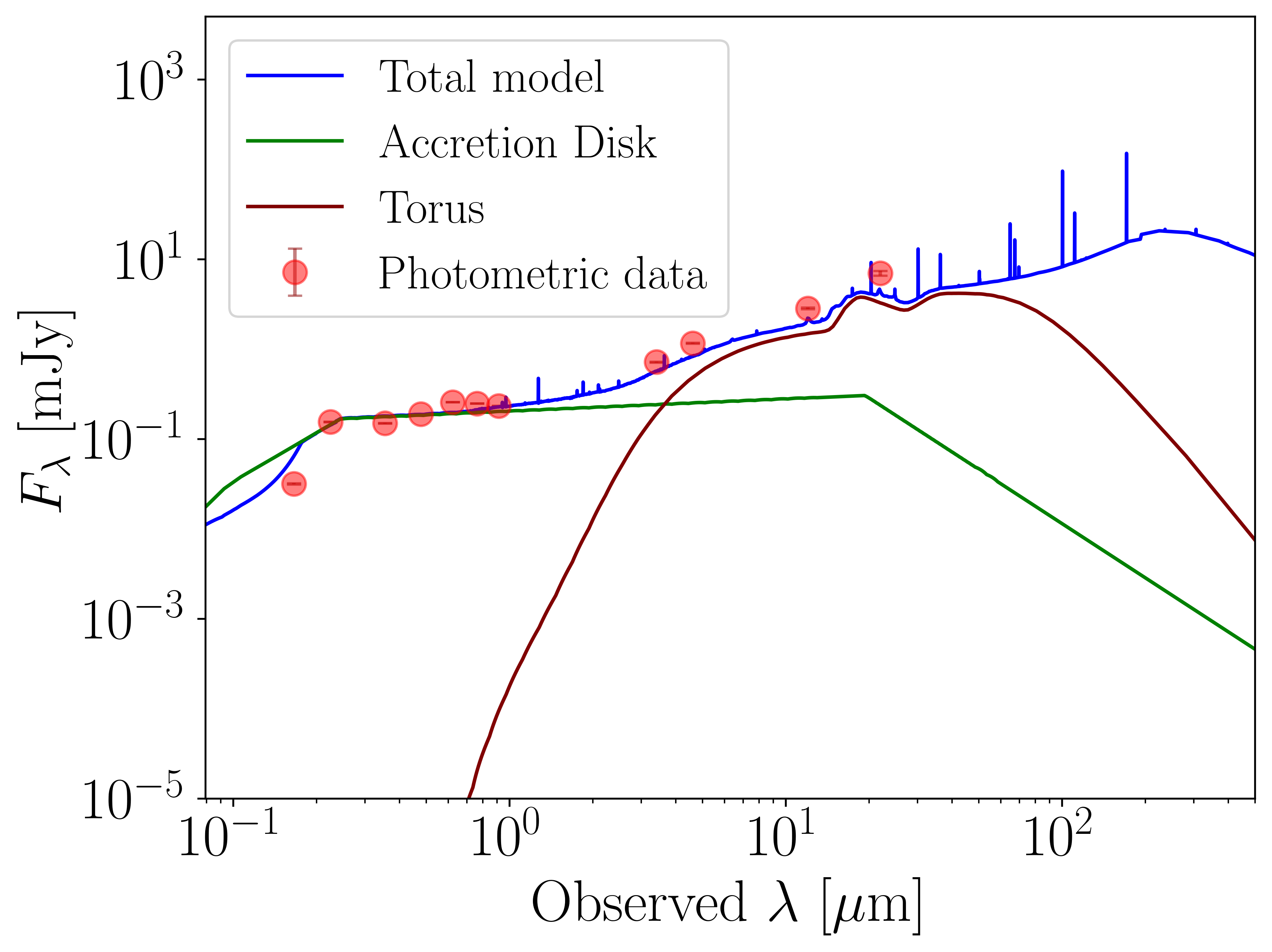}\\
    \includegraphics[width = \columnwidth]{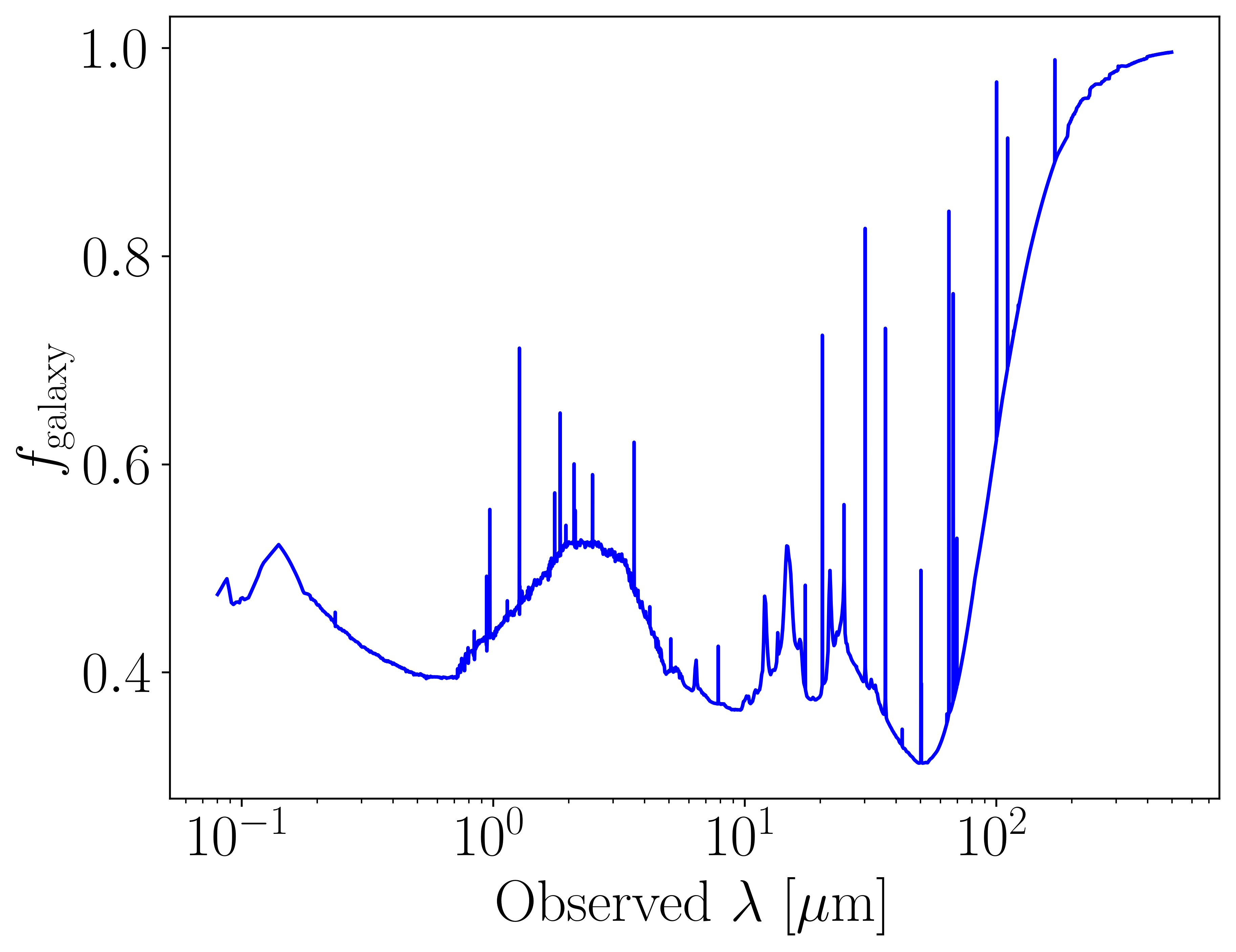}
    \caption{{\bf Top Panel:} The contribution of the principal AGN Components in the UV--IR range, the accretion disk and the torus, in the total SED of the host galaxy. {\bf Bottom Panel:} The relative contribution of the emission from the host galaxy in the IR--UV part of the SED, following Equation \ref{eqn:f_galaxy}. The emission lines are from the full-physics model and are not present in the AGN-only model. The host galaxy contributes about 50\% of the total luminosity in relatively lower wavelengths, and very little at higher wavelengths. See \S \ref{subsec:AGN_SED_simulation} and \S \ref{subsec:discussion} for more discussions.}
    \label{fig:AGN_Contribution_UV_IR}
\end{figure}

    

\subsection{Comparison with SDSS Spectra}
To highlight the overall agreement of the best-fit \texttt{CIGALE v2022.0} SED model with the observed data, we overplot the modeled SED with the input photometry from SDSS as well as the SDSS spectrum for the source from the same epoch (see panels in Figure \ref{fig:spectra}). The best-fit model is quite successful in reproducing multiple features in the observed spectrum, the primary among them is the power-law slope of the spectrum. In addition, the intensities of the narrow emission lines can be recovered, albeit with notably higher intensities from the model - a limitation of the finite grid in the ionization parameter adopted in the \texttt{CIGALE v2022.0} code. The panels in this figure also highlight the importance of the contribution from the broad emission lines (e.g., C\textsc{iv}, Mg\textsc{ii}, and H$\beta$; see e.g., \citealt{Panda_etal_2019FrASS, PozoNunez_etal_2023, 2024ApJS..272...11P}) and the Fe{\sc ii} pseudocontinuum \citep{Panda_etal_2018, Panda_etal_2019, Pandey_etal_2024} that is composed of multiple, overlapping transitions in the spectrum. 

The \texttt{CIGALE v2022.0} modeling does not account for these emission lines given the complexity of reproducing physical conditions for such emitting regions. One also notices minor offsets between the SDSS photometric data and the spectrum. The primary reason behind this is the photometric data results from the convolution of flux over the broad-band filters. Here, for a given redshift and accounting for the broad emission line contribution (there are other contributors such as the Balmer continuum, and FeII pseudocontinuum; see e.g., \citealt{PozoNunez_etal_2023, Czerny_etal_2023, Panda_etal_2023Univ, 2024ApJS..272...11P}) for the source, the minor discrepancies between the photometric data and spectrum can be mitigated. Nonetheless, the underlying power-law continuum of the AGN - the primary ionizing source, is well reproduced by our \texttt{CIGALE v2022.0} models.

\begin{table*} 
    \caption{Mean values of host galaxy properties for RL and RQ quasars of our sample}
    \centering
    \begin{tabular}{|c|c|c|} 
        \hline
            \textbf{Host galaxy} & \multicolumn{ 2}{|c|}{\textbf{Mean value}} \\ 
        \cline{2-3}
            \textbf{property} & \textsc{Radio-Loud Sources} & \textsc{Radio-Quiet Sources} \\
        \hline \hline 
            $M_*$ [$\log_{10}M_{\odot}$] & $11 \pm 0.1$ & $10.9 \pm 0.2$ \\ 
            $L_{\rm dust}$ [$\log_{10}$W] & $38.3 \pm 0.1$ & $38.3 \pm 0.1$ \\ 
            SFR [$M_{\odot}~{\rm yr}^{-1}$] & $199 \pm 68$ & $164 \pm 68$ \\ 
            $\tau$ [Myr] & $2190 \pm 130$ & $2267 \pm 249$ \\ 
            $t_{\rm age}$ [Myr] & $2080 \pm 246$ & $2851 \pm 357$ \\ 
        \hline
    \end{tabular}
    \label{tab:galaxy_properties}
\end{table*}

    

\subsection{AGN SED Simulation}
\label{subsec:AGN_SED_simulation}
As we are interested in extracting the host galaxy contribution and most of the observed photometric data are dominated by the central quasar, it is important to extract the contribution of the host galaxy to the total luminosity. This is necessary to estimate the contribution of the AGN to the SED in each band. Physically, one would expect that, while the AGN would be dominant in radio, optical, and X-ray bands, the host galaxy contribution would be non-negligible in IR and UV bands. To analyze this, we have simulated a pure quasar SED by manually setting the AGN fraction in the IR band, ${\rm frac}_{\rm AGN} = 0.999$, using the simulation parameters described in \citet{yang20}. The contribution of the dominant AGN components in the UV-IR range, the disk, and the torus, is shown in Fig.\  \ref{fig:AGN_Contribution_UV_IR} (top panel). We note that the disk emission dominates the optical range while the emission from the torus is predominant in the IR. However, if we consider the contribution from all wavelengths the contribution of the torus is insignificant compared to the disk emission. The relative fractional contribution of the host galaxy to the total flux density, considering the full-physics and the AGN-only simulations, is calculated
as: 
\begin{equation}
    f_{\rm galaxy} = 1 - \frac{F_{\lambda, {\rm AGN}}}{F_{\lambda, {\rm total}}}
    \label{eqn:f_galaxy}
\end{equation} 
where $F_{\lambda, {\rm AGN}}$ and $F_{\lambda, {\rm total}}$ are the flux densities derived from the AGN-only model (\texttt{skirtor2016}; \citealt{sta16}) and the total model (AGN + host galaxy components), respectively. The dependence of $f_{\rm galaxy}$ on $\lambda$ is shown in Figure \ref{fig:AGN_Contribution_UV_IR} (bottom panel). As seen in the figure we observe substantial contribution from the host galaxy in the SED. Further, we have also checked for changes in the host galaxy properties (from those obtained using all available bands) by systematically removing the X-ray and radio bands from our datasets. We have observed that there is little to no change in the host galaxy properties upon making these changes. 

\section{Discussion}
\label{subsec:discussion}
Supermassive black holes at the center of their host galaxies generate copious amounts of emission from the gravitational potential energy of the accreted material. They can be observed as intense radiation at different wavelengths, e.g., X-ray, UV, infrared, and radio, which is a characteristic signature of AGN. The energy released during the accretion process is also an important source of heating the interstellar and intergalactic medium \citep[e.g.,][]{Morganti17}. The observed correlations between host galaxy properties and the central engine (\textit{e.g.}, $M$--$\sigma$ relation; \citealt{Gebhardt2000, Peterson2008, FerrareseMerritt2000}) suggest a strong connection between the host galaxy and the central AGN. At a broader level quasar activity plays a role in galaxy evolution and, more generally, structure formation in the Universe \citep[e.g.,][]{B&A15}. A huge volume of work exists in the literature, which has probed this connection through spectroscopic studies as well as X-ray and radio imaging \citep[e.g.,][]{Berton2021, LeFevre2019, Giacintucci2019, Fischer2019}. In our current work, we construct the broad-band SEDs of our quasar samples and extract their host galaxy properties via the modeling of those. In particular, we try to address the radio dichotomy issue in quasars by identifying clues in their host galaxy properties, namely, star-formation and stellar mass. 

The main challenge of this work is in separating the quasar and the host galaxy contributions to the SED. As discussed in \S\ref{subsec:AGN_SED_simulation} we perform our feasibility study by simulating a quasar spectrum and extracting the host galaxy contribution from there (see Fig.\ \ref{fig:AGN_Contribution_UV_IR}). In addition, we performed an alternative analysis to check for consistency. Recently \cite{Jalan2023} used a large sample of SDSS DR14 quasars to derive an empirical relation between the host galaxy fraction (ratio of the stellar luminosity to the total continuum luminosity) with the total AGN luminosity and redshift, by extracting the host galaxy contribution with stellar templates \citep[\textit{e.g.},][]{Rakshit2018}, a power-law AGN continuum, iron line templates \citep[see, \textit{e.g.},][]{Barth2015, Kovacevik2010}, and fitting of emission lines. As our parent sample is SDSS DR7, we used these empirical relations both for redshift (although limited to $z \le 0.8$) as well as quasar luminosity. From both the relations, our derived host galaxy fraction is in the range from 20\% to 35\%. 

We have also derived the relative contribution of the host galaxy in the total SED by comparing the relative difference of the SED of a full-physics fitting (\textit{i.e.}, with all components included) with that of an AGN-only fitting where the AGN-fraction in the total luminosity had been set to unity. This relative contribution (of the host galaxy) is shown in Figure \ref{fig:AGN_Contribution_UV_IR}, for only the UV--IR range. We observe that the host galaxy contributes $\sim 50\%$ of the total luminosity at relatively lower wavelengths in this range, and almost the total luminosity at higher wavelengths (around the FIR range). Thus our results on host galaxy contribution appear to be consistent with the estimate of 20\% to 35\% when integrated over all wavelengths in our sample.


There have been several works on the extraction of host galaxy properties through SED modeling of AGN \citep[e.g.,][]{Mountrichas24, Gadallah23, Pouliasis22, Koutoulidis22, Zou22, Zhu23, CutivaAlvarez2023, Mountrichas21, Yamada2023}. Most of these work involve X-ray bright AGN while one study \citep{CutivaAlvarez2023} attempted modeling the quasar SED in the NIR/MIR range. Sources in their sample are well beyond our redshift ranges and they find that there is rapid growth (e-folding time: 750-1000 Myrs) of quasars while the SEDs are degenerate to quasar continuum and starbursts beyond $z > 1.6$. \citet{Mountrichas24} studied X-ray selected AGN spanning a large range of luminosity and redshift and find that the shallower slope of the $M_{\rm stellar}-{\rm SFR}$ relation in X-ray bright AGN host galaxies imply a smaller amount of star-formation in contrast to the galaxy main-sequence. The study shows a slope of $0.21 \pm 0.04$ for AGN-hosting galaxies and $0.48 \pm 0.01$ for non-AGN ones in the redshift range of $0.3 < z < 1.0$. Similar results are observed in the redshift range of $1.0 < z < 2.0$ where the slope for AGN and non-AGN galaxies are $0.50 \pm 0.04$ and $0.88 \pm 0.02$ respectively. As noted in \S\ref{subsec:Host_Galaxy_Properties} we observe similar slopes for our quasar host galaxies, however, the results vary once we include the HBL and {\it AstroSat} sources. \citet{Pouliasis22} perform a similar study with X-ray bright AGN and show that AGN which lie above or within the main-sequence have higher specific accretion rates compared to those below the main-sequence. 

In Figures \ref{fig:S15_noZ} and \ref{fig:MS_z} we observe a hint of dichotomy in the $M_{\star}-{\rm SFR}$ relation. As noted in \citet{chakraborty22}, the Eddington ratios of the sources in the HBL sample are the lowest and they also exhibit lower SFR. This correlation has been observed in previous studies, which have explored it using either the Eddington ratio or its proxy, the specific black hole accretion rate, across the general AGN population \citep[e.g.,][]{Torbaniuk21, Torbaniuk24}, different AGN types (such as Sy2, LINERS, composite, and Compton thick AGN; \citealt{Mountrichas24C}), and in relation to X-ray obscuration (e.g., \citealt{Georgantopoulos23, Mountrichas24C}). In Fig.\ \ref{fig:Eddington_ratio} we further plot the Eddington rates of our quasar sources as a function of their galaxy stellar mass. We observe that the RQ population does not exhibit any notable dichotomy, and it is observed that the Eddington ratios of the quasars follow an increasing trend with the galaxy stellar mass ({\it AstroSat} sources being a deviant). The situation changes with the RL population, where we do see a clear dichotomy (at $\log \lambda_{\rm Edd}$ =-1.5) in the Eddington ratios between the HBL and the {\it AstroSat} sources to that of the non-HBL sources.

We note that across stellar masses, the Eddington ratios remain relatively unchanged (with a slight negative slope). We thus propose that it is likely that the radio jets emanating from low-Eddington ratio systems inhibit star-formation in their host galaxies. However, we require further studies with larger samples to test the hypothesis. It is important to note that none of our HBL or {\it AstroSat} sources are detected in the X-rays implying a lower X-ray threshold for these sources. In the future we wish to perform a wider search, in particular with the eROSITA \citep{Brunner2022_eROSITA, Merloni2024_eROSITA} catalog to discuss the X-ray properties. As noted before, the main aim of this work is to find clues about the larger radio loud fraction for optical quasars with broader emission lines as observed in \citet{chakraborty22}. Although we do not observe any significant difference in host galaxy properties of RL and RQ quasars when seggregated in FWHM, we still observe differences in some host galaxy properties of our sources based on their radio emission. In future, we wish to perform a detailed analysis on the radio SED \citep{Dey22} of our RL sources with a wider range of data. 

\subsection{Summary}
\label{subsec:conclusion}
In the previous work of \cite{chakraborty22} and  \citet{chakraborty21}, the radio-dichotomy of broad-line quasars was studied by looking at their intrinsic differences. The current study aims to examine this dichotomy in the context of host galaxy properties being affected by the presence of radio jets in some systems. For this work, to analyze the host galaxy properties, we compiled a dataset comprising 37 RL and 19 RQ quasars ($0.15<z<1.88$) which have at least one broad emission line (either H$\beta$ or Mg\textsc{ii}) and modeled the SEDs over a wide range of wavelengths from the X-ray to the radio using \texttt{CIGALE v2022.0}. We perform one of the first studies in modeling the broad band, from X-rays to radio, quasar-host galaxy SEDs, classified by their radio emission. The main results of our investigation can be summarized as follows:

 In this work for the first time we construct the multi-wavelength broadband SED (ranging from X-rays to radio) of optically selected broad-line quasars. We further classify them as radio-loud and radio-quiet sources and derive their host galaxy properties by modeling their SEDs. In addition to that we extract the host galaxy properties of our high broad line RL and RQ quasars (FWHM > 15000 kms$^{-1}$) defined as the HBL sample in \citet{chakraborty22} and compare their host galaxy properties with the non-HBL sample. For six of our quasar sources, we obtain FUV data with \textit{AstroSat}/UVIT. We perform analysis of the \textit{AstroSat} observations and employ them for the first time for SED analysis using \texttt{CIGALE v2022.0}. We have introduced a box-car filter for fitting the \textit{AstroSat} FUV data.

 \begin{itemize}

\item We obtained host galaxy properties such as stellar mass, SFR, luminosity absorbed by dust, stellar population age and e-folding time from the SED modeling and we see that the mean values do not exhibit any difference between the RL and RQ populations.\\

\item To validate our results we compare the the best-fit model SED we obtained, input photometric data points from Shen et al. (2011), and the SDSS spectrum of the source from the same epoch. The best-fit model for the optical range is quite successful in reproducing multiple features in the observed spectrum by SDSS. We tested for the consistency between the Bayesian and best-fit values of host galaxy parameters from \texttt{CIGALE v2022.0} using the prescription of \cite{Mountrichas21} and found that 54\% of the RL and 75\% of the RQ satisfy the prescription. \\

\item To ensure the fidelity of the extraction of our host galaxy properties, we simulated a pure AGN SED and compared that with the full SED. Our results show that in the UV-IR range about 50-100\% of the emission can be reconstructed from the host galaxy which is adequate for extracting properties such as stellar mass and SFR. We further use the results from \cite{Jalan2023} to extract the host galaxy fraction and our results show that roughly 20-30\% of the emission in our sources is due to their host galaxies. The result is consistent with our AGN SED simulation when averaged over all wavelengths. \\

\item We obtain the main-sequence relations for our quasars and find that they are way apart from the galaxy main-sequence and tend to follow the high redshift ($z \sim 3$, in Fig.\ \ref{fig:S15_noZ}) main-sequence relation for star-forming galaxies. We note a dichotomy in the stellar mass -SFR relation for our RL sources indicating a quenching of star-formation. When compared with the Eddington ratios of our sources we see that there is some connection between quenched star-formation with the accretion activity of the central engine implying a stronger dichotomy in the Eddington ratio-SFR results. We further note that the redshift of our RL and RQ sources does not have any effect on the observed main-sequence relations. We propose to study further host galaxy correlations in a follow-up study (Chatterjee et al. 2024, in prep). 
 \end{itemize} 

 The radio dichotomy problem in quasars is an outstanding question in extra-galactic astronomy. Our previous work \citet{chakraborty21b, chakraborty22} revealed that the radio loud fraction increases substantially when quasars are classified based on the width of their emission line. In this study, we perform a spectral energy distribution modeling of the host galaxies of our broad line quasars to check for differences between the radio-loud and the radio-quiet populations, as well as the population sampled over their broad line width. Our analysis shows that although the mean values of the host galaxy parameters of our RL and RQ quasars tend to be similar, we see a difference in the main-sequence relation of our quasars indicating a difference in the dynamics of the host galaxy modulated by the presence of radio jets in the central engines. We propose to undertake a similar study with larger statistics in the future to have a more conclusive understanding of these effects.

\section*{Acknowledgements}
The authors wish to thank the anonymous referee whose comments and suggestions have greatly helped in improving the draft. AC wants to thank Evangelos Dimitrios Paspaliaris of INAF – Osservatorio Astrofisico di Arcetri for all his valuable suggestions. SC acknowledges financial support from SERB through the POWER Fellowship (SPF/2022/000084) and the CRG/2020/002064 grant. AC and SC thank Veronique Buat, Denis Burgarella, and Guang Yang for extensive discussion and help with the CIGALE package. SP is supported by the international Gemini Observatory, a program of NSF NOIRLab, which is managed by the Association of Universities for Research in Astronomy (AURA) under a cooperative agreement with the U.S. National Science Foundation, on behalf of the Gemini partnership of Argentina, Brazil, Canada, Chile, the Republic of Korea, and the United States of America. SP also acknowledges the financial support of the Conselho Nacional de Desenvolvimento Científico e Tecnológico (CNPq) Fellowships 300936/2023-0 and 301628/2024-6. MK acknowledges financial support from SERB through the POWER Fellowship (SPF/2022/000084). SJ wants to thank Joseph E. Postma, UVIT calibration manager, for all his help with the CCDLAB Pipeline. RC thanks ISRO for support under the AstroSat archival data utilization program, SERB for a SURE grant, and Presidency University for support under the Faculty Research and Professional Development Fund (FRPDF). SC and RC acknowledge IUCAA for their hospitality and usage of their facilities through the university associateship program. This work uses data from UVIT onboard the AstroSat mission of the Indian Space Research Organisation (ISRO), archived at the Indian Space Science Data Centre (ISSDC). We thank the payload operations center by the Indian Institute of Astrophysics (IIA) for the initial processing of the UVIT data used in this work.

\bibliographystyle{aa}
\bibliography{SED}




\end{document}